\def\boldmu    {\mbox{\boldmath$\mu$}}
\def\boldomega {\mbox{\boldmath$\omega$}}
\def\boldsigma {\mbox{\boldmath$\sigma$}}
\def\boldvarrho{\mbox{\boldmath$\varrho$}}

\def\Nabla{\mbox{\boldmath$\nabla$}}
\def\boldomega{\mbox{\boldmath$\omega$}}
\def\qquad{\hspace{30pt}}
\def\0{\newline}
\def\1{\newline}
\def\2{\par\noindent\newline}

\def\loq{,\kern-0.080em,\kern+0.05em}
\def\hiq{\lq\lq}
\def\bfloq{{\bf,\kern-0.06em,\kern+0.05em}}

\def\footloq{,\kern-0.07em,\kern+0.03em}

\def\quabla{{\raise.7ex\hbox{\boxed{{}}}}}

\def\intii  {\int\!\!\int}

\def\be{\begin{equation}}
\def\ee{\end{equation}}
\def\bea{\begin{eqnarray}}
\def\eea{\end{eqnarray}}
\newcommand{\du}[2]{#1\,\mbox{#2}}     
\newcommand{\sbm}[2]{#1_{#2}}          
\def\ver  {{\it Version 2.01 (30.\,06.\,97)\/}}
\def\bild#1#2#3{
\begin{figure}[htbp]
\epsfxsize=#3cm
\begin{center}
\leavevmode
\epsffile{#1.eps}
\end{center}
\caption[]{#2}
\label{#1}
\end{figure}
}
 \documentstyle[12pt,german,epsf]{article}     
\renewcommand{\baselinestretch}{1.50}
\title{%
{\LARGE\bf Der quantisierte Hall-Effekt}                  \\[-05pt]
{\Large\bf (The Quantized Hall Effect) }                  \\[ 60pt]
      }
\author{
{\it Anleitung zum F-Praktikum: Versuch 504\/}            \\[ 00pt]
{\it (Lab Tutorial. In German.)            \/}            \\[ 00pt]
\ver                                                      \\[ 60pt]
R{\small ALF} D.\ T{\small SCHEUSCHNER}%
\thanks{{\it permanent email address:\/} ralfd@provi.de}  \\[ 00pt]
Angewandte Festk\"orperphysik                             \\[ 00pt]
Ruhr-Universit\"at Bochum                                 \\[ 00pt]
Universit\"atsstra\ss e 150                               \\[ 00pt]
D-44780 Bochum                                            \\[ 60pt]
}
\begin{document}
\maketitle                    
%
\pagebreak 
\vspace*{16cm}\noindent%
Dieses Tutorial ist die vorl\"aufig letzte Version
einer bereits im praktischen Einsatz erprob\-ten
Praktikumsanleitung f\"ur das Experiment des quantisierten
{\sc Hall}-Effekts.
Der Autor hat die Absicht, an diesem Skript keine inhaltlichen
Erg\"anzungen mehr vorzunehmen, w\"a\-re aber dankbar f\"ur
Hinweise auf Druckfehler und sachliche Fehler,
die weiterhin kor\-ri\-giert werden.
Der Autor dankt allen Kommiliton\,{\it inn\/}\,en
f\"ur Anregung und Kritik.
\\
\\
Bochum, den 30.\,06.\,97
%
\pagebreak 
\begin{abstract}
Die Physik niederdimensionaler Elektronensysteme
in Halbleiter-Schichtstruktu\-ren ist ein besonders
aktuelles Arbeitsgebiet der modernen Festk\"orperphysik -
sowohl im Hinblick auf die Grundlagenforschung als auch
im Hinblick auf zuk\"unftige technologische Anwendungen.
\par
In dem vorliegenden Versuch untersuchen wir Systeme
von untereinander als nicht-wechselwirkend angenommenen
Elektronen in zwei Raumdimensionen, sogenannte
2-dimensionale Elektronengase (2DEG).
Diese k\"onnen an Halbleiter-Grenz\-fl\"achen,
so zum Beipiel in einer {\it Metall-Oxid-Halbleiter-Struktur (MOS)\/}
oder an einer Grenzfl\"ache zwischen Halbleitern verschiedener Bandl\"ucke,
einer sogenannten {\it He\-te\-ro\-struk\-tur\/}
(griech.\ {\it heteros\/} = der andere von beiden,
verschieden, anders beschaffen), realisiert werden.
Im ersten Fall ist es die extern angelegte Gatespannung,
im zweiten Fall die geeignete Dotierung
zusammen mit einem Band\-l\"ucken\-sprung,
die eine Bandverbiegung verursacht, welche
zu einem in erster N\"aherung dreiecksf\"ormigen Potentialtopf
f\"uhrt.
Durch die endliche Ausdehnung
des elektronischen Systems in Richtung der dritten Dimension
zeichnen sich die Elektronen (oder L\"ocher) durch quantisierte
Energieniveaux in einer Richtung aus.
Ist nur das unterste Subband besetzt, so haben wir ein
physikalisches System vorliegen, welches sich durch eine
2-dimensionale quantenmechanische Kinematik auszeichnet.
\par
Ein besonderer Impuls in der Erforschung des 2-dimensionalen
Elektronengases ging von der Entdeckung des quantisierten
{\sc Hall}-Effekts aus.
{\sc Von\,Klitzing} fand 1980 in einem MOS-System, da\ss\ bei
gen\"ugend tiefen Temperaturen und in hohen Magnetfeldern der
{\sc Hall}-Widerstand des 2-dimensionalen Elektronengases auf
diskrete durch Naturkonstanten gegebene Plateaux quantisiert ist
($h/ie^2$, mit $h$ = {\sc Planck}sches Wirkungsquantum,
               $e$ = elektrische Elementarladung,
               $i$ = 1,2,3 $\dots$).
F\"ur seine sensationelle Entdeckung erhielt
{\sc von\,Klitzing} 1985 den Nobelpreis.
Die heutige offizielle Definition des
Widerstandsnormals basiert auf diesem Effekt.
\par
Im vorliegenden Praktikumsversuch soll dieses Nobelpreisexperiment
wiederholt werden; allerdings verwenden wir als Materialsystem
eine Al$_x$Ga$_{1-x}$As-GaAs-He\-te\-ro\-struk\-tur. An dieser untersuchen
wir den elektrischen Transport bei niedrigen Temperaturen und
hohen Magnetfeldern. Die Aufgabe besteht in der Realisierung
von klassischem und quantisiertem {\sc Hall}-Effekt, sowie in
der Beobachtung von {\sc Shub\-ni\-kov}-{\sc de\,Haas}-Oszillationen
und ihrem \"Ubergang ins Quanten-{\sc Hall}-Regime. Dazu wird
es notwendig sein, sich mit den Grundlagen der Kryotechnik,
mit der Magnetsteuerung und mit me\ss technischen Fragen
(insbesondere Lock-In-Technik) vertraut zu machen.
Ziel ist die Bestimmung der {\sc von\,Klitzing}-Konstante
$h/e^2$ und ihr Vergleich mit der {\sc Sommerfeld}schen
Feinstrukturkonstante $\alpha=e^2/4\pi\varepsilon_0\hbar c$.
\end{abstract}
\pagebreak 
%
\pagestyle{myheadings}        
%
%
\tableofcontents              
%
%
\newpage
\section{Vorwort}
\markboth{{\rm Der quantisierte {\sc Hall}-Effekt.} \ver}
         {{\rm Der quantisierte {\sc Hall}-Effekt.} \ver}
Die experimentelle Entdeckung des quantisierten {\sc Hall}-Effekts
durch {\sc von\,Klit\-zing},%
\linebreak
{\sc Dor\-da} und {\sc Pepper} besteht in der
{\it Beobachtung von Plateaux\/} im {\sc Hall}-Widerstand
\begin{equation}
R_H=U_H/I
\end{equation}
von sogenannten MOS-({\it metal-oxide-semiconductor-\/})\,Strukturen
bei niedrigen Temperaturen und hohen Magnetfeldern.
\par
{\sc Von\,Klitzing} sah als erster, da\ss\ die {\it exakte\/} H\"ohe
dieser Plateaux gegeben ist durch ganzzahlige Bruchteile von $h/e^2$,
\begin{equation}
R_H = \frac{h}{ie^2},
\phantom{xxx}
i=1,2,3,\,{\dots}
\phantom{xxx},
\end{equation}
wobei - wie \"ublich - $h$ das {\sc Planck}sche Wirkungsquantum
und $e$ die elektrische Elementarladung bezeichnen.
F\"ur seine sensationelle Entdeckung erhielt
{\sc von\,Klitzing} 1985 den Nobelpreis.
Die heutige offizielle Definition des
Widerstandsnormals basiert auf diesem Effekt.
\par
Die Originalarbeit ist zu finden unter
\cite{Klitzing80}.
\bild{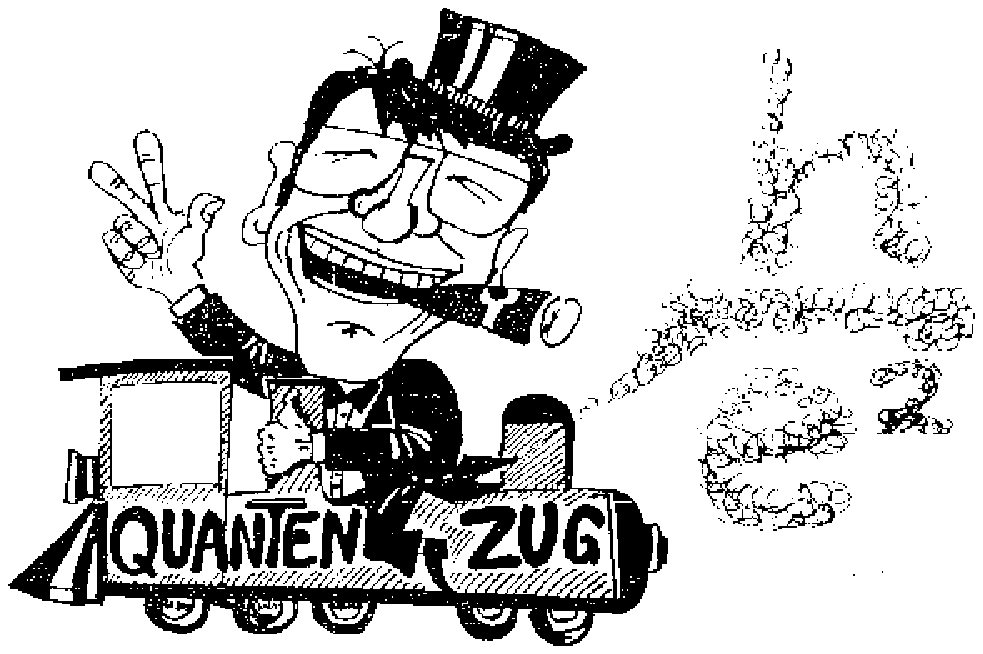}{Der Entdecker des QHE:
                {\sc Klaus von Klitzing}
                \cite{Landwehr90, Karikatur}}{10}
\par
Im vorliegenden Praktikumsversuch wollen wir ein \"ahnliches
Experiment an einer sogenannten Heterostruktur durchf\"uhren.
Dieses Skript soll die f\"ur den Versuch notwendigen Kenntnisse
vermitteln. Es ist so geschrieben, da\ss\ man es in einem Zug
durchlesen und auch verstehen kann, vorausgesetzt, der Leser
bringt eine Reihe von Vorkenntnissen mit, die zum Stoff des
Hauptstudiums Physik geh\"oren:
\begin{itemize}
\item
Klassische Elektrodynamik
(Vektorpotential und lokale Eichinvarianz),
\item
Quantenmechanik
(Quantisierung, Harmonischer Oszillator,
Dichteoperator, Erwartungswert),
\item
Thermodynamik
(chemisches Potential),
\item
Festk\"orperphysik
(Bandstruktur, Halbleiter),
\item
Me\ss technik,
insbesondere Lock-In-Verst\"arker-Technik.
\end{itemize}
\par
Selbstverst\"andlich kann man diesen Versuch auch gleich nach dem Vordiplom
durchziehen, wenn man bereit ist, sich einige Dinge autodidaktisch
beizubringen. Dazu gehe man den Text durch und streiche
diejenigen Begriffe an, deren Bedeutung nicht klar ist.
Ein gutes Physiklexikon
und ein Nachmittag in der Bibliothek helfen meist schon weiter.
Trotzdem habe ich mich bem\"uht, diese Anleitung so
{\it self-contained\/} wie m\"oglich zu halten.
In ihr ist eigentlich alles zu finden,
was zum Verst\"andnis der grundlegenden
Prinzipien ben\"otigt wird.
({\it Tip:\/} Der eilige Leser ben\"otigt zun\"achst
              nur die Kapitel 2.1.-2.3.)
\par
Zwei Dinge sollten die Studenten vor Beginn des Praktikums
erledigen, n\"amlich
\begin{enumerate}
\item
sich eine Kopie der Originalarbeit besorgen \cite{Klitzing80},
\item
sich eine Kopie der Seiten 3-1 und 3-2 aus der Anleitung
des Lock-In-Verst\"arkers SR850, erh\"altlich bei der
Technik der Arbeitsgruppe, aush\"andigen lassen \cite{Stanford}.
\end{enumerate}
\par
Viel Erfolg!
\vfill\eject\noindent%
\section{Theoretische Grundlagen} 
\subsection{Widerstand und Leitf\"ahigkeit}
Nach dem wohlbekannten {\sc Ohm}schen Gesetz
ist der durch einen metallischen Leiter konstanter Temperatur
flie\ss ende elektrische Strom $I$
proportional zur angelegten Spannung $U$
\begin{equation}
I \propto U.
\end{equation}
Ausgedr\"uckt durch den absoluten Widerstand $R$
(gemessen in Ohm $\Omega=V/A$)
bzw.\ durch den absoluten Leitwert $\Gamma$
(gemessen in Siemens $S=A/V$)
k\"onnen wir schreiben
\begin{equation}
I \cdot R = U
\phantom{1234}
\mbox{bzw.}
\phantom{1234}
I = \Gamma \cdot U.
\end{equation}
Im Labor haben wir es \"ublicherweise
mit Proben zu tun, die bestimmte Abmessungen
haben; die Naturgesetze sollten aber
in Termen von {\it Invarianten\/}
ausgedr\"uckt werden.%
\footnote{Dieses entspricht dem Geist der Werke von
          {\sc Helmholtz}, {\sc van\,der\,Waerden} und {\sc Weyl}.}
Diese Forderung legt uns nahe,
die obigen Beziehungen unter Zuhilfenahme
{\it spezifischer Gr\"o\ss en\/},
das hei\ss t abmessungsunabh\"angiger Gr\"o\ss en,
zu formulieren. R\"aumliche Isotropie vorausgesetzt,
schreiben wir
\begin{equation}
{\bf j}\,\varrho = \varrho\,{\bf j} = {\bf E}
\phantom{1234}
\mbox{bzw.}
\phantom{1234}
{\bf j} = \sigma\,{\bf E} \, ,
\end{equation}
wobei die Stromdichte gegeben ist durch
\begin{equation}
{\bf j} = \frac{I}{A} \cdot {\bf u},
\end{equation}
mit $A$ als Betrag der durchflossenen Fl\"ache
und ${\bf u}$ als zugeordnetem Normalenvektor
der L\"ange $1$
(engl.\ {\it unit vector\/}).
$\varrho$ nennen wir den {\it spezifischen Widerstand\/}
und
$\sigma$ die {\it spezifische Leitf\"ahigkeit\/}.
Verstehen wir diese beiden Gr\"o\ss en als skalare Gr\"o\ss en,
so k\"onnen wir setzen
\begin{equation}
\varrho=\frac{R \cdot A}{l}
\phantom{1234}
\mbox{bzw.}
\phantom{1234}
\sigma=\frac{l}{R \cdot A}.
\end{equation}
Im allgemeinen Fall, in dem die Proben sich
in elektromagnetischer Hinsicht anisotrop
verhalten
(zum Beispiel unter Einflu\ss\ eines homogenen
Magnetfeldes oder der Symmetrie-Eigenschaften
der Kristallstruktur),
{\it m\"ussen\/} wir $\varrho$ und $\sigma$
als Matrixoperatoren (oder Tensoren) auffassen.
Wir schreiben sie daher (wie Vektoren) fett:
\begin{equation}
{\bf E} = \boldvarrho \,{\bf j}
\phantom{1234}
\mbox{bzw.}
\phantom{1234}
{\bf j} = \boldsigma  \,{\bf E}.
\end{equation}
Nun hat in $D$ Raumdimensionen der spezifische
Widerstand - definiert als Widerstand mal durchflossene
(im allgemeinen $D$-$1$-dimensionale) Querschnittsfl\"ache
pro (im allgemeinen 1-dimensionale) L\"ange -
die physikalische Dimension
\begin{quote}
\loq Widerstand mal L\"ange hoch $D$$-$$2$\hiq\, ,
\end{quote}
somit der spezifische Leitwert
die physikalische Dimension
\begin{quote}
\loq L\"ange hoch $D$$-$$2$ durch Widerstand\hiq.
\end{quote}
Das bedeutet, da\ss\
in einem idealisierten 2-dimensionalen System,
in der sich der Strom l\"angs einer 2-dimensionalen
Fl\"ache bewegt und die durchflossenen Querschnitte
$1$-di\-men\-si\-o\-nal sind, der spezifische Widerstand
die Einheit eines absoluten Widerstandes,
die spezifische Leitf\"ahigkeit die Einheit
eines absoluten Leitwerts besitzt.
\par
Letzteres allein k\"onnte schon die Vermutung
nahelegen, da\ss\ es ein Leit\-f\"a\-hig\-keits\-ex\-pe\-ri\-ment
an einer quasi-2-dimensionalen Probe
geben k\"onnte, das die {\it spezifischen\/} Gr\"o\ss en
{\it ohne\/} Bezugnahme auf die
{\it tats\"achlichen Abmessungen\/}
der Probe mi\ss t!
\par
Ein weiterer interessanter Aspekt ergibt sich
aus einer einfachen Dimensionsanalyse, die sogar
mit nur rudiment\"aren Kenntnissen quantenmechanischer
Prinzipien erfolgen kann. Einerseits gilt n\"amlich
f\"ur die physikalische Dimension (phys.\,dim.) des
Widerstandes
\begin{eqnarray}
\mbox{phys.\,dim.}\, R &=& \mbox{phys.\,dim.}\, \frac{U}{I}      \nonumber\\
                & & \phantom{+}                                  \nonumber\\
                &=& \frac{\mbox{{\it Spannung\/}}}
                        {\mbox{{\it Stromst\"arke\/}}}
                                                                 \nonumber\\
                & & \phantom{+}                                  \nonumber\\
                &=& \frac{\mbox{{\it Energie pro Ladung\/}}}
                        {\mbox{{\it Ladung pro Zeit\/}}}         \nonumber\\
                & & \phantom{+}                                  \nonumber\\
                &=& \frac{\mbox{{\it Energie mal Zeit\/}}}
                        {\mbox{{\it Ladung zum Quadrat\/}}}      \nonumber\\
                & & \phantom{+}                                  \nonumber\\
                &=& \frac{\mbox{{\it Wirkung\/}}}
                        {\mbox{{\it Ladung zum Quadrat\/}}} \, ;
\end{eqnarray}
andererseits d\"urfen wir zumindestens formal schreiben
\begin{eqnarray}
\mbox{phys.\,dim.}\, R &=& \frac{\mbox{{\it Wirkung\/}}}
                        {\mbox{{\it Ladung zum Quadrat\/}}}      \nonumber\\
                & & \phantom{+}                                   \nonumber\\
                &=& \frac{\mbox{{\it Quantenzahl\/}}}
                        {\mbox{{\it Quantenzahl
                                    zum Quadrat\/}}}
                    \cdot
                    \frac{\mbox{{\it Wirkungsquantum\/}}}
                        {\mbox{{\it Elementarquantum
                                    zum Quadrat\/}}}            \nonumber\\
                & & \phantom{+}                                  \nonumber\\
                &=& \mbox{{\it rationale Zahl\/}}
                    \cdot
                    \frac{h}{e^2},
\end{eqnarray}
wobei
\begin{equation}
\frac{h}{e^2}=25.812\,805\,{\dots}{\rm k}\Omega.
\end{equation}
\par
Allein aus den soeben plausibel gemachten Konzepten
h\"atte man die Spekulation rechtfertigen k\"onnen,
da\ss\ es m\"oglicherweise ein
idealisiertes 2-dimensionales Experiment geben k\"onnte,
in dem Quanteneffekte eine wesentliche Rolle spielen
(also zum Beispiel bei tiefen Temperaturen)
und in dem unabh\"angig von den Abmessungen der Probe
rational quantisierte Widerst\"ande bzw.\
rational oder gar {\it integral quantisierte\/}
Leitf\"ahigkeiten - in Einheiten von $e^2/h$ -
beobachtbar sind. Und vielleicht h\"atte es gar nicht
so fern gelegen, sich vorzustellen, dies k\"onnte
ein {\sc Hall}-Experiment sein.
Schon {\sc Sommerfeld} und {\sc Bethe} spekulierten 1933,
also viele Jahre vor {\sc von\,Klitzing}s Entdeckung,
in Ihrem klassischen Lehrbuch \"uber den Einflu\ss\
der Quantisierung der Elektronenbahnen auf das Verhalten
des (longitudinalen) Magnetowiderstandes hinreichend kalter
Proben \cite{SommerfeldBethe1933}.
\bild{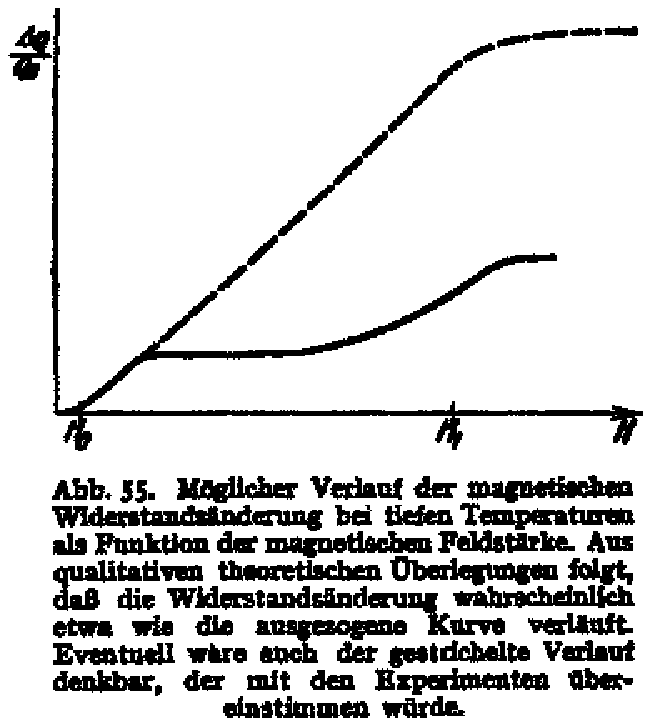}{Vermutetes Verhalten
                nach {\sc Sommerfeld} und {\sc Bethe} 1933
                \cite{SommerfeldBethe1933}}{8}
\par
Noch aufregender sind die Daten von
{\sc Kawaji} {\it et al.\/} aus dem Jahre 1975
\cite{Igarashi75}, die bereits die wesentliche
Struktur sichtbar machen.
\bild{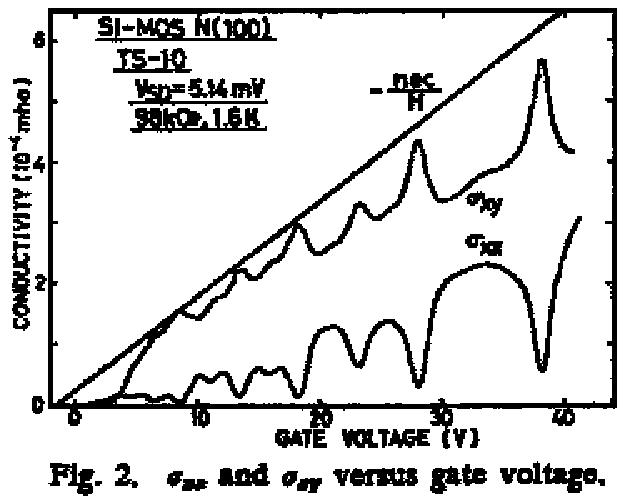}{Die Daten von {\sc Kawaji} {\it et al.\/}
                aus dem Jahre 1975}{8}
Es bedurfte aber einer genialen Interpretation dieser
Stufen, n\"amlich in Termen fundamentaler Naturkonstanten,
und dem Schlu\ss, da\ss\ die Quantisierung {\it exakt\/}
ist, um aus dem Ph\"a\-no\-men eine bahnbrechende Entdeckung
zu machen.
\vfill\eject\noindent%
\subsection{Der klassische Hall-Effekt}
Der klassische {\sc Hall}-Effekt beschreibt
die Wirkung eines Magnetfeldes ${\bf B}$
auf einen elektrischen Strom ${\bf j}$
in einer leitenden Probe mit einer
geeignet gew\"ahlten Geometrie:
Liege der rechteckige leitende Streifen
mit den Abmessungen
$L_x,L_y,L_z$ in x-, y-, z-Richtung
in der xy-Ebene und flie\ss e der Strom in x-Rchtung.
Die Einwirkung eines Magnetfeldes in z-Richtung
f\"uhrt zum Abfall einer Spannung in y-Richtung,
die wir zwischen der oberen und unteren Kante
der Probe abgreifen k\"onnen.
Eine Umkehrung der Stromrichtung oder der Richtung des
senkrecht zur xy-Ebene ausgerichteten Magnetfeldes
\"andert die Polarit\"at der beobachteten Spannung.
\bild{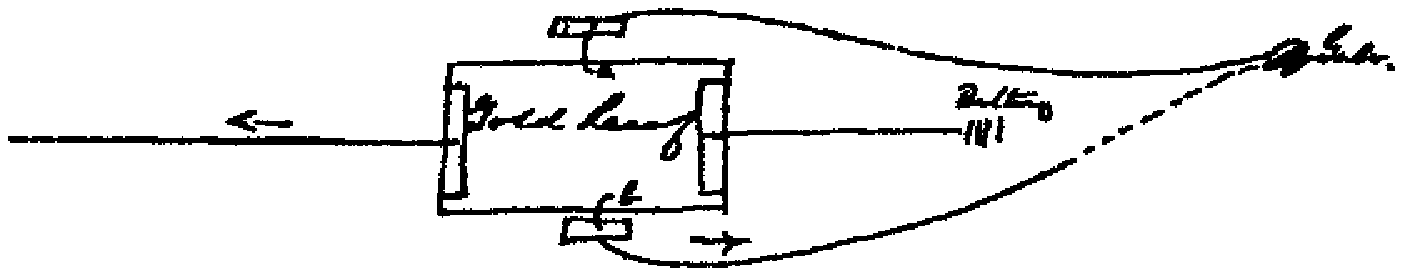}{Das klassische Experiment nach {\sc Hall} \cite{Hall}}{15}
\par
In mathematischen Termen: Sei also
\begin{eqnarray}
{\bf j} &=& (j_x,0,0), \\
{\bf E} &=& (0,E_y,0),
\end{eqnarray}
sowie
\begin{eqnarray}
{\bf B} = (0,0,B_z). 
\end{eqnarray}
Im klassischen Regime ist der {\it longitudinale Magnetowiderstand\/}
\begin{equation}
\varrho_{xx} ({\bf B}) = \frac{E_x}{j_x}
\end{equation}
vom Magnetfeld unabh\"angig,
der {\it transversale Magnetowiderstand\/} 
\begin{equation} 
\varrho_{yx} ({\bf B}) = \frac{E_y}{j_x}
\end{equation}
proportional zum angewendeten Magnetfeld,
so da\ss\ es Sinn macht, die sogenannte {\it Hall-Konstante\/}
\begin{equation}
R_{\mbox{\it Hall}} = \frac{R_H}{B} = \frac{E_y}{j_x\,B}
\end{equation}
zu definieren.
\bild{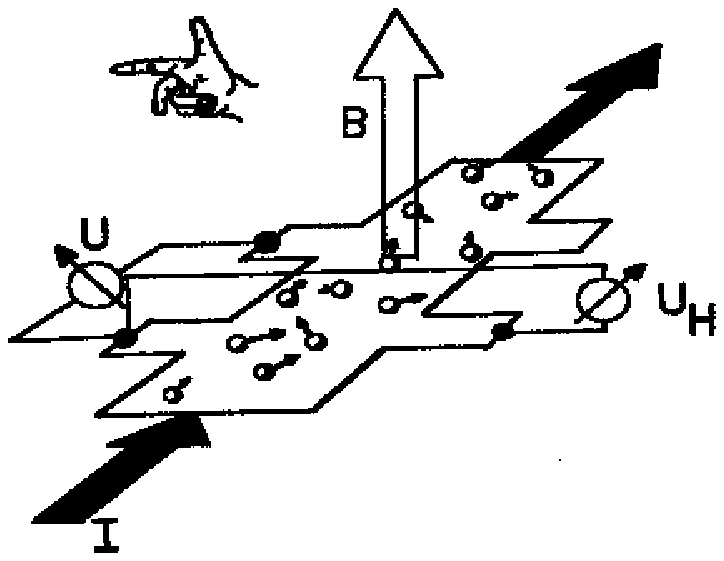}{Typische {\sc Hall}-Geometrie}{8}
\par
Die Ursache dieses sogenannten {\sc Hall}-Effektes liegt
nat\"urlich in der {\sc Lorentz}-Kraft, welche auf die
durch den Festk\"orper sich bewegenden Ladungstr\"ager
wirkt.
Aus ihrer Kenntnis k\"onnen wir das Verh\"altnis
von beobachteter {\sc Hall}-Spannung
zu dem in x-Rich\-tung flie\ss enden Strom,
den sogenannten {\sc Hall}-Widerstand, herleiten.
Diese Vorstellung kann bereits im Rahmen
des relativ einfachen Einteilchen-Bildes entwickelt werden.
\par
Sei ${\bf v}$ die Geschwindigkeit der Elektronen.
Mit $e$ als Elementarladung schreibt sich die
bekannte, universell g\"ultige hydrodynamische Beziehung,
welche die Stromdichte eines Transportph\"anomens
in Relation zu Teilchendichte und -geschwindigkeit setzt,
f\"ur einen Strom von Elektronen {\it in drei Raumdimensionen\/}
als
\begin{equation}
{\bf j}_{3D} = e\,n_{3D}\,{\bf v}.
\end{equation} 
$n_{3D}$ bezeichnet hier die $3D$-Dichte,
das hei\ss t, f\"ur den Fall unserer Probengeometrie ist
(mit $N_e$ als Anzahl der Ladungstr\"ager)
\begin{equation}
n_{3D} = \frac {N_e} { L_x \cdot L_y \cdot L_z }
\end{equation}
und somit 
\begin{equation}
{\bf j}_{3D}
       = e \cdot \frac { N_e } { L_x \cdot L_y \cdot L_z } \cdot {\bf v}.
\end{equation}
\par
{\it Bemerkung:\/} Um uns jetzt und in Zukunft Schreibarbeit zu ersparen,
verwenden wir {\it per conventionem\/} die sogenannte {\it technische
Stromrichtung\/}: Die relevanten Ladungstr\"ager flie\ss en vom
Pluspol zum Minuspol. Mit anderen Worten: Wenn wir nicht ausdr\"ucklich
etwas Gegenteiliges behaupten, tun wir so, als seien die Elektronen positiv
geladen.
\par
Nun entspricht der Betrag $v$ der Geschwindigkeit ${\bf v}$
eines Ladungstr\"agers in x-Richtung gerade dem Quotienten
$L_x/t$ aus der Probenl\"ange $L_x$ und dem Zeitintervall $t$,
die er braucht, um eben diese L\"ange zu durchlaufen, wir
haben also
\begin{equation}
j_{3D}
  =  e \cdot \frac { N_e \cdot v     } { L_x \cdot L_y \cdot L_z }
  =  e \cdot \frac { N_e \cdot L_x/t } { L_x \cdot L_y \cdot L_z }
  =  e \cdot \frac { N_e /t          } {           L_y \cdot L_z }
  =          \frac { I               } {           L_y \cdot L_z },
\end{equation}
wobei wir ausgenutzt haben, da\ss\ der Strom gerade die mit
der Elementarladung $e$ multiplizierte Anzahl $N_e/t$ von
Ladungstr\"agern ist, die pro Zeitintervall $t$ die Probe
durchlaufen haben.
\par
Fassen wir das System als r\"aumlich 2-dimensional auf, so ignorieren
wir die Ausdehnung in z-Richtung. In diesem Fall setzen wir einfach
\begin{equation}
{\bf j}_{2D} = e\,n_{2D}\,{\bf v}.
\end{equation}
Insbesondere gilt f\"ur die von uns gew\"ahlte Geometrie
\begin{equation}
j_{{2D},{x}}=\frac{I}{L_y}=e\,n_{2D}\,v_x.
\end{equation}
\par
Zur\"uck zum {\sc Hall}-Effekt: 
Einen station\"aren Zustand haben wir offensichtlich dann
vorliegen, wenn die {\sc Lorentz}-Kraft ${\bf F}_B$ und die
elektrostatische Kraft ${\bf F}_H$, welche proportional zum
Gradienten der {\sc Hall}-Spannung $U_H$ ist, sich gerade
ausgleichen:
\begin{equation}
{\bf F}_B = - {\bf F}_H
\end{equation}
mit
\begin{eqnarray}
{\bf F}_H &=& e\,  {\bf E}_H                      ,  \\
{\bf F}_B &=& e\, ({\bf v} \times {\bf B})        .
\end{eqnarray}
Einsetzen ergibt sofort
\begin{equation}
({\bf v} \times {\bf B})
=
- {\bf E}_H.
\end{equation}
Da in unserer Geometrie
\begin{eqnarray}
{\bf v} \times {\bf B}
    &=& (v_x, 0, 0) \times (0, 0, B_z) \;=\; (0, -v_x B_z, 0) , \\
{\bf E}_H &=& (0  , E_y, 0  ) \;=\; (0, U_H/L_y, 0) ,
\end{eqnarray}
erhalten wir
\begin{equation}
v_x\,B_z = E_y
\end{equation}
und mit
\begin{equation}
v_x = \frac {I}{L_y} \, \frac{1}{e\,n_{2D}}
\end{equation}
schlie\ss lich
\begin{equation}
\frac{I}{L_y} \, \frac{B_z}{e\,n_{2D}} = \frac{U_H}{L_y}.
\end{equation}
Der Ausdruck f\"ur die {\sc Hall}-Spannung lautet somit
\begin{equation}
U_H =  \frac{B_z}{e\,n_{2D}} \, I =: R_H I,
\end{equation}
das hei\ss t, es ist
\begin{equation}
R_H=\frac{B_z}{e\,n_{2D}}
\end{equation}
und somit
\begin{equation}
\sigma_H=\frac{e\,n_{2D}}{B_z}.
\end{equation}
\par
\par
Wie bereits dargestellt,
normiert man den {\sc Hall}-Widerstand $R_H$
auch auf das angelegte Magnetfeld gem\"a\ss\
\begin{equation}
R_{\mbox{\it Hall}} = \frac{R_H}{B} = \frac{1}{e\,n_{2D}}
\end{equation}
und definiert dadurch die schon erw\"ahnte {\sc Hall}-Konstante.
Als {\sc Edwin Herbert Hall} in den 80er Jahren des vorigen Jahrhunderts
seine Experimente durchf\"uhrte, bemerkte er, da\ss\ das
Vorzeichen der Ladung $e$ - je nach betrachtetem Material -
sowohl positiv als auch negativ sein konnte. Dies war der
erste Hinweis auf den Unterschied zwischen Elektronen- und
L\"ocher-Leitung.
Die im allgemeinen magnetfeldabh\"angige Beziehung
\begin{equation}
{\bf E} = \boldvarrho \,{\bf j},
\end{equation}
wobei $\boldvarrho$ als Tensorgr\"o\ss e (bzw.\
Matrixoperator) zu verstehen ist,
nennen Theoretiker das {\sc Ohm}-{\sc Hall}-Gesetz.
\vfill\eject\noindent%
\subsection{Eine m\"ogliche Quantisierung des Hall-Effekts}
Die klassische Beziehung
\begin{equation}
R_H=\frac{B_z}{en_{2D}}
\end{equation}
kann man nat\"urlich auch anders schreiben, n\"amlich
- unter Verwendung des {\sc Planck}schen Wirkungsquantums -
als
\begin{equation}
R_H=\frac{h}
         {\left(
          { \displaystyle\frac{hn_{2D}}{eB_z} }
          \right)
          e^2
         }.
\end{equation}
Nun erkennen wir in
\begin{equation}
\Phi_0=\frac{h}{e}
\end{equation}
das {\sc London}sche Flu\ss quantum wieder,
eine Gr\"o\ss e, die vielerlei Bedeutung hat (siehe unten),
die aber am einfachsten zu verstehen ist als die quantenmechanische
Realisierung des Konzepts der magnetischen Feldlinie.
Messen wir die St\"arke des Magnetfeldes $B_z$ in
Einheiten von $\Phi_0$, so z\"ahlen wir magnetische Feldlinien ab.
Ihre Anzahl ist gegeben durch
\begin{equation}
n_{\Phi_0} = \frac{B_z}{\Phi_0} = \frac{B_z}{h/e},
\end{equation}
so da\ss\
\begin{equation}
R_H=  \frac{h}
         {\left(
          { \displaystyle\frac{n_{2D}}{n_{\Phi_0}} }
          \right)
          e^2
         }
   =: \frac{h}{\nu e^2}.
\end{equation}
An dieser Stelle k\"onnen wir bereits den Schlu\ss\ ziehen:
Treten in einen idealisierten 2-dimensionalen System
nicht nur die elektrischen Ladungen, sondern auch die
magnetischen Flu\ss linien in Form von elementaren Quanten
auf, so ist der {\it F\"ullfaktor\/}
\begin{equation}
\nu = \frac{n_{2D}}{n_{\Phi_0}}
\end{equation}
eine ganze (bzw.\ rationale) Zahl. Der {\sc Hall}-Widerstand
erf\"ullt dann eine Quantisierungsregel gem\"a\ss\
\begin{equation}
R_H=\frac{h}{\nu e^2}.
\end{equation}
Die zentrale Frage ist, worin die mikroskopische Ursache
f\"ur ein solches Verhalten zu suchen ist.
Zur Diskussion dieser Frage legen wir unser Augenmerk
zun\"achst auf die Kinematik des fraglichen Systems.
\vfill\eject\noindent%
\subsection{Die Zyklotronbewegung der Elektronen}
Springen wir noch einmal zur\"uck zu der im station\"aren
Fall g\"ultigen Beziehung
\begin{equation}
{\bf v} \times {\bf B} = - {\bf E}_H,
\end{equation}
die wir aus Gr\"unden der Mnemonik - unter Weglassung des Index -
jetzt schreiben als
\begin{equation}
0 = e \, ( {\bf E} + {\bf v}\times{\bf B} ).
\end{equation}
Manchmal ist es n\"utzlich, zu einer etwas abstrakteren
und gleichzeitig eleganteren Formulierung \"uberzugehen.
Statt
\begin{equation}
v_x = \frac{1}{e\,n_{2D}} \, \frac{I} {L_y}
\end{equation}
schreiben wir jetzt
\begin{equation}
{\bf v} = \frac{1}{e\,n_{2D}} \cdot {\bf j}_{2D},
\end{equation}
so da\ss\ wir die Gleichung
\begin{equation}
0 = e\,n_{2D}\,U_H - I B_z
\end{equation}
in der Form
\begin{equation}
0 = e\,n_{2D} \cdot {\bf E} + {\bf j}_{2D} \times {\bf B}
\end{equation}
wiedererkennen. Ihre L\"osung kann - ebenso elegant -
geschrieben werden als
\begin{equation}
{\bf j}_{2D}
        = e\,n_{2D} \cdot
          \frac{ {\bf B} \times {\bf E} }{ {\bf B}^2 } ,
\end{equation}
was f\"ur die Absolutbetr\"age impliziert
\begin{equation}
| {\bf j}_{2D} |
    =   \frac{e\,n_{2D}}{B} \, | {\bf E} |
    =   R_H \, | {\bf E} |,
\phantom{xxx} {\bf j}_{2D} \perp {\bf E}.
\end{equation}
\par
Gehen wir jetzt \"uber die station\"are N\"aherung hinaus
und betrachten die volle klassische Dynamik unter Einbeziehung
des Kreisens der Elektronen im Magnetfeld, der sogenannten
{\it Zyklotronbewegung\/}. Ihre physikalische Ursache ist
wieder - die {\sc Lorentz}-Kraft.
\par
Erinnern wir uns an das, was wir in der Vorlesung
\"uber Theoretische Mechanik gelernt haben
\cite{Goldstein80}:
Die Gleichung f\"ur die {\sc Lorentz}-Kraft kann
hergeleitet werden aus der {\sc La\-gran\-ge}-Funktion
\begin{equation}
L({\bf x},\dot{\bf x},t)=
T-U=
\frac{m}{2}\dot{\bf x}^2
+
e\,
{\bf A}({\bf x},t)\cdot\dot{\bf x}
-
e\,
V({\bf x},t)
\end{equation}
vermittels Anwendung der {\sc Lagrange}-Gleichungen
\begin{equation}
\frac{d}{dt}
\frac{\partial L}{\partial\dot x_i}
-
\frac{\partial L}{\partial     x_i}
=
0.
\end{equation}
Man zeigt dieses, indem man die totale Zeitableitung
unter Anwendung der Kettenregel in eine Summe partieller
Ableitungen entwickelt
\begin{equation}
\frac{d}{dt} =
\dot  x \, \frac{\partial}{\partial      x} +
\ddot x \, \frac{\partial}{\partial \dot x} +
           \frac{\partial}{\partial      t}.
\end{equation}
und den resultierenden Ausdruck
in die {\sc Lagrange}-Gleichungen
einsetzt
\begin{equation}
\frac{\partial^2 L}{\partial\dot x_j\partial\dot x_i}\ddot x_j
+
\frac{\partial^2 L}{\partial     x_j\partial\dot x_i} \dot x_j
+
\frac{\partial}{\partial t}
\frac{\partial L}{\partial\dot x_i}
-
\frac{\partial L}{\partial     x_i}
=
0.
\end{equation}
Wir erhalten
\begin{equation}
m\ddot x_i
+
e\,
\frac{\partial A_i}{\partial x_j}\dot x_j
+
e\,
\frac{\partial A_i}{\partial t}
+
\left(
e\,
\frac{\partial V}{\partial    x_i}
-
e\,
\frac{\partial A_j}{\partial x_i}\dot x_j
\right)
=
0
\end{equation}
bzw.\
\begin{eqnarray}
m\ddot x_i
&=&
e\,
\left(
-
\frac{\partial V}{\partial x_i}
-
\frac{\partial A_i}{\partial t}
\right)
+
e\,\dot x_j\,
\left( \frac{\partial A_j}{\partial x_i}
     - \frac{\partial A_i}{\partial x_j} \right)                    \nonumber\\
&=&
e\,
\left(
-
\frac{\partial V}{\partial x_i}
-
\frac{\partial A_i}{\partial t}
\right)
+
e\,\dot x_j\,
\left( ( \delta_{il}\delta_{jm} - \delta_{jl}\delta_{im} ) \,
       \frac{\partial A_m}{\partial x_l} \right)                    \nonumber\\
&=&
e\,
\left(
-
\frac{\partial V}{\partial x_i}
-
\frac{\partial A_i}{\partial t}
\right)
+
e\,\dot x_j\,
\left( ( \varepsilon_{ijk}\varepsilon_{klm} ) \,
       \frac{\partial A_m}{\partial x_l} \right)                    \nonumber\\
&=&
e\,
\left(
-
\frac{\partial V}{\partial x_i}
-
\frac{\partial A_i}{\partial t}
\right)
+
e\,\varepsilon_{ijk}\dot x_j\,
\left( ( \varepsilon_{klm} ) \,
       \frac{\partial A_m}{\partial x_l} \right).
\end{eqnarray}
Dabei haben wir die {\sc Einstein}sche Summenkonvention verwendet,
welche annimmt, da\ss\ grunds\"atzlich \"uber gleiche Indizes
summiert wird.
\par
$\varepsilon_{klm}$ ist der total antisymmetrische
oder {\sc Levy}-{\sc Civita}-Tensor in 3D.%
\footnote{%
$\varepsilon_{123}= 1$,
$\varepsilon_{231}=-1$
und zyklische Vertauschungen $\dots$ alle anderen gleich $0$}
Er erf\"ullt die von uns soeben ausgenutzte Identit\"at
\begin{equation}
\varepsilon_{ijk}\varepsilon_{klm}
=
\delta_{il}\delta_{jm}-\delta_{jl}\delta_{im}
\end{equation}
und definiert das bekannte Vektor- oder {\sc Gibbs}-Produkt durch
\begin{equation}
( {\bf a} \times {\bf b} )_k
=
\varepsilon_{klm} a_l b_m.
\end{equation}
\par
In einer etwas gel\"aufigeren Schreibweise erkennen wir
in der obigen Bewegungsgleichung diejenige f\"ur
ein Teilchen wieder, welches sich unter dem Einflu\ss\
eine {\sc Lorentz}-Kraft bewegt
\begin{eqnarray}
m \dot {\bf v}
&=& e \, ( {\bf E} + {\bf v} \times ( {\bf \Nabla} \times {\bf A} )) \nonumber\\
&=& e \, ( {\bf E} + {\bf v} \times   {\bf B}                      ) ,
\end{eqnarray}
wobei
\begin{equation}
{\bf v} = \dot {\bf x}.
\end{equation}
Es sei daran erinnert, da\ss,
obwohl die homogenen {\sc Lagrange}schen
Gleichungen hier ihre G\"ultigkeit
behalten, das hier betrachtete System
nicht konservativ ist im \"ublichen Sinne:
Die {\sc Lorentz}-Kraft ist nicht der
{\it Gradient\/} eines Potentials
\begin{equation}
{\bf F}\not=-{\bf grad}\;U,
\end{equation}
sondern es gilt
\begin{equation}
{\bf F}=
-{\bf grad}\;U+
\frac{d}{dt}\,{\bf grad}\,_{\dot{\bf x}}\;U,
\end{equation}
mit dem {\it geschwindigkeitsabh\"angigen Potential\/}
\begin{equation}
U=
e\,
V-
e\,
{\bf A}\dot{\bf x}.
\end{equation}
\par
Eine Bewegungsgleichung f\"ur das {\sc Hall}-Problem,
welche \"uber die station\"are N\"aherung hinausgeht,
hat somit die Form
\begin{equation}
m \, \dot {\bf v} = e \,( {\bf E} + {\bf v}\times{\bf B} ),
\end{equation}
oder als Bewegungsgleichung f\"ur den Strom geschrieben
\begin{equation}
\frac{m}{e}\,\frac{d{\bf j}_{2D}}{dt}
=
e\,n_{2D} \cdot {\bf E} + {\bf j}_{2D} \times {\bf B}.
\end{equation}
Ihre L\"osung ist gegeben durch
\begin{equation}
{\bf j}_{2D}(t) = e\,n_{2D} \cdot
                  \frac{ {\bf B} \times {\bf E} }{ {\bf B}^2 }
                  +
                  \exp(t\boldomega_c \times) \, {\bf j}_{2D,0},
\end{equation}
wobei
\begin{equation}
\boldomega_c=(0,0,\omega_c),
\end{equation}
mit der trickreichen als Operatoridentit\"at auffa\ss baren Schreibweise
\begin{equation}
(\boldomega_c\times)=
\left(
\begin{array}{ccc}
0        & -\omega_c & 0 \\
\omega_c & 0         & 0 \\
0        & 0         & 0
\end{array}
\right).
\end{equation}
Der Exponentialoperator ist durch seine Entwicklung
in seine Potenzreihe definiert%
\footnote{%
Man sieht sofort, was gemeint ist, wenn man sich klarmacht, da\ss\
f\"ur drei normierte und orthogonale Einheitsvektoren
${\bf e}_x,{\bf e}_y,{\bf e}_z$ gilt
\begin{equation}
\exp\,(\varphi\,{\bf e}_z\times)\,{\bf e}_x =
(\cos\,\varphi)\,{\bf e}_x +
(\sin\,\varphi)\,{\bf e}_y.
\end{equation}}
und ${\bf j}_{2D,0}$
durch die Anfangsbedingung zu einer festen Zeit gegeben.
Einsetzen der L\"osung in die Gleichung ergibt f\"ur
den Betrag der Winkelgeschwindigkeit oder die sogenannte
{\it Zyklotronfrequenz\/}
\begin{equation}
|\boldomega_c| = \frac{eB}{m}.
\end{equation}
\vfill\eject\noindent%
\subsection{Das Quanten-Regime}
Im Bereich tiefer Temperaturen dominieren Quanteneffekte
(der {\sc Hamilton}-{\it Operator\/} dominiert \"uber $kT$).
Verhalten sich die Ladungstr\"ager in der Probe wie ein
2-dimensionales Elektronengas (also wie ein idealisiertes
System nicht wechselwirkender Elektronen, die sich in einem
2-dimensionalen Raum aufhalten), so ist der entsprechende
Einteilchen-{\sc Hamilton}-Operator durch {\sc Landau}s
Formel gegeben
\cite{LandauLifshitz}:
\begin{equation}
H_L=\frac{1}{2}m{\bf v}^2=\frac{({\bf p}-e{\bf A})^2}{2m},
\end{equation}
wobei ${\bf A}$ hier - wie \"ublich - das Vektorpotential
bezeichnet.
\par
In dem hier betrachteten Problem ist das Magnetfeld
zeitlich konstant, r\"aumlich homogen und in z-Richtung
orientiert. Ein solches Magnetfeld wird am besten durch
die Rotation eines Vektorpotentials in der {\it isotropen Eichung\/}
dargestellt, das hei\ss t, es ist
\begin{equation}
{\bf B}={\bf rot}\;{\bf A}
       ={\bf rot}\;\left(\frac{1}{2}\,{\bf B}\times{\bf r}\right)
\end{equation}
mit
\begin{equation}
{\bf B}=(0,0,B_z),
\phantom{xxx}
{\bf A}=(- B_z y/2, B_z x/2, 0)
\end{equation}
und
\begin{equation}
B=B_z=\mbox{\it const\/}.
\end{equation}
W\"ahrend die Komponenten des kanonischen Impulses ${\bf p}$
die kanonischen Vertauschungsrelationen erf\"ullen,
also untereinander kommutieren,
tun dies die Komponenten des sogenannten kinetischen
Impulses (dem Produkt aus Masse und Geschwindigkeit)
\begin{equation}
K_j=p_j-eA_j=mv_j
\end{equation}
nicht, sondern es gilt in der f\"ur uns relevanten Geometrie
\begin{equation}
\lbrack K_x,K_y \rbrack = i \hbar \cdot eB_z
\end{equation}
mit
\begin{equation}
H=\frac{K_x^2+K_y^2}{2m}
\end{equation}
und
\begin{equation}
B=B_z=\mbox{\it const\/}.
\end{equation}
Diese Situation ist analog zu der des harmonischen Oszillators.
Insbesondere haben wir die Entsprechungen
\begin{eqnarray}
\mbox{2D-{\sc Landau}}
    &\longleftrightarrow&
        \mbox{1D-Oszillator}                                  \nonumber\\
    &\phantom{=}&                                             \nonumber\\
H=\frac{K_y^2}{2m  }+\frac{K_x^2}{2m}
    &\longleftrightarrow&
        H=\frac{P^2  }{2\mu}+\frac{1}{2}\mu \Omega^2 Q^2      \nonumber\\
    &\phantom{=}&                                             \nonumber\\
K_y &\longleftrightarrow& P                                   \nonumber\\
    &\phantom{=}&                                             \nonumber\\
m   &\longleftrightarrow& \mu                                 \nonumber\\
    &\phantom{=}&                                             \nonumber\\
K_x &\longleftrightarrow& \mu\Omega\cdot Q                    \nonumber\\
    &\phantom{=}&                                             \nonumber\\
\lbrack K_x,K_y\rbrack=i \hbar eB_z
    &\longleftrightarrow&
        \lbrack \mu\Omega\cdot Q, P \rbrack=
        \mu\Omega \cdot \lbrack Q,P\rbrack=
        \mu\Omega \cdot i \hbar \, ;
\end{eqnarray}
letzteres hei\ss t nat\"urlich
\begin{eqnarray}
\lbrack Q,P \rbrack = i\hbar.
\end{eqnarray}
F\"ur die beteiligten $c$-Zahlen%
\footnote{$c$-Zahlen sind als {\it classical numbers\/}
          kommutierende Gr\"o\ss en, $q$-Zahlen als
          {\it quantized numbers\/} im allgemeinen nicht-kommutierende
          Gr\"o\ss en, also Operatoren. Diese Sprechweise
          ist in der mathematischen Physik weit verbreitet
          und geht auf fr\"uhe Zeiten der Quantentheorie
          zur\"uck.}
gelten die Zuordnungen
\begin{eqnarray}
\mbox{2D-{\sc Landau}}
                 &\longleftrightarrow& \mbox{1D-Oszillator}      \nonumber\\
                 &\phantom{=}&                                   \nonumber\\
i\hbar\cdot eB_z &\longleftrightarrow& i\hbar\cdot\mu\Omega      \nonumber\\
                 &\phantom{=}&                                   \nonumber\\
m                &\longleftrightarrow& \mu                       \nonumber\\
                 &\phantom{=}&                                   \nonumber\\
i\hbar\cdot \frac{eB_z}{m}
                 &\longleftrightarrow&
                   i\hbar\cdot\Omega
                  =i\hbar\cdot\sqrt{\frac{D}{\mu}},              \nonumber\\
\end{eqnarray}
mit $D$ als Federkonstante.
\par
Mit anderen Worten:
Ein 2-dimensionales mechanisches System
in einem konstanten \"au\ss eren Magnetfeld
entspricht von seiner algebraischen Struktur her einem
ein-di\-men\-si\-o\-na\-len harmonischen Oszillator.%
\footnote{%
Die Nichtkommutativit\"at hat aber in beiden F\"allen
einen unterschiedlichen physikalischen Ursprung. Neuerdings
nennt man in der mathematischen Physik eine ganze Reihe
von nicht-kommutativen Strukturen \loq quantisiert\hiq,
obwohl sie keinen direkten {\it physikalischen\/} Bezug
zur Quantentheorie im Sinne von
{\sc Heisenberg}-{\sc Born}-{\sc Jordan}-{\sc Dirac} haben.}
\par
Entsprechend der Analogie-Beziehung
\begin{equation}
E_n = \left( n+\frac{1}{2} \right) (\hbar eB_z) \left( \frac{1}{m} \right)
\phantom{x}\longleftrightarrow\phantom{x}
E_n = \left( n+\frac{1}{2} \right) \hbar\Omega
\end{equation}
bzw.\
\begin{equation}
E_n = \left( n+\frac{1}{2} \right) \hbar \left( \frac{eB_z}{m} \right)
\phantom{x}\longleftrightarrow\phantom{x}
E_n = \left( n+\frac{1}{2} \right) \hbar\Omega
\end{equation}
haben die Energieniveaux der Ladungstr\"ager
diskrete Werte, die {\it modulo\/} einer gemeinsamen
additiven Konstante, welche den Energienullpunkt
festlegt, allesamt Vielfache der
{\sc Planck}\-schen Konstante $\hbar$ sind.
Die Bedeutung der Bewegungskonstante
\begin{equation}
\omega_c=\frac{eB}{m} \, ,
\end{equation}
wobei wir statt $B_z$ ab jetzt nur noch $B$ schreiben,
wird deutlich wenn wir uns ver\-ge\-gen\-w\"ar\-ti\-gen, da\ss\
die Elektronen in einem konstanten homogenen Magnetfeld
bei verschwindendem \"au\ss erem elektrischen Feld
eine Zyklotronbewegung (engl.\ {\it cyclotron motion\/})
vollf\"uhren, also Kreisbahnen durchlaufen,
f\"ur welche die {\sc Lorentz}-Kraft der Zentralkraft
die Waage h\"alt. Es ist also
\begin{equation}
\frac{mv^2}{r} = evB
\end{equation}
und somit
\begin{equation}
m\omega_c^2r = e \cdot \omega_c r \cdot B,
\end{equation}
woraus obige Beziehung folgt.
\par
F\"ur die quantisierten Energieniveaux
der Zyklotronbewegung folgt
unter Einbeziehung der Tatsache,
da\ss\ der Energienullpunkt im allgemeinen
separat festgelegt werden mu\ss\
\begin{equation}
E_n=\left( n+\frac{1}{2} \right)\,
    \hbar\omega_c+\mbox{\it const}.
\end{equation}
\par
{\it Bemerkung:\/}
Eine nicht zu untersch\"atzende Vorsicht mu\ss\
darin ge\"ubt werden, es nicht zu Verwechselungen
von klassischen und quantenmechanischen Konzepten
kommen zu lassen.
Dies ist nicht anders als beim gel\"aufigen Beispiel
des harmonischen Oszillators:
Klassisch berechnet sich seine Energie gem\"a\ss\
\begin{eqnarray}
E_{cl} &=& E_{kin}+E_{pot}                                  \nonumber\\
       &\phantom{=}&                                        \nonumber\\
       &=& \frac{1}{2} \mu v^2 + \frac{1}{2} Dx^2           \nonumber\\
       &\phantom{=}&                                        \nonumber\\
       &=& \frac{1}{2} \mu v^2 + \frac{1}{2} \mu\Omega^2x^2 \nonumber\\
       &\phantom{=}&                                        \nonumber\\
       &=& \frac{1}{2} \mu\Omega^2A^2(\cos\Omega t)^2 +
           \frac{1}{2} \mu\Omega^2A^2(\sin\Omega t)^2       \nonumber\\
       &\phantom{=}&                                        \nonumber\\
       &=& \frac{1}{2} \mu\Omega^2A^2.
\end{eqnarray}
Quantenmechanisch hingegen ist
\begin{eqnarray}
E_{qu} &=& \left(n+\frac{1}{2}\right)\,\hbar\Omega.
\end{eqnarray}
Klassisch h\"angt seine Energie quadratisch von der Frequenz $\Omega$
und quadratisch von der Amplitude $A$ ab. Quantenmechanisch hingegen
h\"angt seine Energie linear von der Frequenz $\Omega$
({\it modulo\/} einer Nullpunktsenergie)
und linear von der Quantenzahl $n$ ab.
F\"ur hohe
Quantenzahlen allerdings mu\ss\ die quantentheoretische Beschreibung
in die klassische \"u\-ber\-ge\-hen. Dies ist dann der Fall, wenn
\begin{equation}
n\rightarrow\infty
\end{equation}
und asymptotisch gilt
\begin{equation}
n \approx \frac{\mu\Omega A^2}{2\hbar}.
\end{equation}
F\"ur die Zyklotronbewegung ist klassisch
\begin{eqnarray}
E_{cl} &=& \frac{1}{2} mv^2                   \nonumber\\
       &\phantom{=}&                          \nonumber\\
       &=& \frac{1}{2} m\omega_c^2r^2 \; ,
\end{eqnarray}
quantenmechanisch hingegen
\begin{eqnarray}
E_{qu} &=& \left(n+\frac{1}{2}\right)\,\hbar\omega_c.
\end{eqnarray}
Der klassische Grenzfall der quantenmechanischen Beschreibung
ist nun charakterisiert durch
\begin{equation}
n\rightarrow\infty
\end{equation}
mit
\begin{equation}
n \approx \frac{m\omega_c r^2}{2\hbar}.
\end{equation}
\par
F\"ur die klassische Zyklotronbewegung
erhalten wir den Bahnradius aus Gleichsetzen
von {\sc Lorentz}-Kraft und Zentralkraft
\begin{equation}
r_{class}=\frac{mv^2}{evB}
         =\frac{\sqrt{2m \cdot mv^2/2}}{eB}
         =\frac{\sqrt{2mE}}{eB}.
\end{equation}
Setzen wir anstelle der klassischen kinetischen Energie
\begin{equation}
E_{kin} = \frac{1}{2} mv^2/2
\end{equation}
die quantenmechanische Nullpunktsenergie
\begin{equation}
E       = \frac{1}{2} \hbar\omega_c
\end{equation}
in die obige Gleichung ein, definieren wir eine
{\it magnetische L\"ange\/}
\begin{equation}
l_{B} = \frac{\sqrt{2m\hbar\omega_c/2}}{eB}
      = \sqrt{\frac{\hbar}{eB}}.
\end{equation}
Dies ist die fundamentale L\"ange des {\sc Landau}-Problems.
Wir schreiben auch
\begin{equation}
\frac{h}{eB} =
\frac{2\pi\hbar}{eB} =
2\pi l_B^2.
\end{equation}
Das zugeordnete Nullpunktsorbital, welches man
sich {\it nicht\/} als eine Kreisbewegung im klassischen
Sinne vorstellen darf, entspricht  - im Sinne
der von uns betrachteten Analogie - der
Nullpunktsschwingung eines harmonischen Oszillators.
\vfill\eject\noindent%
\subsection{Flu\ss quanten versus Ladungstr\"ager}
Die Gr\"o\ss e
\begin{equation}
\Phi_0=\frac{h}{e}
\end{equation}
nennen wir das einer elektrischen Ladung $e$ zugeordnete
{\it magnetische Flu\ss quantum\/}.
Seine Bedeutung liegt im {\sc Aharonov}-{\sc Bohm}-Effekt
\cite{Franz39, EhrenbergSiday49, AharonovBohm59}:
Wie bereits {\sc Franz} bemerkte \cite{Franz39},
beeinflu\ss t der von einer Elektronenwelle eingeschlossene
magnetische Flu\ss\ deren Wellenzahl ({\it besser:\/} Phase).
{\sc Ehrenberg} und {\sc Siday}
wiesen in ihrer bahnbrechenden elektronenoptischen
Arbeit darauf hin, da\ss\ ein solcher Einflu\ss\ auch dann bestehen bleibt,
wenn die in zwei koh\"arente Teilstrahlen aufgespaltene
Elektronenwelle nirgendwo ein magnetisches Feld durchl\"auft.
{\sc Aharonov} und {\sc Bohm} gaben diesem Effekt schlie\ss lich
Ihren Namen und hoben die physikalische Bedeutung des Vektorpotentials
in der Quantenmechanik hervor.
\par
Das Interferenzmuster in einem
Beugungsexperiment eines elektrisch abgeschirmten magnetischen
Flusses $\Phi$ h\"angt also ab vom
{\it nicht-integrablen Phasenfaktor\/} \cite{WuYang75}
\begin{equation}
\exp \, \frac{ie}{\hbar} \, \oint_{\partial S} {\bf A} \, d{\bf l} =
\exp \, \frac{ie}{\hbar} \, \intii_{S} {\bf rot}\,{\bf A} \, d{\bf F} =
\exp \, \frac{ie}{\hbar} \, \Phi,
\end{equation}
der gerade ein Vielfaches der Eins ist, wenn
\begin{equation}
\Phi = n \cdot \frac{h}{e} =: n \cdot \Phi_0,
\phantom{xxx}
n\in{\bf Z}.
\end{equation}
In diesem Fall ist die Flu\ss linie quantenmechanisch
nicht observabel.
In der {\sc Landau}-{\sc Ginzburg}-Theorie der Supraleitung
ist die zentrale Gr\"o\ss e ein Ordnungsparameter, der
eine gewisse formale \"Ahnlichkeit zu einer quantenmechanischen
Wellenfunktion besitzt. Die Theorie ist allerdings nur dann
mit der mikroskopischen Theorie \`a la {\sc Bardeen}, {\sc Cooper}
und {\sc Schrieffer} vertr\"aglich, wenn die elektrische Ladung
der \loq supraleitenden Elektronen\hiq, der sogenannten
{\sc Cooper}-Paare, statt $e$ genau doppelt so hoch ist,
n\"amlich $2e$. Demzufolge haben die magnetischen
Flu\ss quanten der Supraleitung den Wert $h/2e$.
Sie manifestieren sich als Flu\ss schl\"auche oder
{\sc Abrikosov}-Vortizes, die in das supraleitende
Medium unter bestimmten Umst\"anden eindringen k\"onnen.%
\footnote{Und es ist geradezu eine Ironie der Geschichte,
          da\ss\ in der vollst\"andigen Theorie die
          Flu\ss quanten wieder observabel werden,
          n\"amlich durch Quasiteilchen-Moden im
          Vortexkern, sogenannte chirale Fermionen,
          die man sich als eingefangene Irrl\"aufer
          \loq halber {\sc Cooper}-Paare\hiq\ denken darf.}
\par
Wir wollen nun zeigen, da\ss\ auch im quantisierten
{\sc Hall}-Effekt die Flu\ss quanten eine entscheidende
Rolle spielen.
\par
Wegen der (n\"aherungsweisen) Translationsinvarianz des Systems
(in x- und y-Richtung)
ist jedes dieser {\sc Lan\-dau}-Niveaux hochgradig entartet.%
\footnote{F\"ur die explizite Behandlung der
          {\sc Schr\"odinger}-Gleichung im konstanten \"au\ss eren Magnetfeld
          sei der Leser auf das Lehrbuch von {\sc Landau} und {\sc Lifshitz}
          verwiesen \cite{LandauLifshitz}.}
\footnote{An dieser Stelle sei nur erw\"ahnt, da\ss\
          - wegen der mathematischen Form des Vektorpotentials -
          die Eigenzust\"ande der {\sc Hamilton}-Funktion des
          {\sc Landau}-Problems denen der z-Komponente des
          Bahndrehimpulsoperators
          ${\bf L} =  - i \hbar {\bf r} \times {\bf grad}$
          entsprechen.}
Um exakt zu sein: Im Falle eines unendlich ausgedehnten Systems
ist die Entartung tats\"achlich unendlich; denn aus einer vorgegebenen
L\"osung k\"onnen wir durch Anwendung von x- und y-Translationen
beliebig viele andere erzeugen. Im Falle eines endlich ausgedehnten
Systems, zum Beispiel mit den Abmessungen $L_x \times L_y$
k\"onnte man versuchen, in der \"ublichen Weise
die Zu\-stands\-dich\-te
auszurechnen \cite{Kittel}:
Man stelle sich die Wellenfunktionen als stehende Wellen vor
\begin{eqnarray}
\psi(0,y) &=& \psi(L_x,  y) \;=\; 0 \nonumber\\
\psi(x,0) &=& \psi(  x,L_y) \;=\; 0
\end{eqnarray}
oder fordere - dem Transportproblem angemessener -
wenigstens periodische Randbedingungen
\begin{eqnarray}
\psi(x+L_x,y    ) &=& \psi(x    ,y    ) \nonumber\\
\psi(x    ,y+L_y) &=& \psi(x    ,y    ) \nonumber
\end{eqnarray}
und z\"ahle die Zust\"ande bis hin zu einer
vorgegebenen Grenzenergie ab \cite{Kittel}.
Die nach dieser Energie abgeleitete Anzahl der Zust\"ande ist die
Zustanddichte des 2-dimensionalen Elektronengases in dem betrachteten
endlich ausgedehnten System. Weiter unten werden wir diese Rechnung
explizit durchf\"uhren und auch das richtige Ergebnis erhalten.
\par
Dieser Zugang ist aber nicht wirklich begr\"undet, wenn wir davon ausgehen,
da\ss\ das System sich in einem homogenen Magnetfeld befindet und die
Wellenfunktionen ausgedehnt sind. Die Wahl der Randbedingungen mu\ss\
n\"amlich die lokale Eichinvarianz ({\it engl.\/} local gauge invariance)
erf\"ullen, eines der grundlegenden Prinzipien von Quantenmechanik und
Elektrodynamik.
Eine lokal eichinvariante Randbedingung, welche die Anwesenheit
des Magnetfeldes respektieren w\"urde, h\"atte die Form
\begin{eqnarray}
\psi(x+L_x,y)       &=&\exp\,
                       \left\{ i\,\frac{e}{\hbar}\,\alpha(x,y) \right\}
                       \cdot\psi(x,y)  \nonumber\\
\psi(x    ,y+L_y)   &=&\exp\,
                       \left\{ i\,\frac{e}{\hbar}\,\beta (x,y) \right\}
                       \cdot\psi(x,y)
\end{eqnarray}
und gleichzeitig
\begin{eqnarray}
{\bf A}(x+L_x,y    )&=&{\bf A}(x,y)+{\bf grad}\,\alpha(x,y) \nonumber\\
{\bf A}(x    ,y+L_y)&=&{\bf A}(x,y)+{\bf grad}\,\beta (x,y).
\end{eqnarray}
Somit ist f\"ur ein geeignet gew\"ahltes Koordinatensystem:
($\partial$ steht f\"ur Rand)
\begin{eqnarray}
\Phi &=& B\cdot L_x L_y                                             \nonumber\\
     &\phantom{=}&                                                  \nonumber\\
     &=& \int_{L_x\times L_y} {\bf B}(x,y)\,d{\bf S}                \nonumber\\
     &\phantom{=}&                                                  \nonumber\\
     &=& \int_{L_x\times L_y} {\bf rot}\,{\bf A}(x,y)\,d{\bf S}     \nonumber\\
     &\phantom{=}&                                                  \nonumber\\
     &=& \oint_{\partial(L_x\times L_y)} {\bf A}(x,y)\,d{\bf l}     \nonumber\\
     &\phantom{=}&                                                  \nonumber\\
     &\phantom{=}&                                                  \nonumber\\
     &=& \int_{(0  ,0  )\rightarrow(L_x,0  )}
                                 {\bf A}(x,y)\,d{\bf l}+
         \int_{(L_x,0  )\rightarrow(L_x,L_y)}
                                 {\bf A}(x,y)\,d{\bf l}+            \nonumber\\
     &\phantom{=}&                                                  \nonumber\\
     &\phantom{=}&                                                  \nonumber\\
     & &\phantom{12345678}
        +\int_{(L_x,L_y)\rightarrow(0  ,L_y)}
                                 {\bf A}(x,y)\,d{\bf l}+
         \int_{(0  ,L_y)\rightarrow(0  ,0  )}
                                 {\bf A}(x,y)\,d{\bf l}             \nonumber\\
     &\phantom{=}&                                                  \nonumber\\
     &\phantom{=}&                                                  \nonumber\\
     &=& \int_{(0  ,0  )\rightarrow(L_x,0  )}
                                 {\bf A}(x,y)\,d{\bf l}
        -\int_{(0  ,L_y)\rightarrow(L_x,L_y)}
                                 {\bf A}(x,y)\,d{\bf l}+            \nonumber\\
     &\phantom{=}&                                                  \nonumber\\
     & &\phantom{12345678}
        +\int_{(L_x,0  )\rightarrow(L_x,L_y)}
                                 {\bf A}(x,y)\,d{\bf l}
        -\int_{(0  ,0  )\rightarrow(0  ,L_y)}
                                 {\bf A}(x,y)\,d{\bf l}             \nonumber\\
     &\phantom{=}&                                                  \nonumber\\
     &\phantom{=}&                                                  \nonumber\\
     &=& \int_{(0  ,0  )\rightarrow(L_x,0  )}
                                 {\bf A}(x    ,y    )\,d{\bf l}
        -\int_{(0  ,0  )\rightarrow(L_x,0  )}
                                 {\bf A}(x    ,y+L_y)\,d{\bf l}+    \nonumber\\
     &\phantom{=}&                                                  \nonumber\\
     & &\phantom{12345678}
        +\int_{(0  ,0  )\rightarrow(0  ,L_y)}
                                 {\bf A}(x+L_x,y    )\,d{\bf l}+
        -\int_{(0  ,0  )\rightarrow(0  ,L_y)}
                                 {\bf A}(x    ,y    )\,d{\bf l}     \nonumber\\
     &\phantom{=}&                                                  \nonumber\\
     &\phantom{=}&                                                  \nonumber\\
     &=&   -\int_{(0  ,0  )\rightarrow(L_x,0  )}
                                 {\bf grad}\,\beta (x    ,y    )\,d{\bf l}
           +\int_{(0  ,0  )\rightarrow(0  ,L_y)}
                                 {\bf grad}\,\alpha(x    ,y    )\,d{\bf l}
                                                                    \nonumber\\
     &\phantom{=}&                                                  \nonumber\\
     &=& - \left[\, \beta (L_x,0  ) - \beta (0,0) \,\right]
         + \left[\, \alpha(0  ,L_y) - \alpha(0,0) \,\right] .
\end{eqnarray}
Wenn also $\alpha$ und $\beta$ identisch verschwinden w\"urden
- wie im Falle periodischer Randbedingungen im konventionellen Sinne -
w\"are der Gesamtflu\ss\ durch die Probe identisch Null!
Lokale Eichinvarianz {\it erzwingt\/} die Verwendung von Randbedingungen,
welche die Wellenfunktionen {\it modulo\/} einer Phase festlegen.
Diese ist im allgemeinen wegabh\"angig ({\it griech.\/} anholonom).
\par
Eindeutigkeit der Wellenfunktion verlangt nun,
da\ss\ die Phase $\varphi$ der Wellenfunktion
\begin{equation}
\Psi = \varrho \cdot \exp\,i\varphi
\end{equation}
sich am Rand um ein Vielfaches von $2\pi$ dreht.
Das hei\ss t, die Schleife
\begin{equation}
\varphi(0  ,0  ) \rightarrow
\varphi(L_x,0  ) \rightarrow
\varphi(L_x,L_y) \rightarrow
\varphi(0  ,L_y) \rightarrow
\varphi(0  ,0  )
\end{equation}
mu\ss\ (f\"ur $L_x=L_y$) gehen wie
\begin{equation}
0                  \rightarrow
\frac{1}{4} \cdot 2\pi n \rightarrow
\frac{2}{4} \cdot 2\pi n \rightarrow
\frac{3}{4} \cdot 2\pi n \rightarrow
\frac{4}{4} \cdot 2\pi n
.
\end{equation}
F\"ur unseren Fall bedeutet dies, da\ss\
\begin{eqnarray}
1 &=&   \exp \, \left\{ i \,\frac{e}{\hbar} \cdot
                (\,
                  - \left[\, \beta (L_x,0  ) - \beta (0,0) \,\right]
                  + \left[\, \alpha(0  ,L_y) - \alpha(0,0) \,\right]
                 \,)
                \right\}                                        \nonumber\\
  &=&   \exp \, \left\{ i \,\frac{e}{\hbar} \,\Phi \right\}
\end{eqnarray}
und somit
\begin{equation}
\Phi =  2\pi \, \frac{\hbar}{e} \, n = \frac{h}{e} \, n
     =: n_{\Phi_0} \cdot \Phi_0.
\end{equation}
Mit anderen Worten:
Wenn die Wellenfunktion ausgedehnt ist, mu\ss\ der
Flu\ss\ $\Phi$ in Einheiten des Flu\ss quantums
\begin{equation}
\Phi_0=h/e
\end{equation}
quantisiert sein.
Die Anzahl $n_{\Phi_0}$
der Flu\ss quanten pro Einheitsfl\"ache
berechnet sich aus
\begin{equation}
B = n_{\Phi_0} \cdot \Phi_0.
\end{equation}
Die Probenfl\"ache wird durch die
Anwesenheit der Flu\ss quanten parkettiert.
Je h\"oher das Magnetfeld ist, desto kleiner
werden die Fliesen des Parketts.
Da in einem endlich ausgedehnten System die
Randbedingungen nur noch diskrete Translationen
erlauben, welche die Parkettierung respektieren,
entspricht der Entartungsgrad pro Einheitsfl\"ache
gerade der Anzahl der Flu\ss quanten
pro Einheitsfl\"ache und ist gegeben durch
\begin{equation}
n_{\Phi_0} := \frac{eB}{h}.
\end{equation}
Ein Check der Einheiten zeigt, das alles mit rechten
Dingen zugeht, denn $eB$ hat die Dimension einer
Wirkung pro Fl\"ache und $n_{\Phi_0}$ ist eine
Fl\"achendichte, also ein Ma\ss\ f\"ur die Teilchenzahl
pro Fl\"ache.
\par
Wieviele Ladungstr\"ager pro Fl\"acheneinheit
untergebracht werden k\"onnen, wird nicht nur
durch den Entartungsgrad, sondern auch
durch die Tatsache vorgegeben,
da\ss\ die Ladungstr\"ager Fermionen sind.%
\footnote{Mit anderen Worten: Die Elektronen
          erf\"ullen das {\sc Pauli}-Prinzip.
          Letzteres ist eine Konsequenz der Tatsache,
          da\ss\ die Elektronen Fermionen sind,
          also ununterscheidbare Teilchen,
          die durch eine total antisymmetrische
          Vielteilchen-Wellenfunktion beschrieben werden.}
Dies legt die Anzahl der {\it erlaubten\/} Zust\"ande
pro {\sc Landau}-Niveau und Ein\-heits\-fl\"a\-che fest.
Wenn wir der Einfachheit halber den Spin und m\"ogliche
zus\"atzliche Quantenzahlen au\ss er acht lassen,
kann ein {\sc Landau}-Niveau mit $eB/h$
ununterscheidbaren fermionischen Ladungstr\"agern
pro Einheitsfl\"ache aufgef\"ullt werden.
Wenn $i$ Landau-Niveaus vollst\"andig gef\"ullt sind,
mu\ss\ sich die {\sc Hall}-Leitf\"ahigkeit
ergeben als
\begin{equation}
\sigma_H=\frac{e\,n_{2D} }{B}=i\cdot\frac{e^2}{h},
\end{equation}
wobei $n_{2D}$ - wie oben - die 2-dimensionale
Ladungstr\"agerdichte bezeichnet.
Offensichtlich kann diese Bedingung auch geschrieben werden als
\begin{equation}
n_{2D} = i \cdot \frac{eB}{h}.
\end{equation}
Wenn also - im quantenmechanischen Limes $T\rightarrow 0$ -
das chemische Potential%
\footnote{Das chemische Potential charakterisiert die
          {\sc Fermi}-Verteilung. Am absoluten Nullpunkt
          trennt es die besetzten von den unbesetzten
          Zust\"anden. (Eine detaillierte Erkl\"arung
          folgt weiter unten.)}
genau zwischen zwei {\sc Landau}-Niveaus liegt,
mu\ss\ die {\sc Hall}-Leitf\"ahigkeit quantisiert sein!
Allerdings gibt es keinen Grund zu erwarten,
da\ss\ in einem realen System das chemische Potential
immer genau von der Mitte der einen L\"ucke
zur Mitte der n\"achsten springt.
\vfill\eject\noindent%
\subsection{Beweis f\"ur die Abwesenheit des QHE im idealen 2DEG}
In der Tat, ein mathematisch rigoroser Beweis zeigt,
da\ss\ das ideale freie Elektronengas im unendlich
gro\ss en Volumen {\it keinen\/} quantisierten
{\sc Hall}-Effekt zeigt. Im folgenden sei dieser Beweis kurz
vorgestellt.
\par
Aus der {\sc Lagrange}-Funktion
\begin{equation}
L({\bf x},\dot{\bf x},t)=
T-U=
\frac{m}{2}\dot{\bf x}^2
+
e\,
{\bf A}({\bf x},t)\cdot\dot{\bf x}
-
e\,
V({\bf x},t)
\end{equation}
erhalten wir via {\sc Legendre}-Transformation die {\sc Hamilton}-Funktion
\begin{eqnarray}
H(x,p,t)
  &=& {\bf p}\dot{\bf x} - L({\bf x},\dot{\bf x},t) \nonumber\\
  &=& ( m\dot{\bf x} + e\,{\bf A} ) \,\dot{\bf x}
      - \frac{m}{2} \dot{\bf x}^2
      - e\,{\bf A}\,\dot{\bf x} + e\,V({\bf x},t)
                                                    \nonumber\\
  &=&   \frac{m}{2} \dot{\bf x}^2 + e\,V( {\bf x},t ).
\end{eqnarray}
Im {\sc Landau}-Fall gilt
\begin{eqnarray}
{\bf B}(x,t) &=& {\bf B} \,=\, {\it const\/}                                \\
{\bf A}(x,t) &=& \frac{1}{2} \, {\bf B} \times {\bf x}                      \\
{\bf E}(x,t) &=& {\bf E} \,=\, {\it const\/}                                \\
V      (x,t) &=& \int_0^{\bf x} {\bf E}({\bf x}',t)\,d{\bf x}
                         \,=\, {\bf E} \cdot {\bf x}.
\end{eqnarray}
Die entsprechende {\sc Hamilton}-Funktion bezeichnen wir als $H_{LE}$.
\par
Die {\sc Hamilton}schen Bewegungsgleichungen k\"onnen auch durch
die {\sc Poisson}-Klam\-mern%
\begin{equation}
\lbrace A,B \rbrace
=
\sum_i \,
\frac{\partial A}{\partial q_i} \frac{\partial B}{\partial p_i}
-
\frac{\partial B}{\partial p_i} \frac{\partial A}{\partial q_i}
\end{equation}
ausgedr\"uckt werden
\cite{Goldstein80}.
So ist
\begin{eqnarray}
\dot{\bf x}               &=& \lbrace {\bf x},      H_{LE} \rbrace ,   \\
\dot{\bf v}               &=& \lbrace {\bf v},      H_{LE} \rbrace ,   \\
\frac {d{\bf j}_{2D}}{dt} &=& \lbrace {\bf j}_{2D}, H_{LE} \rbrace .
\end{eqnarray}
Diese Formulierung ist von Nutzen, wenn wir das vorliegende
System quantisieren wollen. Wir ersetzen die klassischen
{\sc Poisson}-Klammern durch die quantenmechanischen
Kommutatoren multipliziert mit einem Faktor $-i/\hbar$. F\"ur
die quantenmechanischen Operatorgr\"o\ss en gilt somit%
\footnote{%
Man beachte, da\ss\ die durch die {\sc Poisson}-Klammern
erzeugte algebraische Struktur im allgemeinen
{\it nicht\/} zu der durch die quantenmechanischen
Kommutatoren erzeugte Struktur isomorph ist,
obwohl die Analogie sehr suggestiv ist und zum
Repertoire vieler einf\"uhrender Vorlesungen
(und professionell arbeitender Wissenschaftler) geh\"ort.
W\"are dem so, w\"urden sich klassische Mechanik und Quantenmechanik
nicht wesentlich unterscheiden. Das Theorem von
{\sc Groenewold} und {\sc van\,Hove} zeigt, da\ss\
eine \loq kanonische Quantisierung\hiq, also die
nach einem Kanon - einer Regel also - erfolgte
Konstruktion eines quantenmechanischen Systems
aus einem klassischen {\it nicht\/} m\"oglich ist
\cite{AbrahamMarsden, Groenwold46, vanHove51}.
Im Grunde genommen ist dies
nicht verwunderlich, ist doch die klassische Physik
nur eine N\"aherung der quantenmechanischen. Und es
ist nicht zu erwarten, da\ss\ man die exakte Theorie
in einer wohldefinierten Weise aus einer approximativen
herleitet: Kanonische Quantisierung ist allenfalls
eine Heuristik wenn nicht die besondere Form eines
akademischen Ratespiels.}%
\begin{eqnarray}
\dot{\bf x}         &=& - \frac{i}{\hbar} \, \lbrack {\bf x}, H_{LE} \rbrack,          \\
\dot{\bf v}         &=& - \frac{i}{\hbar} \, \lbrack {\bf v}, H_{LE} \rbrack,          \\
\frac{d{\bf j}_{2D}}{dt}
                    &=& - \frac{i}{\hbar} \, \lbrack {\bf j}_{2D},
                                                              H_{LE} \rbrack.
\end{eqnarray}
Nun ist es eine wohlbekannte Tatsache, da\ss\ f\"ur den Fall,
da\ss\ die {\sc Hamilton}-Funktion quadratisch in den
Orts- und Impulsvariablen ist, die quantenmechanischen
Bewegungsgleichungen den klassischen entsprechen. Eine
explizite Evaluation der Operatorgleichung verifiziert dies.
So erhalten wir als L\"osung
\begin{equation}
{\bf j}_{2D}(t) = e\,n_{2D} \cdot
             \frac{ {\bf B} \times {\bf E} }{ {\bf B}^2 }
             +
             \exp(t \boldomega_c \times) \, {\bf j}_{2D,0} .
\end{equation}
Um eine Vorhersage \"uber den Me\ss wert des elektrischen
Stroms in unserem {\sc Hall}-Ex\-peri\-ment zu machen, m\"ussen
wir den zeitgemittelten Erwartungswert
\begin{equation}
\overline{<{\bf j}_{2D}(t)>}
=
\lim_{T\rightarrow\infty} \,
\frac{1}{T} \,
\int_0^T \,
<{\bf j}_{2D}(t)>
\end{equation}
des {\sc Heisenberg}-Strom\-ope\-ra\-tors ${\bf j}_{2D}(t)$ ausrechnen.
Dabei ist zu beachten, da\ss\ wir thermodynamische
Gleichgewichtszust\"ande eines quantisierten fermionischen Systems
betrachten m\"ussen.
\par
{\it Zur Erinnerung:\/}
In der ersten Quantenmechanik-Vorlesung f\"angt man mit
{\it reinen Zu\-st\"an\-den\/} an, die durch Zustandsvektoren
$|\Psi>$ beschrieben werden.%
\footnote{%
Meist verwendet man die Spektralamplituden $\psi(q)$ der Zerlegung
\begin{equation}
|\Psi> = \int_Q  dq \, |q> \, <q| \Psi> =: \int_Q dq \, |q> \, \psi(q)
\end{equation}
des abstrakten Zustandsvektors $|\Psi>$
nach Eigenzust\"anden $|q>$ des Ortsperators $q$, im Volksmund
auch {\sc Schr\"odinger}-Wellenfunktionen genannt.}
Der Erwartungswert einer Observablen $A$ ist in diesem Fall gegeben durch
\begin{equation}
<A> = <\Psi|A|\Psi> = \mbox{Spur} \, \lbrace \, |\Psi><\Psi|A \, \rbrace.
\end{equation}
In einem {\it gemischten Zustand}
ist der Erwartungswert der Observable $A$ gegeben durch
\begin{equation}
<A>=\mbox{Spur}\,\lbrace \varrho A \rbrace,
\end{equation}
wobei der sogenannte {\it Dichteoperator\/} oder die {\it Dichtematrix\/}
geschrieben werden kann als
\begin{equation}
\varrho = \sum_i \lambda_i\,|\Psi_i><\Psi_i|,
\end{equation}
mit den statistischen Gewichten
\begin{equation}
\lambda_i > 0, \phantom{xxx} \sum_i \lambda_i = 1.
\end{equation}
Die {\it thermische Dichtematrix\/}
\begin{equation}
\varrho = \exp\,-\beta H
\end{equation}
ist nichts anderes als ein Operator, welcher in Analogie zum
{\sc Maxwell}-{\sc Boltzmann}-Faktor gesehen werden mu\ss\ und
einen gemischten Zustand repr\"asentiert, der ein quantenmechanisches
System im thermodynamischen Gleichgewicht bei endlicher Temperatur
beschreibt.
\par
Ein System von Fermionen wird nun nicht durch den
{\sc Maxwell}-{\sc Boltzmann}-Faktor, sondern durch
die {\sc Fermi}-Verteilungsfunktion
\begin{equation}
f(E) = \frac{1}{1 + \exp\,\beta(E - \mu)}
\end{equation}
charakterisiert, so da\ss\ der Erwartungswert
einer Observable $A$ f\"ur ein durch ein
{\sc Ha\-mil\-ton}-Operator $H$ beschriebenes
fermionisches System mit Volumen $V$ die Form
\begin{equation}
<A>_{\beta,\mu}
=
\mbox{Spur}_V\,\lbrace\,f(H)A\,\rbrace
\end{equation}
hat. $\,\mbox{Spur}_V\,$ deutet an, da\ss\ die Summe
\"uber die Diagonalelemente der betrachteten
Operatoren in Wirklichkeit Integrale \"uber
ein endliches Volumen $V$ sind. Besondere Vorsicht
ist geboten, wenn wir die Systemgr\"o\ss e gegen
unendlich gehen lassen, das hei\ss t den thermodynamischen
Limes durchf\"uhren.
\par
Zur Beantwortung der zu Beginn dieses Abschnitts
gestellten Frage nach dem quantenmechanischen
Pendant des {\sc Hall}-Effekts eines unendlich
ausgedehnten 2-dimensionalen Systems von
Elektronen berechnen wir nun den Zeitmittelwert
des Erwartungswertes des elektrischen Stroms. Es ist
\begin{eqnarray}
\overline{<{\bf j}_{2D}(t)>}
   &=& \lim_{T\rightarrow\infty} \frac{1}{T}
       \int_0^T dt <{\bf j}_{2D}(t)>_{\beta,\mu}                  \nonumber\\
   &\phantom{=}&                                                  \nonumber\\
   &=& e\cdot
       \lim_{V\rightarrow\infty} V^{-1}
       \mbox{Spur}_V\,\lbrace\,f(H_L)\,\rbrace
      \cdot\frac{{\bf B}\times{\bf E}}{{\bf B}^2}                 \nonumber\\
   &=& e \cdot n_{2D} \cdot \frac{{\bf B}\times{\bf E}}{{\bf B}^2},
\end{eqnarray}
was impliziert, da\ss\ die 2-dimensionale Ladungstr\"agerdichte
\begin{equation}
       n_{2D}(\beta,\mu)
       = \lim_{V\rightarrow\infty} \,
         \mbox{Spur}_V \lbrace \, f(H_L) \, \rbrace
\end{equation}
 eine glatte Funktion in $\beta$ und $\mu$ ist.
\par
Fazit: Wir reproduzieren das klassische Resultat.
Da es keinen Grund f\"ur die Erwartung gibt,
da\ss\ in einem realen System das chemische Potential $\mu$
immer zwischen den {\sc Landau}-Niveaus hin- und herspringt,
insbesondere von L\"uckenmitte zu L\"uckenmitte,
schlie\ss en wir aus dem Ergebnis, da\ss\
mit dem Auff\"ullen der {\sc Landau}-Niveaus die
{\sc Hall}-Leitf\"ahigkeit {\it kontinuierlich\/} steigen
m\"u\ss te. Wir m\"ussen also nach einem {\it zus\"atzlichen\/}
Mechanismus suchen, wenn wir eine befriedigende Erk\"arung
des beobachteten Quanteneffektes fin\-den wol\-len.
\par
Die Untersuchung dieses Problems definiert ein aktuelles
Forschungsgebiet. Selbst heute sind noch viele Fragen offen
\cite{Janssen94}, vielleicht aber ist auch das grundlegende
Prinzip noch gar nicht verstanden.
Bevor man sich im Detail mit diesen Fragen auseinandersetzt,
sollte man zun\"achst die experimentelle Realisierung der
zugeh\"origen physikalischen Systeme verstehen.
\vfill\eject\noindent%
\section{Von der Theorie zur Messung} 
\subsection{Experimentelle Realisierung
            von 2-dimensionalen Elektronengasen}
2-dimensionale Elektronen- oder L\"ochergase
k\"onnen an Halbleiter-Grenz\-fl\"a\-chen,
so zum Beipiel in einer
{\it Metall-Oxid-Halbleiter-Struktur (MOS)\/}
oder an einer Grenzfl\"ache
zwischen Halbleitern verschiedener Bandl\"ucke,
einer sogenannten {\it Heterostruktur\/},
realisiert werden
\cite{AndoFowlerStern82, Sze81}.
Im ersten Fall ist es die extern angelegte Gatespannung,
im zweiten Fall die geeignete Dotierung
zusammen mit einem Bandl\"uckensprung,
die eine Bandverbiegung verursacht, welche
zu einem in erster N\"aherung dreieckf\"ormigen Potentialtopf
f\"uhrt.
Durch die endliche Ausdehnung
des elektronischen Systems in Richtung der
dritten Dimension zeichnen sich die Elektronen (oder L\"ocher)
durch quantisierte Energieniveaus in einer Richtung aus
- ganz \"ahnlich wie im Fall des in den Quantenmechanik-Lehrb\"uchern
diskutierten Kastenpotentials. Im betrachteten Fall des
Dreieckpotentials sind die Energie-Eigenfunktionen
allerdings {\sc Airy}-Funktionen.
\bild{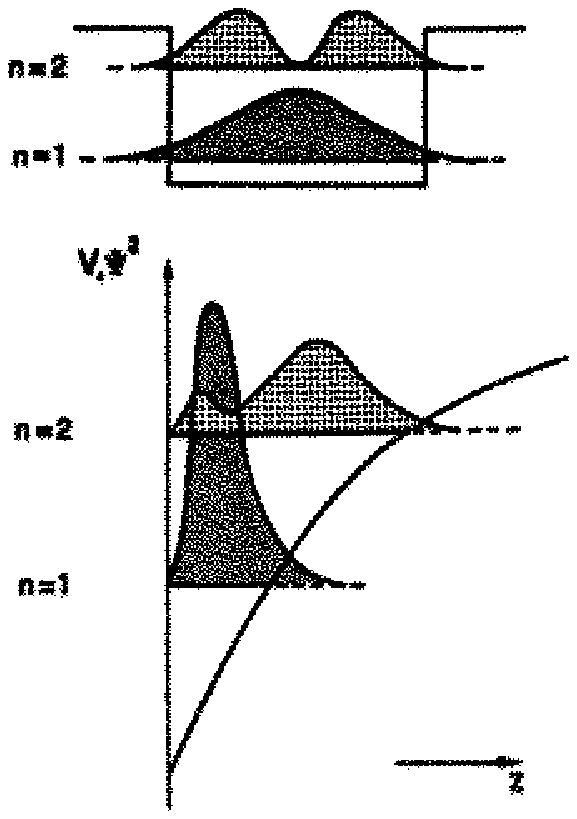}{Kastenpotential versus Dreieckspotential}{6}
\par
W\"ahrend die Elektronen (oder L\"ocher) in Richtung der z-Achse
durch die dargestellten Randbedingungen eingeschr\"ankt sind,
k\"onnen sie sich in den anderen beiden Richtungen frei bewegen.
Somit ist ihr Wellenzahlvektor ${\bf k}$ eine gute Quantenzahl
lediglich f\"ur die zwei Dimensionen x und y,
nicht aber f\"ur die dritte, die z-Richtung.
Somit erhalten wir eine Reihe von sogenannten
{\it Subb\"andern\/} $0,1,\dots$ f\"ur jeden
Energieeigenwert $E_0,E_1,\dots\;$.
In der N\"ahe des absoluten Nullpunktes ist nur
das unterste Niveau oder Subband besetzt.
Das System wird somit physikalisch
{\it exakt\/} 2-dimensional.
\bild{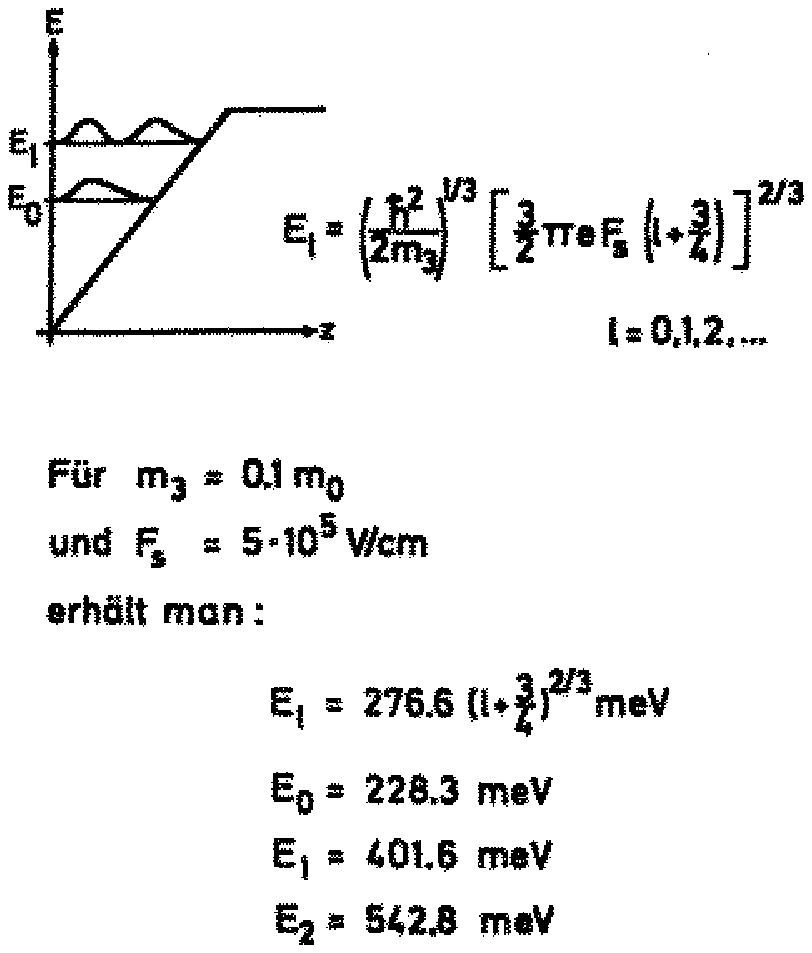}{Energieniveaus im Dreieckspotential}{9}
\par
Die diesen Plateaux zugeordneten
{\sc Hall}-Leitf\"ahigkeiten sind quantisiert gem\"a\ss
\begin{equation}
\sigma_H=i\cdot\frac{e^2}{h}.
\end{equation}
Den Plateaux entsprechen Nullstellen in der longitudinalen
Leif\"ahigkeit $\sigma_{xx}$.
Es soll hier nur erw\"ahnt werden, da\ss\ sp\"atere Experimente
von {\sc Tsui}, {\sc St\"ormer} und {\sc Gossard}
in Heterostrukturen hoher Elektronenbeweglichkeit
unter sehr starken Magnetfeldern sogar
Plateaux zu gebrochenen Werten (haupts\"achlich
mit ungeradem Nenner) zeigten
\cite{Tsui82}.
Dieser fraktionell-quantisierte
{\sc Hall}-Effekt ist ein Ph\"a\-no\-men f\"ur sich und soll hier
nicht im Detail diskutiert werden
\cite{Chakraborty1988}.
\par
Es mu\ss\ hervorgehoben werden, da\ss\ die Kinematik des
{\sc Hall}-Effektes zusammen mit der {\sc Landau}-Quantisierung
die Existenz der Plateaux {\it nicht\/} erkl\"art. Ein zus\"atzlicher
Aspekt mu\ss\ hinzukommen, n\"amlich ein physikalischer Mechanismus,
aus dem f\"ur die {\sc Hall}-Leit\-f\"a\-hig\-keit bei kontinuierlicher
und gleichm\"a\ss iger Erh\"ohung des Magnetfeldes der Wechsel
von Ansteigen und Verharren auf dem quantisierten Wert folgt.
Es herrscht \"Uber\-einstimmung darin, da\ss\ dieser Mechanismus
mit der Existenz lokalisierter Elektronenzust\"ande zwischen
zwei {\sc Landau}-Niveaus zusammenh\"angt. Das Auff\"ullen dieser
lokalisierten Zust\"ande ver\-\"an\-dert den Transport-Koeffizienten
nicht. Hierdurch werden die Punkte, f\"ur welche die
Quantisierungsbedingung erf\"ullt ist,
zu einer plateauf\"ormigen Linie ausgezogen.
\par
Elektronen-Lokalisierung h\"angt mit Unordnung in
kondensierter Materie zusammen: Etwas salopp formuliert,
ist der integral-quantisierte {\sc Hall}-Effekt ein
\loq Dreckeffekt\hiq, das hei\ss t, er ben\"otigt
die Pr\"asenz von Streuzentren bzw.\ St\"orstellen.
In dieser Hinsicht steht er im Gegensatz zum
fraktionell-quantisierten {\sc Hall}-Effekt,
der nur in ultra-reinen Proben zu beobachten ist.
Die schwierige physikalische Frage besteht nun darin,
wie es kommt, da\ss\ sich Lokalisierung und Magnetotransport
so arrangieren k\"onnen, da\ss\ wir das Ph\"a\-no\-men des
integral-quantisierten {\sc Hall} beobachten.
Die Genauigkeit der Quantisierung und die M\"oglichkeit,
im Rahmen eines Festk\"orper-Experiments vermittels
Bestimmung der sogenannten {\sc von\,Klitzing}-Konstante
\begin{equation}
R_{vK}=\frac{h}{e^2}=25.812\,805\,{\dots}{\rm k}\Omega
\end{equation}
durch \loq simple Widerstandsmessungen\hiq\
die {\sc Sommerfeld}sche Feinstrukturkonstante
\begin{equation}
\alpha = \frac{1}{\hbar c}\cdot\frac{e^2}{4\pi\varepsilon_0}
       = \frac { \mu_0 c e^2 } {2h }
\end{equation}
{\it unabh\"angig\/} von Feinheiten der Probengeometrie
bestimmen zu k\"onnen, deutet auf einen
{\it topologischen Quantisierungs\-mechanismus\/} hin -
ganz \"ahnlich der Quantisierung des magnetischen Flusses
durch einen supraleitenden Ring
\cite{Kinoshita96}.
Solche Quantisierungen sind
er\-fah\-rungs\-ge\-m\"a\ss\
sehr robust, m\"u\ss ten also zum Beispiel
unsensibel gegen\"uber der Art und Form der Verunreinigung
sein. Interessanterweise scheint aber die {\it Existenz\/}
der letzteren gerade eine {\it Voraussetzung\/} f\"ur das
Zustandekommen des Effektes zu sein.
\par
Unter Experimentalphysikern und Ph\"anomenologen
gilt der quantisierte {\sc Hall}-Effekt heute
als verstanden: Empirisch gesehen wei\ss\ man
ziemlich genau, was passiert, und kann auch bei
komplizierten Probengeometrien und -topologien
verl\"a\ss liche Vorhersagen machen.
Der Stand der theoretischen Forschung allerdings ist der,
da\ss\ es eine vollst\"andig ausgearbeitete
und gleichzeitig rigorose Theorie des quantisierten
{\sc Hall}-Effektes im Rahmen der Quantentransporttheorie
(der Quantenfeldtheorie des Nichtgleichgewichts) nicht gibt.
Strenggenommen gibt es nicht einmal einen Beweis daf\"ur,
ob die {\sc von\,Klitzing}-Konstante tats\"achlich mit der
{\sc Sommerfeld}schen Feinstrukturkonstante
({\it modulo\/}
einer Umrechnungsvorschrift) gleichzusetzen ist
(man denke an Abweichungen von dem Typ eines {\sc Lamb}-Shifts).
Aber es sind gerade die topologischen Argumente,
die - im Rahmen fundierter semiph\"anomenologischer
Theorien - darauf hindeuten, da\ss\ die
{\sc von\,Klitzing}\-sche
Quantisierungsregel tats\"achlich ein fundamentales Naturgesetz
darstellt.
Die Situation ist vielleicht vergleichbar mit der in der
Quantenelektrodynamik: Obwohl die Theorie noch Fehlstellen und
Inkonsistenzen beinhaltet, z\"ahlt sie zu den erfolgreichsten
Konzepten der modernen Physik. Im Literaturverzeichnis ist
eine Liste von Fachartikeln und Lehr\-b\"u\-chern
\"uber die quantisierten {\sc Hall}-Effekte zu finden,
auf die der interessierte Leser (und vielleicht der
zuk\"unftige Forscher auf diesem Gebiet) verwiesen sei.
\vfill\eject\noindent%
\subsection{MOS-(metal-oxide-semiconductor-)Strukturen}
Eine MOS-(metal-oxide-semiconductor-)Struktur ist
- wie der Name schon andeutet - ein Schichtsystem
mit der Abfolge Metall-Siliziumdioxid-Silizium.
Letzteres ist entweder eine Si-MOS-Struktur in
p-dotierten Bereichen
(man spricht auch von {\it p-Substraten\/} oder {\it p-Wannen\/})
oder in n-dotierten Bereichen
({\it n-Substraten\/} oder {\it n-Wannen\/}).
Die Metall-Schicht wird als sogenannte Gate-Elektrode
verwendet, da\ss\ hei\ss t, die an sie angelegte Spannung $V_G$
bestimmt Bandstruktur und Ladungstr\"agerverteilung.
Die Einstellung der Besetzungsgrenze der Zust\"ande im Metall
durch $V_G$ bestimmt die Verbiegung der Bandstruktur im Silizium.
\par
Bekanntlich unterscheiden wir in dotierten Halbleitern
zwischen Majorit\"atsladungs\-tr\"a\-gern (das sind diejenigen,
welche durch die Dotierung in der Mehrzahl auftreten) und
Minorit\"atsladungstr\"agern (das sind diejenigen,
welche infolge der thermischen Anregung von Elektronen-Loch-Paaren
als Minderheit immer vorhanden sind).
Beschr\"anken wir uns auf die Betrachtung von p-MOS-Systemen.
In diesem Fall sind die Majorit\"atsladungstr\"ager die L\"ocher,
die Minorit\"atsladungstr\"ager die Elektronen.
Wir unterscheiden drei F\"alle:
\begin{list}
   {
    }                                                                           
   {\setlength{\leftmargin}{3.75cm} 
    \setlength{\labelsep}{0.75cm}   
    \setlength{\labelwidth}{2.75cm} 
    \setlength{\itemsep}{0cm}       
    }
\item[$V_G < 0$]
Die Valenzbandkante $E_V$ r\"uckt an die {\sc Fermi}-Energie $E_F$,
und es ist f\"ur die L\"ocher g\"unstiger, sich nahe der
Grenzfl\"ache anzuordnen. Wenn $E_V>E_F$ wird, entsteht
an der $\mbox{\rm Si-SiO}_2$-Grenzfl\"ache eine positive Raumladung
von L\"ochern. Dieser Zustand wird Anreicherung (Akkumulation) genannt.
({\it Achtung\/}:
Wir d\"urfen noch einmal daran erinnern,
da\ss\ wir gerade den Fall eines p-Substrats behandeln.)
\item[$V_G = V_{FB} \approx 0$]
Ist $V_G=0$, so verlaufen die B\"ander nicht flach, da im Oxid
Ladungen eingelagert sind, welche die B\"ander verbiegen. Durch eine
angelegte Spannung, der sogenannten Flachbandspannung $V_{FB}$,
kann man diese Verbiegung kompensieren. Wird $V_G<V_{FB}$, tritt
zun\"achst Verarmung (Depletion) auf, da es f\"ur die L\"ocher
energetisch g\"unstiger ist, sich von der Oberfl\"ache entfernt
anzuordnen. ($V_{FB}$ ist nahe null Volt, daher wird hier
$V_{FB} \approx 0$ angenommen.)
\item[$V_G > 0$]
Die Leitungsbandkante $E_L$ wird unter die {\sc Fermi}-Energie $E_F$%
\linebreak
gedr\"uckt, und Elektronen aus dem Valenzband fallen in
die so entstehende dreieckf\"ormige Potentialtasche und
besetzen somit die Zust\"ande nahe der Grenzschicht. Es
entsteht ein 2-dimensionales Elektronengas.
Der Wert von $V_G$ bei Einsetzen dieser sogenannten Inversion nennt
man Threshold-Spannung $V_{th}$.
\end{list}
\bild{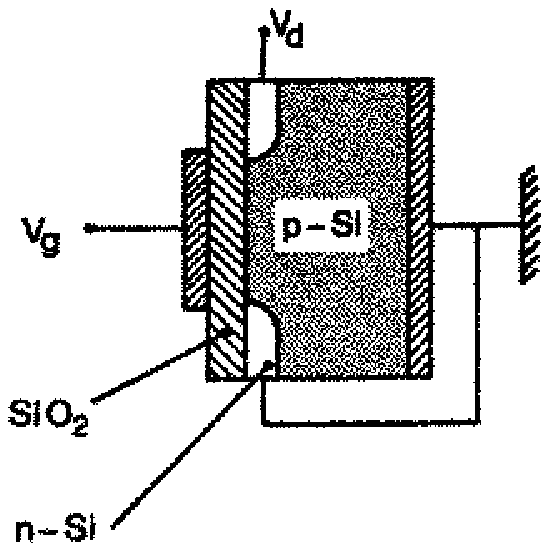}{p-MOS-Struktur}{6}
\par
F\"ur eine Si-MOS-Struktur auf einem n-Substrat gilt
entsprechendes; das hei\ss t, die Ver\-h\"alt\-nis\-se drehen
sich um. Insgesamt haben wir also sechs physikalisch
unterschiedliche F\"alle zu unterscheiden.
\par
Eine Si-MOS-Struktur kann als Kondensator der Fl\"ache $A$
und Dicke $d$ gef\"ullt mit einem Dielektrikum der Permittivit\"at
$\varepsilon_{{\rm SiO}_2}$ aufgefa\ss t werden, so da\ss\ wir
f\"ur die Kapazit\"at setzen d\"urfen
\begin{equation}
C=\frac{\varepsilon_{{\rm SiO}_2}\varepsilon_0 A}{d},
\end{equation}
mit $\varepsilon_0=8.8542\cdot 10^{-12}\,C/Vm$
und $\varepsilon_{{\rm SiO}_2}=3.8$.
F\"ur die 2-dimensionale Ladungstr\"agerdichte
$n_{2D}$
ergibt sich die sogenannte
{\it Kondensatorformel\/}
\begin{equation}
n_{2D}=\frac{C}{eA}\cdot(V_G-V_{th})
      =\frac{\varepsilon_{{\rm SiO}_2}\varepsilon_0}{ed}
       \cdot(V_G-V_{th}),
\end{equation}
die insbesondere im Falle der Inversion von Bedeutung ist.
\par
Silizium ist ein indirekter Halbleiter:
Das absolute Minimum der Leitungsbandkante
liegt nicht bei ${\bf k}=0$ sondern bei endlichen k-Werten.
In [100]-Ober\-fl\"a\-chen\-ori\-en\-tie\-rung haben wir
zwei T\"aler
(engl.\ {\it valleys\/}),
die - obwohl im k-Raum nicht wegzusammenh\"angend - energetisch
\"aquivalent sind. Somit sind die energetischen Zust\"ande in
[100]-ober\-fl\"a\-chen\-ori\-en\-tier\-tem Silizium zweifach entartet.
So ergibt sich f\"ur die Entartung eines Landauniveaus f\"ur
[100]-Silizium
\begin{equation}
n_{L} = g_s g_v \cdot \frac{eB}{h}=4 \cdot \frac{eB}{h}
\end{equation}
mit
\begin{equation}
g_s=2
\end{equation}
als Spinentartungsfaktor und
\begin{equation}
g_v=2
\end{equation}
als Valleyentartungsfaktor.
Die Valleyentartung wird nur in sehr
starken Magnetfeldern me\ss bar aufgehoben.
\vfill\eject\noindent%
\subsection{Heterostrukturen}
Eine Heterostruktur ist ein Schichtsystem aus verschiedenen Materialien
(griech.\ {\it heteros\/} = ver\-schieden),
hier aus Halbleitern verschiedener Bandstruktur.
Heterostrukturen werden realisiert durch Molekularstrahlepitaxie,
einem programmgesteuerten Pr\"azisionsverfahren f\"ur das Wachstum
wohldefinierter Kristalle
\cite{Herman89}.
Die gemeinsame Verwendung von Galliumarsenid und
Aluminium-Gallium-Arsenid erm\"oglicht die Kombination
von Halbleitern verschiedener Bandl\"ucke bei gleicher
Gitterkonstante, mit anderen Worten: ein Aufwachsen
von verschiedenen Schichten ohne Fehlanpassung
(engl.\ {\it mismatch\/}).
\bild{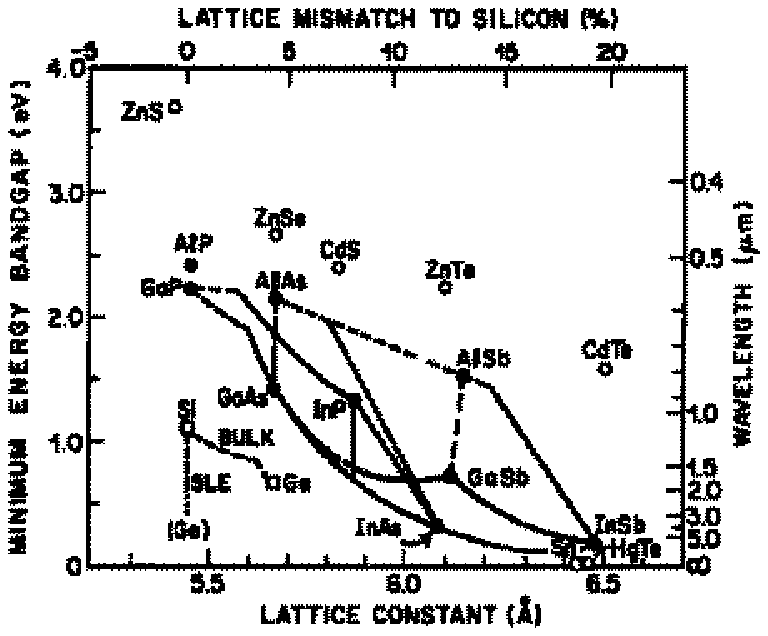}{Bandl\"ucken versus Gitterkonstanten von
                III-V-Halbleitern}{9}
\par
Eine typische Al$_x$Ga$_{1-x}$As-GaAs-Heterostruktur sieht wie folgt aus:
Auf einem semi-isolierenden Substrat aus Galliumarsenid
werden nacheinander aufgewachsen
\begin{enumerate}
\item $1-4\,\mu{\rm m}$            GaAs,
\item $10-40\,nm$ undotiertes      Al$_x$Ga$_{1-x}$As
                                   (der sogenannte {\it Spacer\/}),
\item $20-50\,nm$ Si-(n-)dotiertes Al$_x$Ga$_{1-x}$As,
\item $10-20\,nm$                  GaAs
                                   (die sogenannte {\it Cap\/}).
\end{enumerate}
Im thermischen Gleichgewicht mu\ss\ die {\sc Fermi}-Energie $E_F$
\"uber die verschiedenene Grenz\-fl\"a\-chen hinweg konstant sein.
Die Erf\"ullung dieser Bedingung erzwingt die Verbiegung
der Leitungsbandkante zur Grenzfl\"ache hin.
Der energetische Unterschied der Bandl\"ucken
von GaAs und Al$_x$Ga$_{1-x}$As ist so gro\ss,
da\ss\ die Leitungsbandkante des GaAs-Puffers an der
$\mbox{Al$_x$}$$\mbox{Ga$_{1-x}$}$$\mbox{As}$-GaAs-Grenzfl\"ache
bis unter die {\sc Fermi}-Energie $E_F$ gedr\"uckt wird.
Dort bildet sich ein n\"aherungsweise dreieckf\"ormiges
Potential unterhalb von $E_F$ aus, welches vergleichbar ist
mit dem oben beschriebenen Potentialverlauf in MOS-Inversionsschichten.
Somit besetzten die Elektronen nahe an
der Al$_x$Ga$_{1-x}$As-GaAs-Grenzfl\"ache
Zust\"ande im Leitungband und bilden
durch ihre Beweglichkeit in x- und y-Richtung
eine leitf\"ahige Schicht, das 2-dimensionale
Elektronengas.
\bild{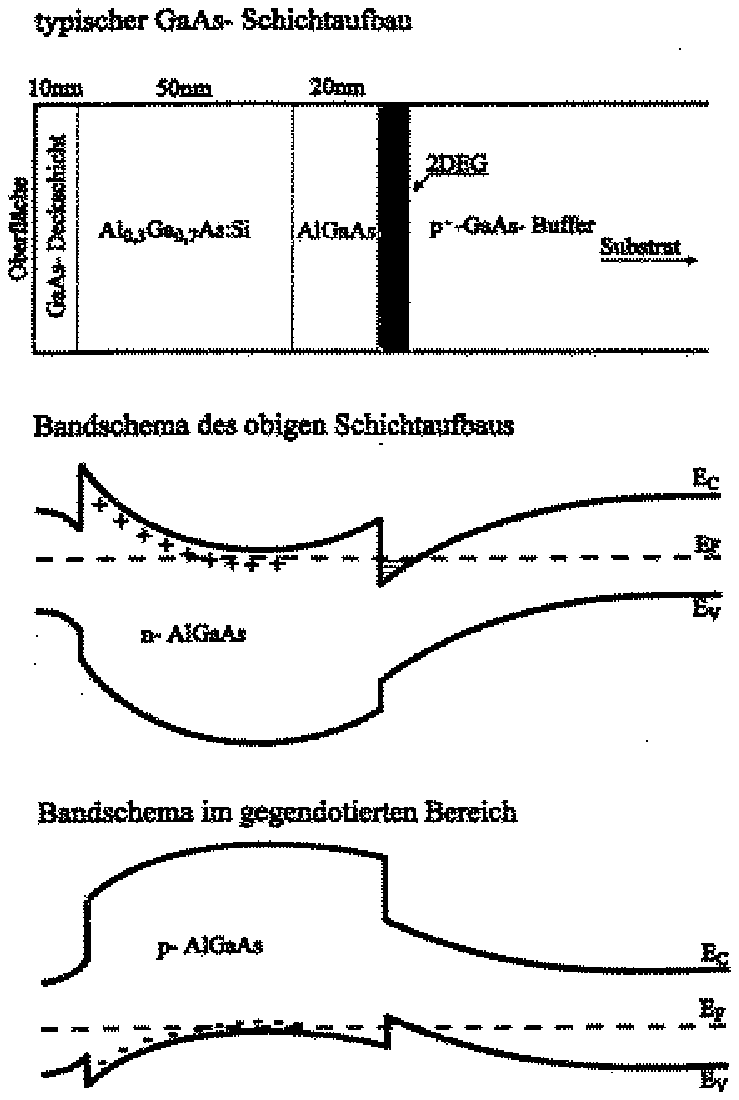}{Heterostruktur}{9}
\par
Die Beweglichkeit der Elektronen wird
durch die Streueffekte im Kristall eingeschr\"ankt.
Der Spacer wird eingef\"ugt, um die Elektronen
in der Grenzschicht m\"oglichst weit von den
ionisierten Donatoratomen (hier Silizium als
IV-Element auf IIIer-Pl\"atzen) in der
dotierten Al$_x$Ga$_{1-x}$As-GaAs-Schicht
zu trennen.
\par
Da Galliumarsenid ein direkter Halbleiter ist,
gibt es hier keine Valleyentartung; die
{\sc Fermi}-Fl\"ache ist die Oberfl\"ache einer Kugel.
\vfill\eject\noindent%
\subsection{Vom 2-dimensionalen zum 3-dimensionalen
            Elek\-tro\-nen\-gas: Sub\-b\"an\-der}
Der {\sc Hamilton}-Operator eines quantenmechanischen
Systems freier Elektronen in drei Raumdimensionen besitzt
ein nach unten beschr\"anktes und nach oben unbeschr\"anktes
kontinuierliches Spektrum. Die Energieeigenwerte
lassen sich als quadratische Funktion des kontinuierlichen
Wellenzahlvektors ${\bf k}$ auffassen und haben die Form
\begin{equation}
E({\bf k})=\frac{\hbar^2{\bf k}^2}{2m}
          =\frac{\hbar^2k_x^2+\hbar^2k_y^2+\hbar^2k_z^2}{2m}.
\end{equation}
Ist die Bewegung in eine der drei Raumrichtungen (\"ublicherweise
die z-Richtung) ein\-ge\-schr\"ankt, so wird das Spektrum der zugeordneten
Komponente des Wellenzahlvektors (z.\,B.\ $k_z$) diskret.
Im Idealfall wird eine solche Einschr\"ankung durch das Potential
eines Kastens mit unendlich hohen W\"anden definiert.
Die Betrachtung der realistischen Situation
eines Potentials mit n\"aherungsweise dreieckf\"ormigen Verlauf
unterscheidet sich davon qualitativ nicht, wohl aber quantitativ,
das hei\ss t, hinsichtlich der Abst\"ande der Energieniveaus und
der Ausdehnung der Wellenfunktionen. Auf jeden Fall k\"onnen wir
schreiben
\begin{equation}
E^j(k_x,k_y)=\frac{\hbar^2k_x+\hbar^2k_y}{2m}+E^j_z.
\end{equation}
Bei Diskretisierung einer Raumdimension also zerf\"allt
der Raum der Energieeigenwerte des 3-dimensionalen Systems
in eine diskrete Summe von Unterr\"aumen von Energie-Eigen\-werten,
die zu einem idealisierten 2-dimensionalen System geh\"oren.
\par
Nun haben wir in einem 3-dimensionalen Festk\"orper
infolge der Gitterperiodizit\"at nach dem Theorem von
{\sc Bloch} nicht einfach ein nach unten beschr\"anktes
und nach oben unbeschr\"anktes Energiekontinuum, sondern
Energieb\"ander vorliegen. Durch die Ein\-schr\"an\-kung
einer Raumdimension zerfallen diese somit in eine
abz\"ahlbare Menge 2-di\-men\-si\-o\-na\-ler Un\-ter\-b\"an\-der,
sogenannter Subb\"ander. F\"ur kleine Wellenzahlen gilt
\begin{equation}
E^j(k_x,k_y)=\frac{\hbar^2k_x+\hbar^2k_y}{2m}+E^j_z,
\end{equation}
wobei
\begin{equation}
m = 0.067 \, m_e
\end{equation}
die hier in der freien Elektronenmasse $m_e$ ausgedr\"uckte
effektive Masse des Elektrons in Galliumarsenid ist.
\par
Wenn man die St\"arke der Dotierung in
Al$_x$Ga$_{1-x}$As so w\"ahlt, da\ss\
die Anzahl der Elektronen in der Grenzschicht
nur zur Besetzung des untersten Subbandes $j=0$ ausreicht,
k\"onnen wir davon ausgehen, da\ss\ im Limes
verschwindender Temperatur tats\"achlich nur
das unterste Subband besetzt ist. Zwar hat
das betrachtete physikalische System noch eine
Ausdehnung in z-Richtung, aber es besteht keine
kinematische Freiheit in dieser Richtung mehr;
das System verh\"alt sich,
\loq wie wenn es\hiq\ (lat.\ {\it quasi\/})
exakt 2-dimensional w\"are: Wir sprechen daher auch von einem
quasi-2-dimensionalen Elektronengas.
\par
Erlauben wir uns an dieser Stelle einen kleinen Exkurs und
berechnen die Zu\-stands\-dich\-te f\"ur ein Elektronengas in
$n$ Raumdimensionen. Das Volumen einer $n$-dimensionalen Kugel
oder $n$-{\it Disk\/} ${\bf D}^n$ mit Radius $r$ ist bekanntlich
gegeben durch
\begin{equation}
\mbox{{\rm Vol}}({\bf D}^n)=
\frac{\pi^{n/2}}{\Gamma(\frac{n}{2}+1)}\,r^n.
\end{equation}
Wir erinnern daran, da\ss\ die $\Gamma$-Funktion
die Eigenschaften
\begin{eqnarray}
\Gamma(n+1) &=& n!\,,  \phantom{1234}
                       \mbox{{\rm f\"ur}}
                       \phantom{x}
                       n=0,1,2,\dots,         \\
\Gamma(x+1) &=& x\,\Gamma(x),                 \\
\Gamma(1/2) &=& \sqrt{\pi}
\end{eqnarray}
erf\"ullt. Insbesondere ist
\begin{eqnarray}
\mbox{{\rm Vol}}({\bf D}^3)  &=& \frac{4}{3} \pi \cdot r^3,  \\
                             &\phantom{=}&                   \nonumber\\
\mbox{{\rm Vol}}({\bf D}^2)  &=&             \pi \cdot r^2,  \\
                             &\phantom{=}&                   \nonumber\\
\mbox{{\rm Vol}}({\bf D}^1)  &=& 2               \cdot r  .
\end{eqnarray}
Damit erhalten wir f\"ur eine $n=3,2,1$-dimensionale
{\sc Fermi}-Kugel
\begin{eqnarray}
\mbox{{\rm Vol}}({\bf D}^3_F)  &=&
       \frac{4}{3} \cdot \pi k^3_F  \;=\;
       \frac{4}{3} \cdot \pi \cdot \frac{(2mE_F)^{3/2}}{\hbar^3},  \\
                                     & & \nonumber           \\
\mbox{{\rm Vol}}({\bf D}^2_F)  &=&
                   \pi k^2_F  \;=\;
                   \pi \cdot \frac{ 2mE_F       }{\hbar^2},  \\
                               & & \nonumber           \\
\mbox{{\rm Vol}}({\bf D}^1_F)  &=&
                      2k_F    \;=\;
                       \frac{2{(2mE_F)}^{1/2}}{\hbar}.
\end{eqnarray}
In einem spin- und valleyentarteten $n$-dimensionalen System
(mit einem w\"urfelf\"ormigen Volumen der Kantenl\"ange $L$)
ist die Gesamtzahl der erlaubten Zust\"ande%
\footnote{%
Man beachte, da\ss\ diese Formel von verschwindenden bzw.\
periodischen Randbedingungen f\"ur die Wellenfunktionen
ausgeht \cite{Kittel}.}
gegeben durch
\begin{equation}
n_{nD} (E_F) = g_s g_v
               \cdot
               \frac{ {\rm Vol}({\bf D}^n_F) } { (2\pi/L)^n },
\end{equation}
wobei wir einen Faktor $L^n$ unterdr\"ucken, wenn wir uns
- wie \"ublich - auf ein Einheitsvolumen beziehen wollen.
F\"ur die Herleitung dieser f\"ur die Festk\"orperphysik
fundamentalen Formel sei der Leser zum Beispiel auf
das Lehrbuch von {\sc Kittel} verwiesen \cite{Kittel}.
\par
F\"ur die Zu\-stands\-dich\-ten in $n=3,2,1$ Raumdimensionen erhalten
wir somit
\begin{eqnarray}
    & &                                \nonumber\\
D_{3D}(E) \;=\; \frac{d n_{3D} (E)}{dE}
     &=& g_s g_v \cdot \frac{4\pi m\sqrt{2mE}}{h^3},            \\
     & &                                \nonumber\\
D_{2D}(E) \;=\; \frac{d n_{2D} (E)}{dE}
     &=& g_s g_v \cdot \frac{2\pi m}          {h^2},            \\
     & &                                \nonumber\\
D_{1D}(E) \;=\; \frac{d n_{1D} (E)}{dE}
     &=& g_s g_v \cdot \frac{1}{h}\sqrt{\frac{2m}{E}},
\end{eqnarray}
wobei wir f\"ur $n$$=$$3$ einfach $E$$=$$E_F$ setzen,
in den niederdimensionalen F\"allen $n$$=$$2,1$ dagegen
\begin{equation}
E=E_F-E^j_z
\end{equation}
f\"ur ein Subband $j$.
\par
Wir erinnern noch einmal daran,
da\ss\ wir f\"ur den Spinentartungsfaktor
\begin{equation}
g_s=2
\end{equation}
setzen.
Der Valleyentartungfaktor $g_v$ ist nur f\"ur indirekte
Halbleiter wie Silizium von Bedeutung. F\"ur Galliumarsenid
setzen wir diesen Faktor gleich $1$.
\par
Damit sei unserer Exkurs \"uber
die Zu\-stands\-dich\-te in $n$ Dimensionen beendet.-
Wesentlich ist die Beobachtung,
da\ss\ die Subband-Zustanddichte
in zwei Raumdimensionen konstant ist (!).
F\"ur die 2-dimensionale Ladungstr\"agerdichte
erhalten wir im Falle der Besetzung des untersten Subbandes:
\begin{equation}
n_{2D} = \int_{E^0_z}^{E_F} D_{2D}(E)\,dE
       = g_s g_v \cdot \frac{2\pi m}{h^2} \cdot E,
\end{equation}
mit $E=E_F-E^0_z$.
\vfill\eject\noindent%
\subsection{2-dimensionales Elektronengas im Magnetfeld}
Wir betrachten nun den Fall, in dem ein Magnetfeld senkrecht
zur Grenzfl\"ache, an der sich das 2-dimensionale Elektronengas
ausgebildet hat, angelegt ist. Diese Anordnung wird auch
{\sc Faraday}-Geometrie genannt
(im Gegensatz zur {\sc Vogt}-Geometrie
$B\parallel\mbox{{\rm Fl\"ache}}$).
\par
Die klassische Kinematik folgt
aus der Bedingung, da\ss\ sich {\sc Lorentz}-Kraft und Zentralkraft
die Waage halten m\"ussen. Dies ist genau dann der Fall, wenn
sich die Elektronen auf Kreisbahnen mit der {\it Zyklotronfrequenz\/}
\begin{equation}
\omega_c=\frac{eB}{m}
\end{equation}
bewegen. Die Projektion einer solchen Bewegung entspricht
der eines harmonischen Oszillators. Es ist daher naheliegend
zu vermuten, da\ss\ die quantenmechanische Behandlung
der vorliegenden Situation zu einer Quantisierungsregel
f\"uhrt, welche an die quan\-ten\-me\-cha\-ni\-sche Behandlung
des harmonischen Oszillators erinnert. In der Tat zeigt
man durch L\"o\-sen der zugeordneten {\sc Schr\"odinger}-Gleichung,
da\ss\ die Elektronen quantisierte E\-ner\-gie\-ei\-gen\-wer\-te
mit dem Abstand $\hbar\omega_c$ einnehmen, die sogenannten
{\sc Landau}-Niveaus
\cite{LandauLifshitz}.
F\"ur die Gesamtenergie der
Elektronen erh\"alt man\begin{equation}
E^j_n=\hbar\omega_c(n+\frac{1}{2})+E^j_z,
\end{equation}
mit $n=0,1,2\dots\;$.
Da sich die Elektronen nur noch auf {\sc Landau}-Niveaus
aufhalten d\"urfen, ist die Zu\-stands\-dich\-te nicht mehr
gleichm\"a\ss ig auf die Energien bis hin zu $E_F$ verteilt,
sondern kondensiert in gleichm\"a\ss iger Weise auf den
Werten der {\sc Landau}-Niveaus. F\"ur die Anzahl der
Zust\"ande auf einem {\sc Landau}-Niveau, dem sogenannten
{\it Entartungsgrad\/}, erhalten wir
\begin{equation}
n_{L} = \int_{E^j_n-\frac{1}{2}\hbar\omega_c}
            ^{E^j_n+\frac{1}{2}\hbar\omega_c}
              D_{2D}(E)\,dE
      = g_s g_v \cdot \frac{2\pi m}{h^2} \cdot \hbar\omega_c
      = g_s g_v \cdot \frac{eB}{h}.
\end{equation}
Der beschriebene Typ von Entartung kann auch als Drehimpulsentartung
aufgefa\ss t werden.
Die Anzahl gef\"ullter {\sc Landau}-Niveaus berechnet sich
aus der Anzahl $n_{2D}$ der Ladungstr\"ager pro Einheitsfl\"ache
bezogen auf den
Entartungsgrad $n_{L}$ eines {\sc Landau}-Niveaus:
\begin{equation}
\mbox{{\rm Anzahl der gef\"ullten {\sc Landau}-Niveaus}}
=  \frac{ n_{2D} }{ n_{L} }
=  \frac{1}{g_sg_v} \cdot \frac{h n_{2D} }{eB}
=: \frac{1}{g_sg_v} \cdot \nu.
\end{equation}
Der F\"ullfaktor $\nu$ gibt die Anzahl gef\"ullter
spin- und valley-aufgespalteter Niveaus an. Die
vollst\"andige F\"ullung eines {\sc Landau}-Niveaus
entspricht also einem F\"ullfaktor von zum Beispiel
$\nu=4$ in Silizium und einem F\"ullfaktor von
$\nu=2$ in Galliumarsenid.
\par
Beziehen wir die {\sc Zeemann}-Aufspaltung der Elektronen
im Magnetfeld mit ein, so erhalten wir f\"ur die Gesamtenergie
der Elektronen eines 2-dimensionalen Elektronengases im
Magnetfeld:
\begin{equation}
E^j_{n,s}
=   \hbar\omega_c(n+\frac{1}{2})
  + g^* \mu_B B \cdot b + E^j_z,
  \end{equation}
mit $s=\pm 1/2$ und $n=0,1,2,\dots\;$.
Durch Streuprozesse an Verunreinigungen und Defekten im Kristall
werden die im Idealfall scharf definierten {\sc Landau}-Niveaus
verbreitert. Diese Verbreiterung spielt eine bedeutende Rolle
im Verst\"andnis des quantisierten {\sc Hall}-Effektes.
\par
{\it Bemerkung:\/} Auch wenn wir von falschen Randbedingungen
f\"ur die Wellenfunktionen ausgegangen sind, erhalten wir
das richtige Ergebnis f\"ur die Entartung eines {\sc Landau}-Niveaus.
Die \"Ubereinstimmung erkl\"art sich aus der Tatsache, da\ss\
sich die {\sc Fermi}-Energie bzw.\ das chemische Potential
beim Einschalten des Magnetfeldes nicht \"andert. Durch Vergleich
mit der oben vorgestellten eichinvarianten Formulierung haben
wir dies hiermit sogar bewiesen.
\vfill\eject\noindent%
\subsection{Drude-Modell f\"ur den klassischen Hall-Effekt}
In der Einleitung haben wir den klassischen {\sc Hall}-Effekt
im Einteilchen-Bild hergeleitet. Im folgenden wollen wir
die Herleitung im Rahmen des {\sc Drude}-Bildes wiederholen.
Der Vorteil dieses Rahmens besteht darin, da\ss\ er mikroskopische
Gr\"o\ss en wie die Relaxationszeit zwischen zwei St\"o\ss en
in Relation zu den makroskopischen Gr\"o\ss en wie spezifische
Leit\-f\"a\-hig\-keit und spezifischen Widerstand setzt.
\par
An dieser Stelle ist es n\"utzlich, die grundlegenden Annahmen
des {\sc Drude}-Modells zu rekapitulieren:
\begin{enumerate}
\item
Elektronen bewegen sich als freie Teilchen, wenn wir von den
St\"o\ss en mit den als harte Kerne dargestellten Ionen
(engl.\ {\it hard core ions\/})
absehen.
Sowohl die Elektron-Elektron-Wechselwirkung zwischen den
St\"o\ss en wird vernachl\"assigt
(engl.\ {\it independent electron approximation\/}),
als auch die Elektron-Ion-Wechselwirkung
(engl.\ {\it free electron approximation\/}).
\item
Ohne irgendwelche Annahmen \"uber den detailierten Mechanismus
der Streuprozesse zu machen, wird die Arbeitshypothese
aufgestellt, da\ss\ die Elektronen von Zeit zu Zeit
an den {\it hard core ions\/} streuen.
Durch einen derartigen Streuproze\ss\
\"andert sich die Impulsrichtung eines gestreuten Elektrons.
Die Streung kann als elastisch an\-ge\-nom\-men werden.
\item
Die Wahrscheinlichkeit einer Streuung bzw.\ die
\"Anderung der Impulsrichtung eines Elektrons wird
durch den Kehrwert einer Beruhigungszeit, der sogenannten
{\sc Drude}-Impulsrelaxationszeit $\tau$,
quantifiziert.
\item
Es wird angenommen, da\ss\ die Elektronen das thermische
Gleichgewicht (in Bezug auf ihre Umgebung) ausschlie\ss lich
durch die beschriebenen Streuprozesse erreichen.
\end{enumerate}
Eine kritische W\"urdigung der {\sc Drude}-Theorie
f\"uhrt unmittelbar zu fundamentalen noch offenen
Grundlagenfragen der Thermodynamik (mikroskopische
Reversibilit\"at versus makroskopische Irreversibilit\"at,
{\sc Boltzmann} versus {\sc Gibbs}, Me\ss proze\ss\ usw.).
Eine Einbeziehung der Quantenkinematik zeigt, da\ss\ sogar
die weiterf\"uhrende {\sc Boltzmann}sche Transporttheorie
f\"ur die vollst\"andige Beschreibung der Ph\"a\-no\-me\-ne,
die f\"ur uns von Interesse sind, unzureichend ist.
\par
Im folgenden soll lediglich plausibel gemacht werden,
da\ss\ die {\sc Drude}-Theorie f\"ur die dynamischen Gleichungen
der makroskopischen Transportgr\"o\ss en Reibungsterme liefert.
Ein wesentlicher Punkt ist, da\ss\ die Streuung im {\sc Drude}-Bild
zwar als elastisch an\-ge\-nom\-men werden darf, da\ss\ sie aber
aus Sicht der Quantenmechanik als inkoh\"arent betrachtet
werden mu\ss\ und daher schon ein Element der Irreversibilit\"at
(Entropiezunahme) auf mikroskopischer (oder sollten wir sagen
mesoskopischer?) Ebene einf\"uhrt. Daher ist es nicht verwunderlich,
da\ss\ sich die makroskopische Beschreibung auf eine effektive
dissipative Ki\-ne\-ma\-tik/Dy\-na\-mik reduziert.
\par
Es ist n\"utzlich, sich eine Analogie zwischen
der klassisch-mechanischen Dynamik und der Dynamik des elektrischen
Transports zu vergegenw\"artigen.
Das {\sc Newton}sche Gesetz
\begin{equation}
{\bf F} = m \, \frac{d{\bf v}}{dt}
\phantom{xxxxx}
\mbox{({\rm falls $m=const.$})}
\end{equation}
setzt die auf einen K\"orper ausge\"ubte Kraft ${\bf F}$
in Beziehung zu seiner Beschleunigung $\dot{\bf v}$ und
steht zum Reibungsgesetz
\begin{equation}
{\bf F}=\kappa{\bf v}
\end{equation}
in genau derselben Relation wie die
einen idealen Leiter beschreibende
2.\ {\sc London}sche Gleichung
\begin{equation}
{\bf E}=\mbox{{\it const\/}} \, \frac{d{\bf j}_{2D}}{dt}
\end{equation}
zum {\sc Ohm}schen Gesetz
\begin{equation}
{\bf E}=\varrho\,{\bf j}_{2D}.
\end{equation}
\par
{\it Bemerkung:\/}
Die 2.\ Londonsche Gleichung wird zuweilen auch
Beschleunigungsgleichung genannt. Sie beschreibt
einen idealen Leiter. Um das Ph\"a\-no\-men der
Supraleitung zu beschreiben (ideale Leitf\"ahigkeit
bei Verdr\"angung des Magnetfeldes) reicht sie allein
nicht aus!
\par
Heben wir noch einmal den wesentlichen Punkt hervor:
In einem realen Leiter bewegen sich die Elektronen
von Sto\ss\ zu Sto\ss\ ballistisch, werden also
auf der dazwischen liegenden Strecke beschleunigt.
Die mittlere Zeit dieser freien Bewegung ist die
{\sc Drude}-Im\-puls\-re\-la\-xa\-ti\-ons\-zeit $\tau$.
\"Ubrigens ist sie theoretisch sehr schwer zu bestimmen.
\par
Angenommen, ein
Elektron h\"atte nach einem Sto\ss\ gerade die
Geschwindigkeit ${\bf v}_0$
(fett\-ge\-schrie\-be\-ne Buchstaben stehen f\"ur Vektoren!),
so beschleunigt das
\"au\ss ere stromtreibende Feld ${\bf E}$ dieses
bis zum n\"achsten Sto\ss\ von ${\bf v}_0$ auf
${\bf v}_0-(e{\bf E}/m)t$, wobei $t$ die
verstrichene Zeit bezeichnet. Wenn wir annehmen,
da\ss\ alle m\"oglichen ${\bf v}_0$ in etwa genauso
h\"aufig vorkommen, so erhalten wir f\"ur die
mittlere Driftgeschwindigkeit
\begin{equation}
{\bf v}_{mean}=-\frac{e{\bf E}\tau}{m},
\end{equation}
mit $\tau$ als Zeitmittel und
f\"ur die makroskopische Stromdichte
in dem uns hier interessierenden Fall von zwei Raumdimensionen
\begin{equation}
{\bf j}_{2D}
        = - n_{2D} e{\bf v}_{mean}
        = \left(\frac{ n_{2D} e^2\tau}{m}\right)\,{\bf E}
        = \sigma {\bf E},
\end{equation}
wobei $n_{2D}$ - wie oben -
die 2-dimensionale Ladungstr\"agerdichte bezeichnet.
\par
{\it Bemerkung:\/}
Den Quotienten aus dem Betrag der mittleren Driftgeschwindigkeit
und dem Betrag der angelegten elektrischen Feld\-st\"ar\-ke
\begin{equation}
\mu=
\frac{|{\bf v}_{mean}|}{{\bf E}}=
\frac{1}{e}\cdot\frac{\sigma}{n_{2D}}=
\frac{e\tau}{m}
\end{equation}
nennen wir {\it Beweglichkeit\/} oder {\it Mobilit\"at\/}.
\par
Die im {\sc Ohm}schen Gesetz
\begin{equation}
{\bf j}_{2D}=\sigma{\bf E}
\end{equation}
durch die
{\sc Drude}-Impulsrelaxationszeit
ausgedr\"uckte Gr\"o\ss e
\begin{equation}
\sigma_0=\frac{n_{2D}e^2\tau}{m}
\end{equation}
nennen wir die {\sc Drude}-Leitf\"ahigkeit.
\par
Mit anderen Worten: Messen wir im Experiment
die spezifische Leitf\"ahigkeit $\sigma$ bzw.\
den spezifischen Widerstand $\varrho$, so k\"onnen
wir bei Kenntnis der Ladungstr\"agerdichte die
{\sc Drude}-Impulsrelaxationszeit $\tau$
bestimmen.
\par
Die Analogie zwischen dissipativer Mechanik und
Transporttheorie wird noch sichtbarer, wenn wir
die {\sc Newton}-Gleichung mit Sto\ss term herleiten.
Schreiben wir das {\sc Newton}\-sche Gesetz als
\begin{equation}
\frac{d{\bf p}(t)}{dt}={\bf f}(t),
\end{equation}
so erhalten wir f\"ur die Impuls\"anderung
\begin{equation}
d{\bf p}(t)={\bf f}(t)dt,
\end{equation}
was sich in endlicher N\"aherung liest als
\begin{equation}
\Delta{\bf p}={\bf p}(t+\Delta t)-{\bf p}(t)\approx{\bf f}(t)\Delta t
\end{equation}
bzw.\ als
\begin{equation}
{\bf p}(t+\Delta t)=
{\bf p}(t)+{\bf f}(t)\Delta t+O(\Delta t^2).
\end{equation}
Offensichtlich ist die Kollisionswahrscheinlichkeit
$\Delta t/\tau$, die Wahrscheinlichkeit f\"ur
das dazu komplement\"are Verhalten, das
\loq {\sc Newton}-Verhalten\hiq, gerade $1-\Delta t/\tau$.
Letzteren Term f\"uhren wir als Gewichtsfaktor in
die Absch\"atzung f\"ur den inkrementierten Impuls ein,
so da\ss\ wir setzen d\"urfen
\begin{eqnarray}
{\bf p}(t+\Delta t)
&=& \left( 1 - \frac{\Delta t}{\tau} \right)
    ({\bf p}(t)+{\bf f}(t)\Delta(t) + O(\Delta t^2)) \nonumber\\
&=& {\bf p}(t) + {\bf f}(t)\Delta t
    - \frac{\Delta t}{\tau} {\bf p}(t) + O(\Delta t^2).
\end{eqnarray}
F\"ur das Impuls-Inkrement erhalten wir somit
\begin{equation}
\Delta{\bf p}
= {\bf p}(t+\Delta t)-{\bf p}(t)
= \Delta t \left( {\bf f}(t)-\frac{{\bf p}(t)}{\tau} \right)
 + O(\Delta t^2),
\end{equation}
so da\ss\ die {\sc Newton}-Gleichung mit Sto\ss term
die folgende Form haben mu\ss:
\begin{equation}
\frac{d{\bf p}(t)}{dt}
=
{\bf f}(t)-\frac{{\bf p}(t)}{\tau}.
\end{equation}
Beschreibt ${\bf f}$ die {\sc Lorentz}-Kraft, so wird
die obige Gleichung zu
\begin{equation}
\frac{d{\bf p}(t)}{dt} =
e \left( {\bf E}+ \frac{{\bf p}}{m}\times{\bf B} \right)
- \frac{{\bf p}}{\tau}.
\end{equation}
Ein station\"arer Zustand stellt sich ein,
wenn die linke Seite dieser Gleichung verschwindet,
das hei\ss t, es ist in Komponenten geschrieben
\begin{eqnarray}
0 &=& e E_x + \omega_c p_y - \frac{p_x}{\tau}, \nonumber \\
0 &=& e E_y - \omega_c p_x - \frac{p_y}{\tau},
\end{eqnarray}
mit der {\it Zyklotronfrequenz\/}
\begin{equation}
\omega_c=eB/m.
\end{equation}
Multiplikation mit $n_{2D}e\tau/m$ und umarrangieren
der Summanden ergibt
\begin{eqnarray}
\sigma_0 E_x   :=   \left( \frac{n_{2D}e^2\tau}{m} \right) \, E_x
             &  = &          -  \omega_c \tau j_y + j_x,  \nonumber \\
\sigma_0 E_y   :=   \left( \frac{n_{2D}e^2\tau}{m} \right) \, E_y
             &  = & \phantom{+} \omega_c \tau j_x + j_y,
\end{eqnarray}
wobei $\sigma_0$ die {\sc Drude}-Leitf\"ahigkeit f\"ur $B \equiv 0$
bezeichnet. Unter der Bedingung, da\ss\ kein transversaler Strom
${\bf j}_{2D,y}$ flie\ss t, erhalten wir f\"ur das {\sc Hall}-Feld
\begin{equation}
E_y = \left( \frac{\omega_c\tau}{\sigma_0} \right) j_x
    = \left( \frac{B}{n_{2D}e} \right) j_x.
\end{equation}
F\"ur den nicht-diagonalen Anteil
des spezifischen Widerstandes erhalten wir somit
\begin{equation}
\varrho_{yx} =  \frac{E_{y,H}}{j_x}
             =  \frac{B}{n_{2D}e}.
\end{equation}
Es ist n\"utzlich, das {\sc Ohm}-{\sc Hall}-Gesetz
in einer Matrix- bzw.\ Tensor-Schreibweise zu notieren:
\begin{equation}
\left(
\begin{array}{c}
E_x \\
E_y
\end{array}
\right)
=
\left(
\begin{array}{cc}
\varrho_{xx} & \varrho_{xy} \\
\varrho_{yx} & \varrho_{yy}
\end{array}
\right)
\,
\left(
\begin{array}{c}
j_x \\
j_y
\end{array}
\right),
\end{equation}
explizit geschrieben als
\begin{equation}
\left(
\begin{array}{c}
E_x         \\
\phantom{=} \\
E_y
\end{array}
\right)
=
\left(
\begin{array}{cc}
  \displaystyle{   \frac{m}{n_{2D}e^2\tau}} &
  \displaystyle{ - \frac{B}{n_{2D}e}        } \\
  \phantom{=}                            &
  \phantom{=}                            \\
  \displaystyle{   \frac{B}{n_{2D}e}        } &
  \displaystyle{   \frac{m}{n_{2D}e^2\tau}}
\end{array}
\right)
\,
\left(
\begin{array}{c}
j_x         \\
\phantom{=} \\
j_y
\end{array}
\right)
=
\sigma_0^{-1}\cdot
\left(
\begin{array}{cc}
  \displaystyle{   1   } &
  \displaystyle{ - \omega_c\tau      } \\
  \phantom{=}                            &
  \phantom{=}                            \\
  \displaystyle{   \omega_c\tau      } &
  \displaystyle{   1   }
\end{array}
\right)
\,
\left(
\begin{array}{c}
j_x         \\
\phantom{=} \\
j_y
\end{array}
\right)
.
\end{equation}
{\it Klassisch\/} (und nur klassisch!) gilt also
\begin{equation}
\frac{d\varrho_{yx}}{dB}
= \frac{1}{n_{2D}e}
= \frac{e\tau}{m} \cdot \frac{m}{n_{2D}e^2\tau}
= \mu \varrho_{xx},
\end{equation}
kompakt geschrieben als
\begin{equation}
\frac{d\varrho_{yx}}{dB} = \mu \varrho_{xx}
\phantom{xxx}
\mbox{(klassisch).}
\end{equation}
Der {\sc Hall}-Konstante $R_{Hall}$ ist {\it per definitionem\/}
der auf das Magnetfeld $B$ normierte nicht-diagonale
Anteil des spezifischen Widerstandes:
\begin{equation}
R_{Hall}:=\frac{R_H}{B}
         =\frac{1}{B} \cdot \frac{U_y}{I_x}
         =\frac{1}{B} \cdot \frac{E_y}{j_{2D,x}}
         =\frac{1}{B} \cdot \varrho_{yx}.
\end{equation}
\par
{\it Bemerkung:\/}
Beachte, da\ss\ $R_{Hall}$ eine andere
Einheit hat als $\varrho_{yx}$.
\par
{\it Bemerkung:\/}
Beachte ferner, da\ss\ der spezifische Widerstand
sich aus dem absoluten Widerstand der Probe
in drei Dimensionen anders berechnet als in zwei Dimensionen.
Im ersteren Fall haben wir ihre Dicke zu ber\"ucksichtigen,
im zweiten Fall nicht. Seien die Abmessungen der
Probe in drei Dimensionen $L_x \cdot L_y \cdot L_z$,
in zwei Dimensionen $L_x \cdot L_y$, dann sind die
{\sc Hall}-Widerst\"ande
\begin{equation}
R_{Hall} = \frac{R_H}{B}
         = \frac{1}{B} \cdot \frac{U_y}{I_x}
         = \frac{1}{B} \cdot \frac{E_y \cdot L_y}{j_{2D,x} \cdot L_y}
         = \frac{1}{B} \cdot \frac{E_y}{j_{2D,x}}
\end{equation}
bzw.\
\begin{equation}
R_{Hall} = \frac{R_H}{B}
         = \frac{1}{B} \cdot \frac{U_y}{I_x}
         = \frac{1}{B} \cdot\frac{U_y\cdot L_z}{I_x},
         = \frac{1}{B} \cdot \frac{E_y \cdot L_y}{j_{3D,x} \cdot L_y \cdot L_z}
         = \frac{1}{B} \cdot \frac{E_y}{j_{3D,x} \cdot L_z},
\end{equation}
wobei
im 2D-Fall $j_{2D,x}$  eine Stromdichte bezogen auf einen
linienf\"ormigen Querschnitt bezeichnet,
im 3D-Fall $j_{3D,x}$  eine Stromdichte bezogen auf einen
fl\"achenf\"ormigen Querschnitt bezeichnet.
\par
Die Inversion
\begin{equation}
\left(
\begin{array}{c}
j_x \\
j_y
\end{array}
\right)
=
\left(
\begin{array}{cc}
\sigma_{xx} & \sigma_{xy} \\
\sigma_{yx} & \sigma_{yy}
\end{array}
\right)
\,
\left(
\begin{array}{c}
E_x \\
E_y
\end{array}
\right)
\end{equation}
des {\sc Ohm}-{\sc Hall}-Gesetzes folgt aus elementarer Matrixalgebra.
Wir erhalten
\begin{eqnarray}
\sigma_{xx}
&=& \frac{ \varrho_{xx}}{\varrho_{xx}^2+\varrho_{xy}^2}
 =  \sigma_0 \, \frac{1           }{1+\omega_c^2\tau^2},     \nonumber\\
\sigma_{xy}
&=& \frac{-\varrho_{xy}}{\varrho_{xx}^2+\varrho_{xy}^2}
 =  \sigma_0 \, \frac{\omega_c\tau}{1+\omega_c^2\tau^2}.
\end{eqnarray}
Im Limes gro\ss er Magnetfelder und damit gro\ss er
{\sc Hall}-Spannungen wird
$E_y \gg E_x$ und $\varrho_{yx} \gg \varrho_{xx}$.
Wir erhalten f\"ur die Komponenten der Leitf\"ahigkeiten
die N\"aherungen
\begin{eqnarray}
\sigma_{xx} &\approx&
            \phantom{+} \frac{\varrho_{xx}}{\varrho_{xy}^2}, \nonumber\\
\sigma_{xy} &\approx&
                     -  \frac{1           }{\varrho_{xy}  }.
\end{eqnarray}
Es ist also m\"oglich, da\ss\ man gleichzeitig
\begin{eqnarray}
\sigma_{xx}  &=&  0 , \\
\varrho_{xx} &=&  0 ,
\end{eqnarray}
im Gegensatz zu jeglicher Intuition.
\vfill\eject\noindent%
\subsection{Beobachtung des quantisierten Hall-Effekts}
Im klassischen Regime erhalten wir somit die folgenden Plots:
\begin{enumerate}
\item
$\varrho_{yx}$ {\it versus\/} $B$: linear-monoton ansteigend;
\item
$\varrho_{xx}$ {\it versus\/} $B$: konstant;
\item
$\varrho_{yx}$ {\it versus\/} $n_{2D}$: hyperbolisch abfallend;
\item
$\varrho_{xx}$ {\it versus\/} $n_{2D}$: hyperbolisch abfallend.
\end{enumerate}
\bild{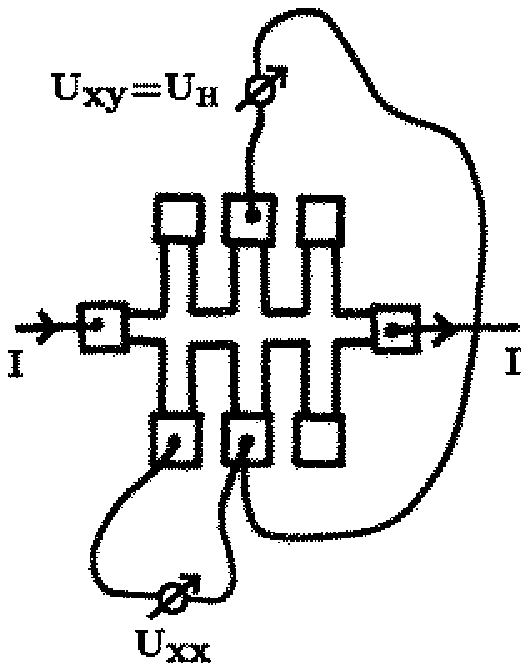}{QHE-Messung (schematisch)}{6}
\par
Im Rahmen der Tieftemperatur-Transportexperimente
am 2-dimensionalen Elektronengas beobachten wir hingegen f\"ur
\begin{enumerate}
\item
$\varrho_{yx}$ {\it versus\/} $B$
eine vor\"ubergehend ansteigende Funktion mit
breiten sogenannten {\sc Hall}-Plateaux
bei den Werten
\begin{equation}
\varrho_{yx}=
\frac{B}{e\,n_{2D}}=
\frac{B}{e}\cdot\frac{h}{\nu eB}=
\frac{h}{\nu e^2},
\end{equation}
mit $\nu=g_sg_v \cdot i$ und $i=1,2,3\dots\;$.
Sie stehen in Korrespondenz zur Bedingung
\begin{equation}
n_{2D}=\frac{\nu eB}{h},
\end{equation}
mit $n_{2D}$ als 2-dimensionale Ladungstr\"agerdichte.
\par
Diese Bedingung beschreibt
$\nu$ vollst\"andig gef\"ullte
{\sc Landau}-Niveaus.
Bei hohen Magnetfeldern wird die Spinentartung
($g_s$$=$$2$) und bei sehr hohen Magnetfeldern
die Valleyentartung (im Falle von Halbleitern
mit Valleys) aufgehoben. Bei vollst\"andiger
Aufhebung aller Entartungen entspricht der
F\"ullfaktor $\nu$ der ganzen Zahl $i$;
\item
$\varrho_{xx}$ {\it versus\/} $B$
bei steigendem $B$
eine zun\"achst konstante bis leicht abfallende Funktion,
die bei mittleren Magnetfeldern Oszillationen,
sogenannte {\sc Shubnikov}-{\sc de\,Haas}-Os\-zil\-la\-ti\-o\-nen
zeigt, deren Amplitude mit zunehmenden Magnetfeld anw\"achst,
bis sie schlie\ss lich bei den Magnetfeldern, bei denen
$\varrho_{yx}$ Plateaux zeigt, identisch verschwindet;
\item
$\varrho_{yx}$ {\it versus\/} $n_{2D}$ eine vor\"ubergehend
abfallende Funktion mit Plateaux bei den Werten
\begin{equation}
\varrho_{yx}=
\frac{B}{e\,n_{2D}}=
\frac{B}{e}\cdot\frac{h}{\nu eB}=
\frac{h}{\nu e^2},
\end{equation}
mit $\nu=g_sg_v \cdot i$ und $i=1,2,3\dots\;$;
\item
$\varrho_{xx}$ {\it versus\/} $n_{2D}$
eine Funktion, die Ihre endlichen Werte lediglich
bei den Magnetfeldern hat,
an denen $\varrho_{yx}$ seinen Wert \"andert,
dagegen bei den Magnetfeldern verschwindet,
an welchen letztere ihre Plateaux besitzt.
\end{enumerate}
F\"ur die Leitf\"ahigkeiten $\sigma_{xx}$ und $\sigma_{xy}$
gelten die entsprechenden Ergebnisse.
\bild{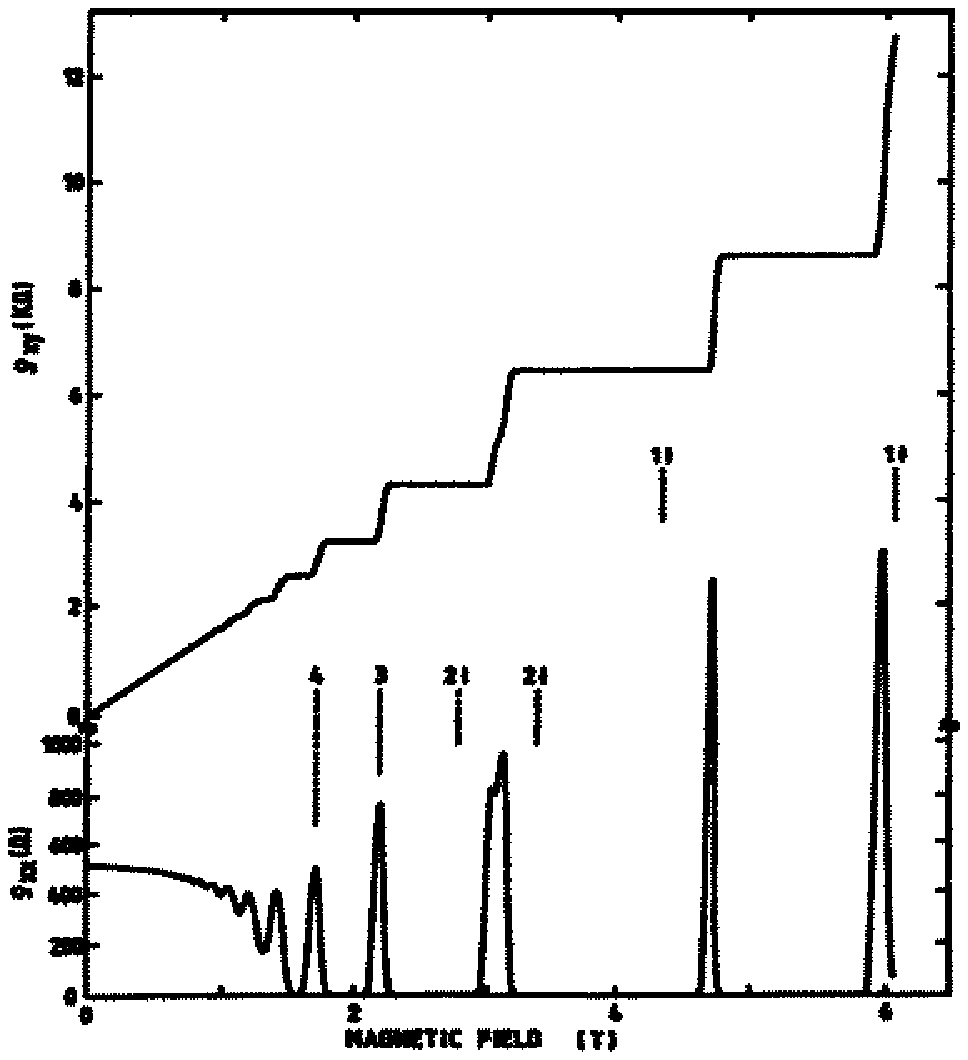}{QHE versus Magnetfeld (typische Me\ss kurve)}{12}
\bild{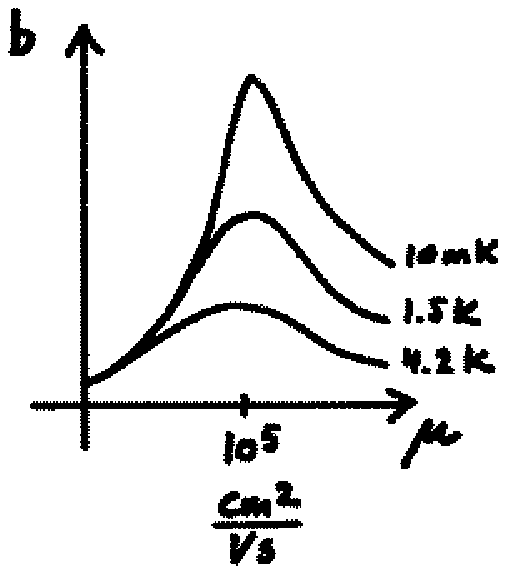}{Breite der QHE-Plateaux vs.\ Beweglichkeit (schematisch)}{6}
\vfill\eject\noindent%
\subsection{Das chemische Potential}
In der Physik der Halbleiter werden h\"aufig die Begriffe
\loq {\sc Fermi}-Energie\hiq\
und
\loq chemisches Potential\hiq\
durcheinandergebracht.
Wir heben daher an dieser Stelle hervor \cite{AshcroftMermin}:
\begin{itemize}
\item
Die {\sc Fermi}-{\it Energie\/}
oder das {\sc Fermi}-{\it Niveau\/}
$E_F$
(engl.\ {\it Fermi level\/})
ist {\it per definitionem\/} die Energie,
welche die besetzten von den unbesetzten Einteilchen-Zust\"anden
trennt.
\item
Das chemische Potential $\mu$ ist die freie Energie pro Teilchenzahl.
Dabei setzen wir stillschweigend voraus, da\ss\ die Teilchenzahl
eine ladungsartige, erhaltene Gr\"o\ss e ist.%
\footnote{Photonen z.\,B.\ tragen {\it keine\/} ladungsartige Quantenzahl
          und haben somit {\it kein\/} chemisches Potential.}
Es ist also eine thermodynamische Gr\"o\ss e, die zum Beipiel
in Analogie zum elektrischen Potential (Energie pro Ladung)
gesehen werden kann. (Man vergegenw\"artige sich, das die freie
Energie die um den Summanden $TS$ (mit $S$ als Entropie)
verminderte Energie eines makroskopischen Systems ist.)
\end{itemize}
Betrachten wir zun\"achst ein Metall. Wir werden dann die Frage
stellen, wie sich die Konzepte auf den Fall des Halbleiters
\"ubertragen.
\par
Bei einem Metall fallen {\sc Fermi}-Niveau und chemisches Potential
am absoluten Nullpunkt der Temperaturskala zusammen. Die
Verteilungsfunktion ist eine Stufen-Funktion, die an der
Stelle $E=\mu$ auf Null abf\"allt. Bei endlichen Temperaturen
$T>0$ wird diese Stufe abgerundet, da einige Elektronen unterhalb
von $\mu$ auf Niveaus oberhalb von $\mu$ thermisch angeregt sind.
H\"aufig bezeichnet man auch den Wert des chemischen
Potentials am absoluten Nullpunkt als {\sc Fermi}-{\it Energie\/},
also
\begin{equation}
E_F:=\mu\mid_{T=0} \, .
\end{equation}
Man beachte, da\ss\ dies eine andere Definition der
{\sc Fermi}-Energie ist als die oben vogestellte.
Durch die Verwendung verschiedener Konventionen
kann es leicht zu Verwechselungen kommen,
zumal sich das chemische Potential $\mu$
(und nat\"urlich auch die {\sc Fermi}-Energie -
im eigentlichen Sinne definiert als
die Grenze zwischen besetzten und unbesetzten
Einteilchen-Zust\"anden)
in Abh\"angigkeit von der Temperatur \"andert.
\par
Die Angelegenheit wird komplexer,
wenn wir Systeme kondensierter Materie mit E\-ner\-gie\-l\"ucke
(engl.\ {\it gap\/})
betrachten. Sind zum Beispiel alle Zust\"ande unterhalb des Gaps
besetzt und oberhalb des Gaps unbesetzt (wie bei einem sogenannten
intrinsischen Halbleiter, z.\,B.\ ideal reinem Silizium am absoluten
Nullpunkt), so erf\"ullt {\it jedes\/} Energieniveau in der Energiel\"ucke
die Definition einer {\sc Fermi}-Energie im eigentlichen Sinne (!). Wenn
Halbleiterphysiker von \loq der\hiq\ {\sc Fermi}-Energie eines
intrinsischen Halbleiter sprechen, meinen sie das chemische Potential,
welches f\"ur endliche Temperaturen wohldefiniert ist und im
Limes verchwindender absoluter Temperatur sich bei undotierten
Halbleitern in der Mitte des Gaps befindet.
\par
Auch wir wollen zuweilen dem allgemeinen
Sprachgebrauch folgen und das chemische Potential
\loq {\sc Fermi}-Niveau\hiq\ nennen;
denn gerade bei Systemen mit L\"ucken im Energiespektrum kann
nur ersteres gemeint sein!
\par
Wir heben noch einmal hervor: Man ist stets auf der sicheren Seite,
wenn man den thermodynamisch sauber definierten Begriff des
chemischen Potentials verwendet. Halbleiterphysiker sind es gewohnt,
dieses (nicht ganz korrekt) als {\sc Fermi}-Energie zu bezeichnen.
\vfill\eject\noindent%
\subsection{Shubnikov-de\,Haas-Oszillationen}
Ein System mit L\"ucken im Energiespektrum ist das von uns
betrachtete System von Elektronen in einem konstanten Magnetfeld.
Im Idealfall besteht es aus einer \"aquidistanten
Menge von hochentarteten Niveaus, die im realen Fall durch Streuprozesse
an Verunreinigungen verbreitert sind. Sowohl die Distanz als auch
der Entartungsgrad (im idealisierten Fall) sind proportional zum
\"au\ss eren Magnetfeld. Im Grenzfall niedriger Temperaturen
treten beide in Konkurrenz, wenn die Elektronen sich
bei dem sich ver\"andernden Mag\-net\-feld umverteilen.
\par
Betrachten wir ein Beispiel:
Das Magnetfeld und damit der Entartungsgrad seien so stark,
da\ss\ alle Elektronen im untersten {\sc Landau}-Niveau
kondensiert sind.
Vermindern wir nun langsam die St\"arke des Magnetfeldes, so
kommt irgendwann der Zeitpunkt, zu dem der Entartungsgrad sich
so vermindert hat, da\ss\ einige der Elektronen auf das
n\"achsth\"ohere {\sc Landau}-Niveau ausweichen. Dieser
Vorgang wiederholt sich entsprechend f\"ur die h\"oheren
Niveaus. Man kann es auch so sehen:
Die {\sc Fermi}-Energie $E_F$ durchwandert die {\sc Landau}-Niveaus.
In der Realit\"at sind diese nun verbreitert, d.\,h.\ wir finden
eine schwankende Zu\-stands\-dich\-te bei $E_F$. Theoretische Analysen,
die \"uber die {\sc Drude}-Transporttheorie hinausgehen, zeigen,
da\ss\ auch die
{\sc Drude}-Impulsrelaxationszeit $\tau$
schwanken kann.
Hierbei spielt die Dynamik sogenannter {\it Screening-Effekte\/}
(das sind Abschirmungseffekte von St\"orstellen) eine wesentliche
Rolle. Schwankt aber die
{\sc Drude}-Impulsrelaxationszeit,
so schwankt auch die Beweglichkeit der Elektronen. Eine kleine
Zu\-stands\-dich\-te an der {\sc Fermi}-Kante entspricht somit einer
kleinen Beweglichkeit.
\par
Bei vollst\"andiger F\"ullung eines {\sc Landau}-Niveaus,
wenn also das chemische Potential genau zwischen zwei
{\sc Landau}-Niveaus liegt,
finden wir auf der Hauptdiagonalen der Tensoren
ein Minimum der diagonalen Leitf\"ahigkeit $\sigma_{xx}$
und damit auch
ein Minimum des spezifischen Widerstandes $\varrho_{xx}$.
Die Schwankungen der Probenleitf\"ahigkeit
in Abh\"angigkeit des \"au\ss eren Magnetfeldes
sind die schon oben genannten
{\sc Shubnikov}-{\sc de\,Haas}-Os\-zil\-la\-ti\-o\-nen.
Die {\sc Shubnikov}-{\sc de\,Haas}-Minima liegen periodisch
in $1/B$. Berechnen wir die Periode:
Sei der F\"ullfaktor im allgemeinsten Fall
(mit Valley- und Spin-Entartung)
\begin{equation}
\nu = g_s g_v \cdot i,
\end{equation}
so k\"onnen wir f\"ur die Anzahl der Ladungstr\"ager schreiben:
\begin{equation}
n_{2D} = i \cdot n_{L} = i \cdot g_s g_v \cdot \frac{eB_{(i)}}{h},
\end{equation}
das hei\ss t
\begin{equation}
\frac{1}{B_{(i)}} = \frac{ig_sg_ve}{h n_{2D} }
\end{equation}
und damit
\begin{equation}
\frac{1}{B_{(i+1)}} = \frac{(i+1)g_sg_ve}{h n_{2D} }.
\end{equation}
Die Differenz beider Terme ergibt die Periode der Oszillationen:
\begin{equation}
\Delta(\frac{1}{B})
=
\frac{1}{B_{(i+1)}}-\frac{1}{B_{(i)}}
=
g_sg_v\cdot\frac{e}{h n_{2D} }.
\end{equation}
\bild{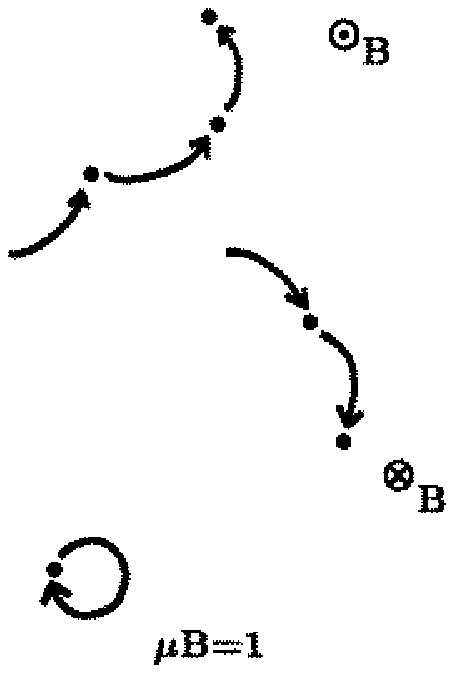}{Zustandekommen des Magnetowiderstandes}{6}
\par
Voraussetzung f\"ur die Beobachtung
der {\sc Shubnikov}-{\sc de\,Haas}-Oszillationen sind
\begin{enumerate}
\item
die Tatsache, da\ss\ das Produkt aus
\"au\ss erem Magnetfeld $B$ und Beweglichkeit $\mu=e\tau/m$ der
Ladungstr\"ager m\"oglichst hoch ist
\begin{equation}
\mu B \gg 1,
\end{equation}
was  \"aquivalent ist zur Bedingung
\begin{equation}
\omega_c\tau \gg 1.
\end{equation}
\par
Die Elektronen sollten also mindestens einmal,
m\"oglichst mehrmals kreisen, ohne an einem Streuproze\ss\
teilzunehmen. Diese Bedingung bestimmt den Einsatz der
{\sc Shubnikov}-{\sc de\,Haas}-Oszillationen und ist
erst bei niedrigen Temperaturen relevant. In Zahlen:
Sei
\begin{equation}
T=3\,{\rm K},
\end{equation}
dann ist
\begin{equation}
4\,kT \approx 1\,{\rm meV}.
\end{equation}
Gleichzeitig ist
\begin{equation}
\hbar\omega_c = 1.65\,{\rm meV}\,B(T),
\end{equation}
mit der effektiven Masse f\"ur Elektronen in Galliumarsenid
\begin{equation}
m = 0.07\,m_e \, ;
\end{equation}
\item
die Tatsache, da\ss\ die thermische Aufweichung der {\sc Fermi}-Kante
deutlich schmaler ist als der energetische Abstand zweier
{\sc Landau}-Niveaus
\begin{equation}
\hbar\omega_c\gg k_B\cdot T,
\end{equation}
typisch
\begin{equation}
\hbar\omega_c \ge 4k_BT,
\end{equation}
das hei\ss t, die verwendeten Temperaturen sollten m\"oglichst
niedrig sein.
\end{enumerate}
\newpage%
\vspace*{5cm}
\bild{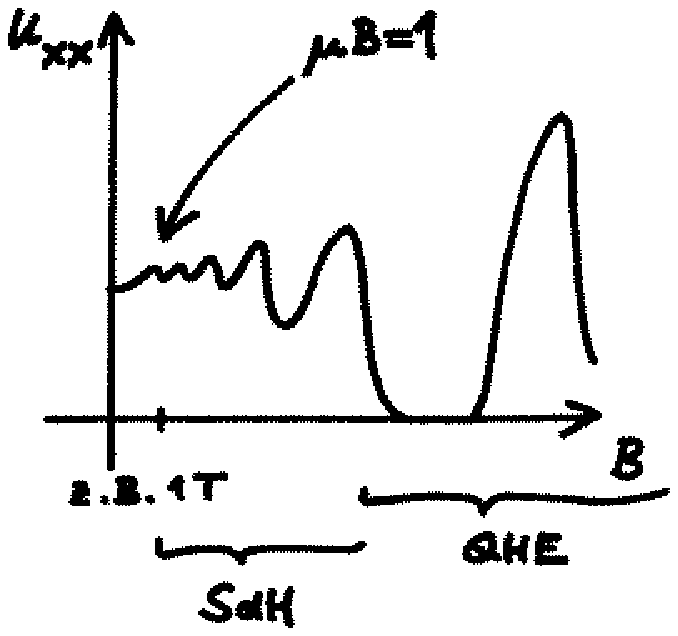}{{\sc Shubnikov}-{\sc de\,Haas}-Oszillationen (schematisch)}{8}
\newpage%
\section{Deutung des Effektes} 
\subsection{Eichtheoretisches Argument nach Laughlin}
Der quantisierte {\sc Hall}-Effekt geh\"ort
offensichtlich in die Kategorie
makroskopischer Quanteneffekte bzw.\
topologischer Quantisierungseffekte;
siehe hierzu die beigef\"ugte Tabelle.
\begin{table}\vspace{0.5cm}
\begin{center}
%
\begin{tabular}{|c|c|c|c|}                                       \hline
Konzept & Ph\"anomen                      & & Entdeckungsjahr \\ \hline\hline
        & $\bullet$ &
Beobachtung der Supraleitung in Hg        & 1911            \\ \hline
$\bullet$ &         &
{\sc Dirac}-Monopol (hypothetisch)        & 1931            \\ \hline
        & $\bullet$ &
{\sc Meissner}-{\sc Ochsenfeld}-Effekt    & 1933            \\ \hline
$\bullet$ &         &
{\sc London}-Theorie                      & 1935            \\ \hline
        & $\bullet$ &
Suprafl\"ussiges Helium-4                 & 1938            \\ \hline
$\bullet$ &         &
{\sc Ginzburg}-{\sc Landau}-Theorie       & 1950            \\ \hline
$\bullet$ &         &
BCS-Supraleitung                          & 1956/1957       \\ \hline
$\bullet$ &$\bullet$  &
{\sc Aharonov}-{\sc Bohm}-Effekt          & 1959            \\ \hline
$\bullet$ &$\bullet$  &
Flu\ss quantisierung                      & 1961            \\ \hline
$\bullet$ & $\bullet$ &
{\sc Josephson}-Effekt                    & 1962            \\ \hline
        & $\bullet$ &
Suprafl\"ussiges Helium-3                 & 1972            \\ \hline
        & $\bullet$ &
Quantisierter {\sc Hall}-Effekt           & 1981            \\ \hline
        & $\bullet$ &
Hochtemperatur-Supraleitung               & 1986            \\ \hline
        & $\bullet$ &
Supraleitung in Fullerenen                & 1991            \\ \hline
\end{tabular}
\normalsize
\end{center}
\vspace{0.5cm}
\caption{Makroskopische Quanteneffekte und topologische Quantisierungseffekte}
\vspace{0.75cm}\end{table}
\par
Es war {\sc F.\ London}, der als erster die Idee hatte, das von
{\sc Weyl} vorgeschlagene Eichfeldkonzept, welches das
elektromagnetische Vektorpotential in Relation zu einer
spekulativen L\"angen\"anderung von Vektoren unter Parallelverschiebung
im Rahmen der Allgemeinen Relativit\"atstheorie (!) setzt,
neu zu interpretieren, n\"amlich als den Generator einer
Drehung der Phase der quantenmechanischen Wellenfunktion
\cite{Yang79}.
{\sc F.\ London} war auch der erste, der \"uber die
Existenz quantisierter magnetischer Fl\"usse spekulierte
\cite{Yang70}.
\par
In diese Kategorie von Argumenten f\"allt auch das Eichargument
von {\sc Laughlin}, das auf einem Gedankenexperiment beruht
\cite{Laughlin81}:
Unter der Annahme, da\ss\ die Wellenfunktion in der
{\sc Hall}-Probe ausgedehnt ist und als makroskopische
Gr\"o\ss e eindeutig ist, erhalten wir eine der Flu\ss quantisierung
analoge Bedingung, wenn wir die Probe zu einem Ring so zusammenbiegen,
da\ss\ Strominjektions-
(engl.\ {\it  source\/})
und Stromextraktionskontakt
(engl.\ {\it drain\/})
miteinander identifiziert sind.
Der quantisierte Flu\ss\ durch den Ring sei mit $\Phi$ bezeichnet
und ist von dem magnetischen Flu\ss, der dem {\sc Hall}-Magnetfeld
entspricht, welches an jeder Stelle normal zur Probenfl\"ache
ausgerichtet ist, wohl zu unterscheiden.
Man nennt ihn auch den {\it fiktiven magnetischen Flu\ss\/}.
\bild{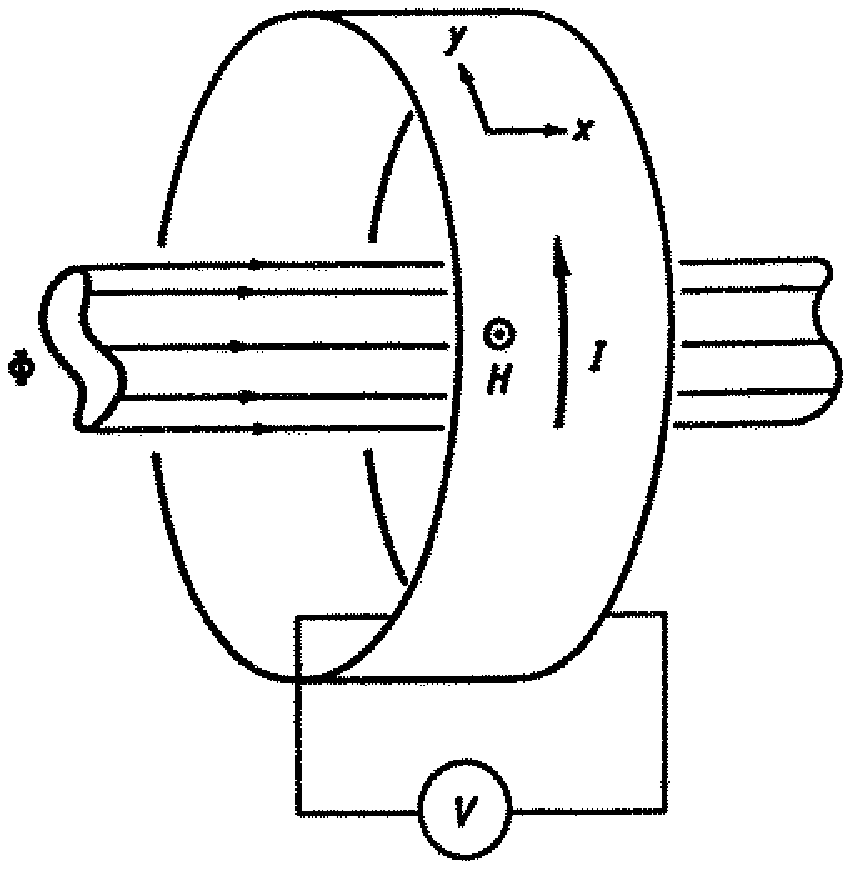}{{\sc Laughlin}s Eichargument}{10}
\par
Uns interessiert die Beziehung
von totalem Strom $I$ durch die Probe zum Spannungsabfall
zwischen den Seitenkanten am Rande des Transportweges der Ladungstr\"ager.
Wir k\"onnen sie aus dem {\sc Faraday}schem Induktionsgesetz herleiten.
Letzteres schreiben wir nicht, wie \"ublich als
\begin{equation}
U_{ind}=-n\cdot\frac{\Delta\Phi}{\Delta t} \, ,
\end{equation}
sondern als
\begin{equation}
I=\frac{\partial E_{ww}}{\partial\Phi}.
\end{equation}
\"Uberpr\"ufen wir diese Formel: Man \"uberzeugt sich leicht,
da\ss\ die Einheiten stimmen: Energie wird in Volt-Ampere-Sekunde
gemessen und der magnetische Flu\ss\ in Volt-Sekunde.
Andererseits ist die Wechselwirkungsenergie im Falle
der hier gew\"ahlten Probengeometrie (mit den Kantenl\"angen $a$,$b$,$c$)
\begin{equation}
E_{ww}=\int {\bf j}_{2D}{\bf A} \, d^3 x
      =|{\bf j}_{2D}| \, b \cdot |{\bf A}| \, c
      = I \cdot \Phi.
\end{equation}
Differentiation nach $\Phi$ reproduziert das ge\-w\"unsch\-te Ergebnis.
Wegen der Quantisierung des fiktiven magnetischen Flusses m\"ussen
wir die partielle (adiabatische) Ableitung durch eine Differenz ersetzen.
F\"ur $i$ Elektronen, die
von der einen zur gegen\"uberliegenden Seitenkante
transportiert werden,
erhalten wir die Energie- bzw.\ Strombilanz
\begin{equation}
I=\frac{\Delta E_{ww}}{\Delta\Phi}
 =i\cdot\frac{\mbox{Elektronenenergie}}{\mbox{Flu\ss quantum}}
 =\frac{ieU_H}{h/e}
 =\frac{ie^2}{h}\,U_H .
\end{equation}
Das hei\ss t, da\ss\ wir unter der Voraussetzung, da\ss\
die dem superstrom-\"ahnlichen {\sc Hall}-Strom zugeordnete
Wellenfunktion die Eichinvarianz der Theorie in der
Ebene respektiert, die ge\-w\"unsch\-te Quantisierung erhalten.
\par
Wie wir schon oben gesehen haben,
liefert das Denken in Termen von
Flu\ss quanten tiefe Einsichten
in den fundamentalen Charakter des quantisierten {\sc Hall}-Effektes:
Offensichtlich k\"onnen wir die St\"arke des Magnetfeldes
durch die auf eine Fl\"acheneinheit bezogene Anzahl
$n_{\Phi_0}$ von Flu\ss quanten $\Phi_0$ ausdr\"ucken;
das hei\ss t, es ist
\begin{equation}
B = n_{\Phi_0} \cdot \Phi_0 =  n_{\Phi_0} \cdot \frac{h}{e}.
\end{equation}
Somit erkennen wir in der Gr\"o\ss e
\begin{equation}
g_sg_v \cdot n_{\Phi_0} = g_sg_v \cdot \frac{eB}{h} = n_{L}
\end{equation}
den Entartungsgrad eines {\sc Landau}-Niveaus wieder.
Wir k\"onnen auch schreiben
\begin{eqnarray}
\mbox{Anzahl der gef\"ullten {\sc Landau}-Niveaus}
&  = &  \frac{1}{g_sg_v} \cdot
        \frac{\mbox{Anzahl der Ladungstr\"ager}}
             {\mbox{Anzahl der Flu\ss quanten}}       \nonumber\\
& \phantom{=} &                                       \nonumber\\
&  = &  \frac{1}{g_sg_v} \cdot \frac{h n_{2D} }{eB}
\, = \, \frac{1}{g_sg_v}\cdot\nu
\, = \, \frac{ n_{2D} }{ n_{L} }
\; .
\end{eqnarray}
Wenn wir der Einfachheit halber $g_s=g_v=1$ annehmen,
haben wir f\"ur $\nu=1$ gerade einen Zustand
vorliegen, in dem sich in der Probe genauso
viele Ladungstr\"ager wie Flu\ss quanten befinden.
Nach {\sc Kivelson}, {\sc Lee} und {\sc Zhang}
liegt dann ein makroskopischer Quantenzustand vor,
in dem bosonische Quasiteilchen, welche man sich als
Bindungszust\"ande von je einer elektrischen
Ladung und je einem magnetischen Flu\ss quant
vozustellen hat, in einen durch eine
makroskopische Wellenfunktion zu beschreibenden
Quantenzustand kondensiert sind
\cite{Kivelson92, Kivelson96, Zhang89, Zhang92}.
Dieser Zustand ist vergleichbar mit einem
supraleitenden Grundzustand geladener Bosonen.
(Analoges gilt f\"ur die h\"oheren F\"ullfaktoren.)
\par
So suggestiv diese Vorstellungen auch sein m\"ogen -
ihre Rechtfertigung erg\"abe sich erst aus einer
{\it mikroskopischen\/} Beschreibung des beobachteten Effektes.
Die Situation ist hier ganz \"ahnlich wie in der
gew\"ohnlichen Supraleitung:
So erm\"oglicht die in Termen einer
makroskopischen Wellefunktion formulierte
{\sc Landau}-{\sc Ginzburg}-Theorie eine
ziemlich treffende Beschreibung
(i) des Phasen\"ubergangs vom normalleitenden zum supraleitenden Zustand,
(ii) des Verhaltens gegen\"uber \"au\ss eren Magnetfeldern sowie
(iii) der Eigenschaften von Flu\ss schl\"auchen;
eine mikroskopische Erkl\"arung liefert aber erst
die Theorie von {\sc Bardeen}, {\sc Cooper} und
{\sc Schriefer}, ausgehend von der Idee einer
von Phononen vermittelten schwach attraktiven
Wechselwirkung von Elektronen und der Bildung von
sogenannten {\sc Cooper}-Paaren, die - als Quasibosonen -
schlie\ss lich in den supraleitenden Grundzustand kondensieren.
\par
F\"ur den integral-quantisierten {\sc Hall}-Effekt
bleibt also die Frage: Was ist die fundamentale
Elektron-Elektron-Wechselwirkung, welche den
makroskopischen Quantenzustand herbeif\"uhrt?
Diese Frage steht im klaren Gegensatz zur
weitverbreiteten Lehrmeinung, der integral-quantisierte
{\sc Hall}-Effekt lasse sich im Rahmen einer Theorie
nicht-wech\-sel\-wir\-ken\-der Elektronen vollst\"andig verstehen.
Eine befriedigende Antwort steht bis heute noch aus.
\vfill\eject\noindent%
\subsection{Lokalisierungsbild (bulk states)}
Durch unterschiedlich starke Streuung der Elektronen an St\"orstellen
wird die hochgradige Entartung der {\sc Landau}-Niveaus (teilweise)
aufgehoben, so da\ss\ diese sich zu B\"andern verbreitern. Wenn wir
den quantisierten {\sc Hall}-Effekt erkl\"aren wollen, m\"ussen
wir annehmen,
\begin{itemize}
\item
da\ss\ die Elektronen, die im Zentrum eines solchen Bandes
liegen, nicht gestreut werden, und somit ausgedehnte Zust\"ande
bilden, die zum Stromtransport beitragen,
\item
da\ss\ Elektronen, die in einer gewissen Umgebung des Zentrums liegen,
so gestreut werden, da\ss\ sie lokalisierte Zust\"ande bilden.
\end{itemize}
\bild{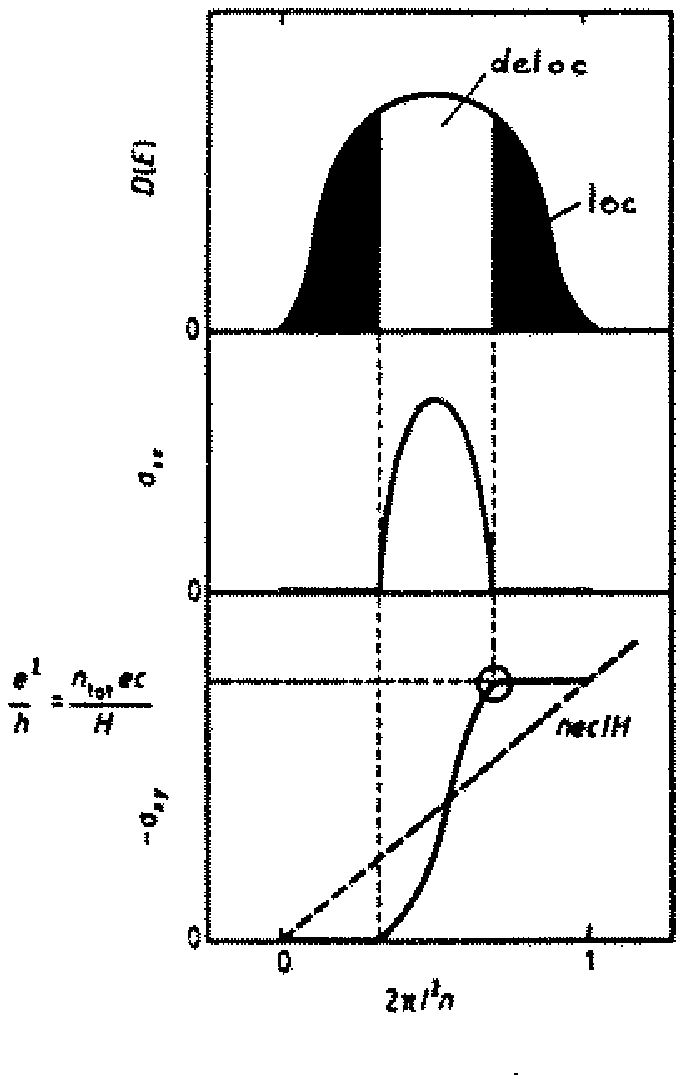}{Lokalisierung versus Delokalisierung}{8}
\par
Die St\"orstellen in der Probe bilden ein \loq Potentialgebirge\hiq.
Die elektronische Wellenfunktion neigt dazu, sich in dessen
T\"alern zu lokalisieren. In Konkurrenz dazu neigen Tunneleffekte
zwischen benachbaren Minima dazu, die Wellenfunktion zu delokalisieren.
Zus\"atzlich versucht das Magnetfeld, die Elektronen auf Kreisbahnen
zu zwingen. Da\ss\ sich nun alle Effekte zusammen so arrangieren,
da\ss\ der beschriebene
Lo\-ka\-li\-sie\-rungs-De\-lo\-ka\-li\-sie\-rungs-\"Ubergang
mit einer so robusten Quantisierung vertr\"aglich ist, mu\ss\ in
einer mikroskopischen Theorie erst einmal gezeigt werden,
und zwar rigoros, nicht nur vermittels Computer-Simulation.
In dem vorliegenden Rahmen wollen wir uns darauf beschr\"anken,
dies als plausible Annahme gelten zu lassen.
\par
Entscheidend f\"ur das Auftreten des quantisierten {\sc Hall}-Effektes
ist nun, wie sich das System beim Durchfahren des chemischen Potentials
durch die verbreiterten {\sc Landau}-Niveaus verh\"alt.
F\"ur den Fall eines ansteigenden Magnetfeldes beobachten wir:
\begin{itemize}
\item
Solange Bereiche lokalisierter Zust\"ande durchfahren werden,
k\"onnen Zust\"ande, die mit Elektronen be- oder entv\"olkert werden,
nichts zur Leitf\"ahigkeit der Probe beitragen. In diesem Bereich
\"andert sich nichts am {\sc Hall}-Widerstand der Probe und der
quantisierte Wert bleibt erhalten. Die longitudinale Leitf\"ahigkeit
(und damit der longitudinale Widerstand) verschwindet.
\item
Werden Bereiche ausgedehnter Zust\"ande durchfahren,
liefern die Zust\"ande, die mit Elektronen be- oder entv\"olkert werden,
einen Beitrag zur Leitf\"ahigkeit. In diesem Bereich \"andert sich
die {\sc Hall}-Leitf\"ahigkeit der Probe. Die longitudinale
Leitf\"ahigkeit ist gerade die Steigung dieser \"Anderung versus
\"au\ss erem Magnetfeld. (Letztere Aussage ist im Sinne eines
ph\"anomenologischen Fits zu verstehen; einen einfachen formalen
Beweis, der von einfachen physikalischen Voraussetzungen ausgeht,
gibt es leider nicht.)
\end{itemize}
\bild{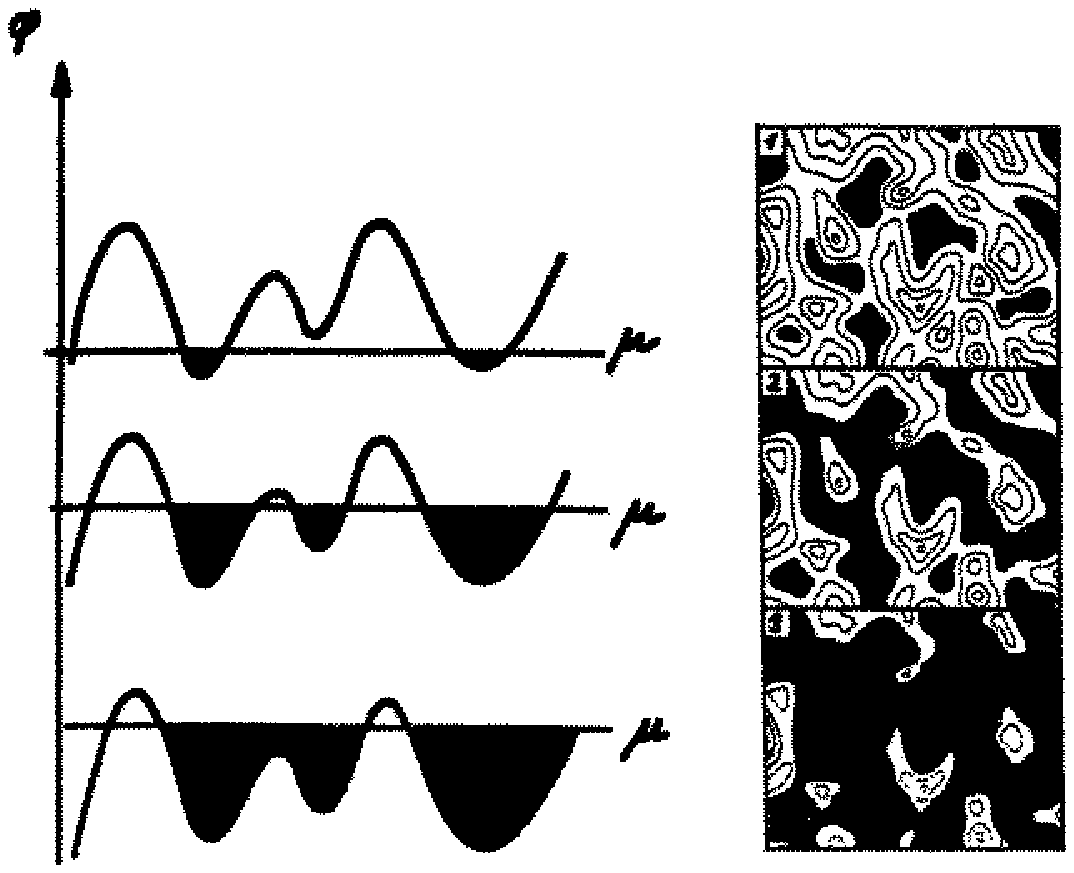}{Lokalisierung-Delokalisierung und Percolation}{12}
\par
Die bisherigen experimentellen Fakten legen nahe,
da\ss\ der quantisierte {\sc Hall}-Effekt
un\-ab\-h\"an\-gig von Geometrie und Gr\"o\ss e der Probe ist.
Eine mikroskopische Theorie, welche den Effekt
an sich erkl\"art, sollte auch diese Tatsache erkl\"aren.
Wir erwarten daher, da\ss\ die Theorie wesentlich
auf Renormierungsgruppen-Argumente, also auf Analysen
von Skalenverhalten einer
{\it Quantentheorie eines ungeordneten Systems im Magnetfeld\/}
bauen mu\ss. In der Tat ist dies genau der Ansatz,
von dem in modernen quantenfeldtheoretischen Zug\"angen
zum quantisierten {\sc Hall}-Effekt ausgegangen wird.
Der interessierte Leser sei auf den hervorragenden,
gut zu lesenden einf\"uhrenden Artikel von {\sc Khurana}
in {\it Physics today\/} hingewiesen, der auch einen Hinweis
auf interessante Originalarbeiten gibt
\cite{Khurana8809}.
\par
An dieser Stelle seien kurz die wesentlichen Ideen
der bahnbrechenden Arbeiten von
{\sc Levine}, {\sc Libby}, {\sc Pruisken}
und {\sc Khmel'nitzkii} skizziert
\cite{Levine84, Pruisken84, Khmelnitzkii83}.
\par
Wie {\sc Khurana} in seinem einf\"uhrenden Artikel
hinweist, gibt es Analogien zur Quantenfeldtheorie
der Elementarteilchen. In der Transporttheorie spielen
Leitf\"ahigkeiten die Rolle inverser Kopplungkonstanten
in der Quantenfeldtheorie. Die Untersuchung der allgemeinen
Struktur einer Elementarteilchentheorie f\"uhrt stets auf
eine Renormierungsgruppenanalyse: Wie skalieren die
Kopplungskonstanten beim Skalieren des Impuls\"ubertrages?
Im Falle der nichtabelschen Eichtheorien beobachtet man zum
Beispiel {\it asymptotische Freiheit\/} (das Verschwinden
der Wechselwirkungen der Quarks im Hochenergie-Limes),
in der Sprache der Renormierungsgruppe ultraviolett stabiler
Fixpunkt genannt. Umgekehrt ist das Verhalten der Kopplungskonstante
im Infrarot-Limes so, da\ss\ die Wechselwirkung immer st\"arker
wird: Quark Confinement. F\"ur eine Diskussion all dieser Fragen
siehe \cite{Itzykson80}.
\par
Im Falle des Stromtransports entspricht
dies dem Fall verschwindender Leitf\"ahigkeit bei wachsender
Systemgr\"o\ss e. Dies ist exakt der Fall, den wir im
Verschwinden metallischen (ohmschen) Verhaltens beim Vorliegen
von {\it Lokalisierung\/} in niederen Dimensionen haben.
\par
Alle beschriebenen Systeme lassen sich in einem
feldtheoretischen {\sc Lagrange}schen Rahmen beschreiben.
Dabei kann man aus den klassischen Bewegungsgleichungen
nur einen ganz kleinen Teil der Physik ablesen;
die wesentlichen Strukturen werden im allgemeinen
erst im Rahmen der Quantisierungsprozedur
({\sc Green}-Funktionen, {\sc Feynman}-Graphen etc.)
vermittels oft m\"uhseliger Rechnungen sichtbar.
\par
Bestimmte Strukturen haben kein klassisches Analogon:
So ist aus der Quan\-ten\-feld\-the\-o\-rie wohlbekannt, da\ss\
man Terme zur {\sc Lagrange}-Funk\-ti\-on hinzuaddieren kann,
welche die Bewegungsgleichungen nicht \"andern, wohl aber
die quantenmechanische Phase drehen, die ja proportional
zu $\exp\,iS$ ist, mit $S=\int L\,dt$ als Wirkung.
(In der klassischen Me\-cha\-nik sind dieses die Terme
proportional zu einer totalen Zeitableitung.)
In der \"ublichen Elektrodynamik in drei Raumdimensionen
w\"are dies ein Zusatzterm proportional zu ${\bf E}\cdot{\bf B}$,
ein sogenannter {\sc Chern}-{\sc Pontryagin}-Term
(im Gegensatz zum \"ublichen {\sc Maxwell}-Term
proportional zu ${\bf E}^2-{\bf B}^2$),
in der 2-dimensionalen Version ein sogenannter
{\sc Chern}-{\sc Simons}-Term proportional ${\bf A}\times{\bf B}$.
Fordern wir, da\ss\ die Wellenfunktionen
$\psi\sim\exp\,iS$ der betrachteten
Quantisierung eindeutig sind, so m\"ussen die diesen
topologischen Termen zugeordneten Kopplungskonstanten
quantisiert sein.
\par
Es sind {\sc Lagrange}-Funktionen genau dieses
Typs, welche in der Magnetotransporttheorie
eine Rolle spielen und die sich durch zwei
Kopplungsparameter, einem konventionellen ($\sigma_{xx}$)
und einem topologischen ($\sigma_{xy}$), auszeichnen.
Das entsprechende {\it two-parameter scaling\/}
zeigt, da\ss\ bei wachsender Systemgr\"o\ss e die topologische
Kopplungskonstante (die gerade der {\sc Hall}-Leitf\"ahigkeit
entspricht) quantisierte Werte annimmt, die gew\"ohnliche
Kopplungskonstante hingegen das beobachtete oszillatorische
Verhalten zeigt.
Diese \loq Auf\-bl\"at\-te\-rung\hiq\ von Quantenfeldtheorien mit
topologischen Termen ist ein universelles Ph\"a\-no\-men und
den Theoretikern wohlbekannt
\cite{Cardy82a, Cardy82b}.
(Die Abbildung zeigt die entsprechenden
Renormierungsgruppenfl\"usse, wobei die Pfeile in Richtung
wachsender Systemgr\"o\ss e zeigen
\cite{Khmelnitzkii83}.)
\bild{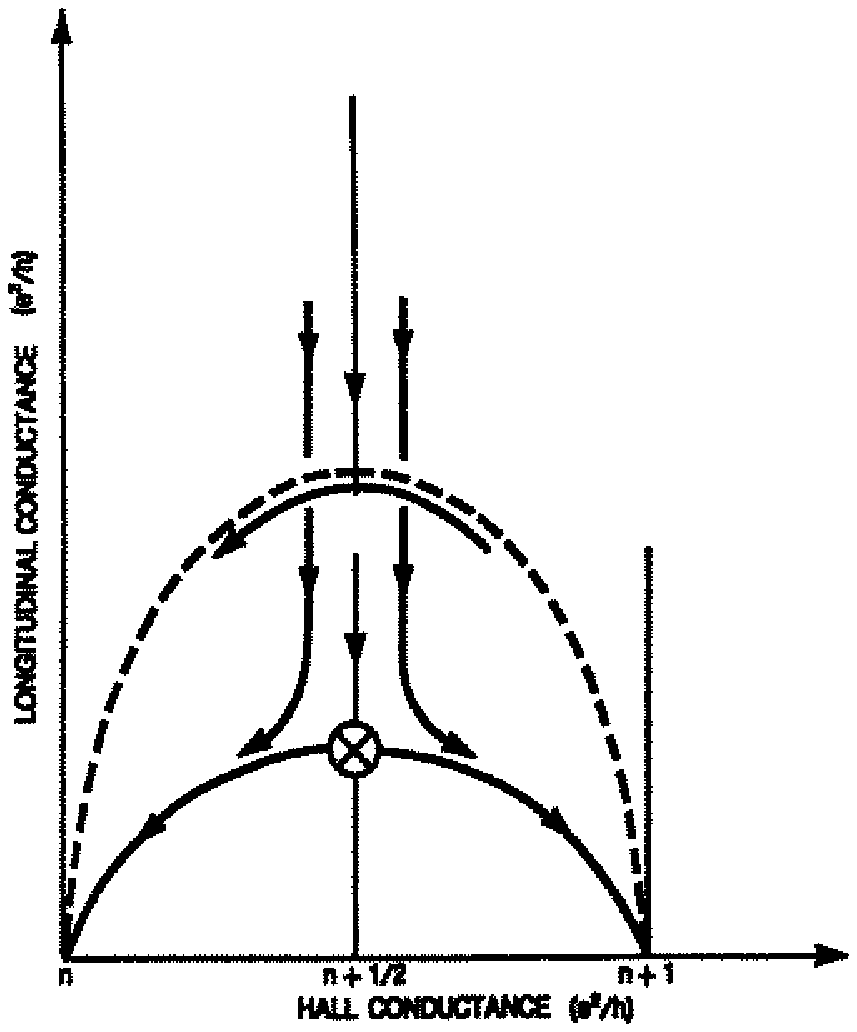}{{\sc Pruisken}-{\sc Khmel'nitzkii} two-parameter scaling}{10}
\par
Interessant dabei ist, da\ss\ die
aus der Lokalisierungstheorie hervorgegangenen
mikroskopischen Quantenfeldtheorien
des quantisierten {\sc Hall}-Effektes
gerade vom ersten Typus
\cite{Pruisken84},
die ph\"anomenologischen Theorien
hingegen vom zweiten Typus sind
\cite{Frohlich93}.
Die zugeordneten {\sc Lagrange}-Dichten haben stets die Form
\begin{equation}
{\cal L}=\sigma_{xx}\cdot{\cal L}_0+\sigma_{xy}\cdot{\cal L}_{top}.
\end{equation}
Dies erkl\"art schlie\ss lich, warum die Quantisierung exakt sein mu\ss.
Die Sache hat nur zwei Haken: Zum einen steckt in der Wahl der geeigneten
{\sc Lagrange}-Dichte - gleichg\"ultig, auf welchem Level sie erfolgt -
immer eine Modell-Annahme; zum anderen ist es sehr schwer,
die Quantenfeldtheorie dieses Problems rigoros zu l\"osen.
Letzteres gilt aber auch f\"ur die experimentell
am erfolgreichsten verifizierte Theorie schlechthin,
die Quan\-ten\-e\-lek\-tro\-dy\-na\-mik.
\par
{\it Bemerkung:\/} Man k\"onnte an dieser Stelle
vielleicht einwenden, da\ss\ die hier nur angerissenen
Formalismen ein wenig zu akademisch sind
und da\ss\ alles auch viel einfacher gehen m\"u\ss te.
Unter Experten besteht aber durchaus ein Konsensus darin,
da\ss\ die Anwendung quantenfeldtheoretischer Methoden
f\"ur den Lokalisierungs-Delokalisierungs-\"Ubergang
im \"au\ss eren Magnetfeld der geeignete Rahmen ist
(siehe z.\,B.\
\cite{Janssen94}).
\vfill\eject\noindent%
\subsection{Randkanal-Bild (edge states)}
Die Physik des quantisierten {\sc Hall}-Systems ist ein ausgezeichnetes
Beipiel f\"ur die Anwendung einer quantenelektrodynamischen Theorie
auf ein reales System der kondensierten Materie. Diese mu\ss\ nat\"urlich
die lokale Eichinvarianz, die ein sehr fundamentales physikalisches Prinzip
darstellt, respektieren.
Wenn das betrachtete physikalische System, die Probe,
eine endliche Ausdehnung und damit einen Rand besitzt,
mu\ss\ die Eichinvarianz auf dem Rand
(engl.\ {\it edge\/})
kompatibel zur Eichinvarianz im Hauptteil
(engl.\ {\it bulk\/})
sein.
H\"atten wir zum Beispiel ein Modell
des quantisierten {\sc Hall}-Systems in einem
{\sc Lagrange}schen Rahmen formuliert und w\"urden durch Variation
der Wirkung die klassischen Bewegungsgleichungen ableiten,
so m\"u\ss ten wir bei der Variation Randterme ber\"ucksichtigen,
die im unendlich ausgedehnten Fall meistens wegdiskutiert werden.
\par
Es ist eine wohlbekannte Tatsache aus der mathematischen Physik,
da\ss\ Begriffe wie chemisches Potential und Strom ihren Ursprung
in der Eichinvarianz haben
\cite{Araki77}.
Eichinvarianz impliziert die Existenz von Ladungen
und zugeordneten Erhaltungsgesetzen. Meist ist die
Ladung in elementarer Weise mit den Teilchen verkn\"upft,
so da\ss\ wir auch von Erhaltung der Teilchenzahl $N$ sprechen
d\"urfen. (Dies ist aber nicht der allgemeinste Fall.)
Thermodynamische Gleichgewichtszust\"ande solcher
Systeme werden nicht nur durch die inverse Temperatur
$\beta=1/kT$, der kanonisch konjugierten Variablen zur
Energie, sondern durch ein chemisches Potential $\mu$,
der kanonisch konjugierten Variablen zur Teilchenzahl
(im allgemeinen Fall zur Gesamtladung), gekennzeichnet.
Zwischen zwei Kontakten, die ja die Verbindungen der Probe
zu Reservoirs, die sich im thermodynamischen Gleichgewicht
befinden, darstellen,
mu\ss\ die Differenz des chemischen Potentials endlich sein,
wenn wir einen elektrischen Stromtransport haben wollen.%
\footnote{Bei Bosonen ist die Endlichkeit
          des chemischen Potentials Voraussetzung
          f\"ur die M\"oglichkeit der sogenannten
          {\sc Bose}-Kondensation,
          mit deren Hilfe man den im Praktikumsexperiment
          beobachtbaren $\lambda$-\"Ubergang (nach der
          Form des Verlaufs der spezifischen W\"arme)
          des fl\"ussigen Helium-4 bei 2.18 K
          in den suprafl\"ussigen Zustand erkl\"aren kann.
          Es gibt auch eine Theorie
          des quantisierten {\sc Hall}-Effekts, welche sich
          auf das Prinzip der {\sc Bose}-Kondensation bezieht.
          Hier sind die Bosonen gedachte Bindungszust\"ante
          aus Flu\ss quanten und Elektronen, welche bei
          ganzzahligen (oder auch bestimmten rationalen)
          F\"ullfaktoren kondensieren
          (siehe auch
          \cite{Kivelson92, Kivelson96, Zhang89, Zhang92}).
          \par
          Im Falle von {\sc Bose}-Gasen unbeschr\"ankter
          Teilchenerzeugung, oder etwas pr\"aziser gesprochen,
          im Falle von Gasen von {\sc Bosonen} ohne ladungsartige
          erhaltende Quantenzahl (wie Photonen und Phononen)
          gibt es kein chemisches Potential und somit auch
          {\it nicht\/} die M\"oglichkeit der {\sc Bose}-Kondensation.
          Da\ss\ man das Photonengas in einem Hohlraum
          lediglich durch einen Parameter, n\"amlich
          $\beta=1/kT$, charakterisieren kann, ist gerade
          einer der Eckpfeiler der ber\"uhmten Hypothese
          {\sc Planck}s und seiner Entdeckung der nach ihm
          benannten Konstante
          (vgl.\ \cite{Weidlich, Haag}).
          }
\bild{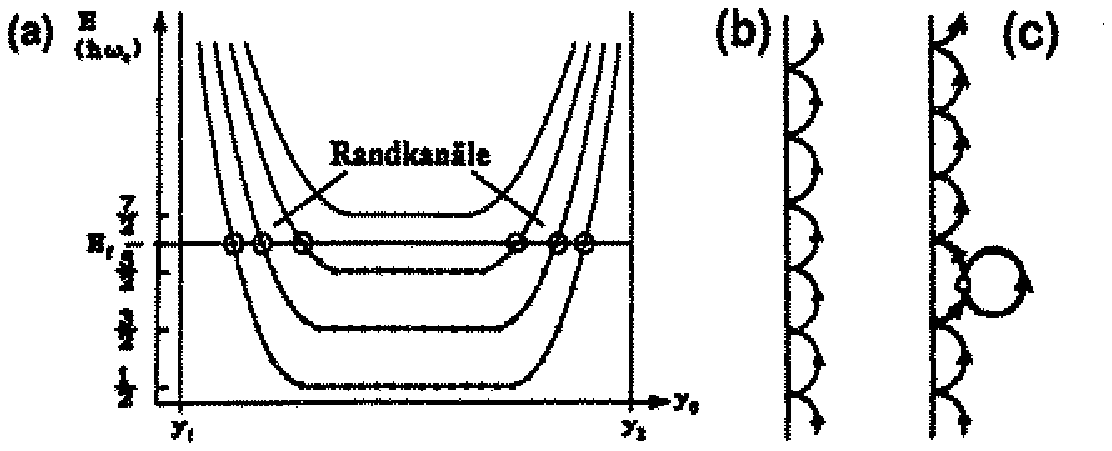}{Randzust\"ande versus Volumenzust\"ande}{12}
\par
Nun besteht elektrische Leitf\"ahigkeit immer dann, wenn sich die Elektronen
zugunsten einer bestimmten Bewegungs- oder Vorzugsrichtung im $k$-Raum
(auf den Parabeln der Subb\"ander) umverteilen k\"onnen. Dabei sind lediglich
die Elektronen nahe der {\sc Fermi}-Kante wesentlich, da alle energetisch
tieferliegenden Zust\"ande besetzt sind und ihre Umbesetzung durch die
Ununterscheidbarkeit der Elektronen trivial ist, das hei\ss t nichts zum
Transport beitr\"uge.
Ferner m\"ussen freie Zust\"ande
einer Vorzugsrichtung im $k$-Raum vorhanden sein,
d.\,h.\ die Zu\-stands\-dich\-te
bei der {\sc Fermi}-Energie darf nicht verschwinden. Wie wir gesehen haben,
wird diese Bedingung gerade in der Umgebung der {\sc Landau}-Niveaus erf\"ullt.
\par
Der wesentliche Punkt ist nun der folgende:
Neben dem schon vorgestellten Szenario, da\ss\
die {\sc Fermi}-Energie ein r\"aumlich starr vorgegebenes Spektrum von
{\sc Landau}-B\"andern durchl\"auft, d\"urfen wir davon ausgehen, da\ss\
letztere sich selbst an dem Rand der Probe so nach oben biegen, da\ss\ sie
ihrerseits eine global konstante {\sc Fermi}-Kante durchlaufen. Denn die
Tatsache, da\ss\ der Probenrand f\"ur die Elektronen ein un\"uberwindliches
Hindernis darstellt, kann so gedeutet werden, da\ss\ die Elektronen der Probe
in einem Potentialtopf oder einsperrendem Potential
(engl.\ {\it confining potential\/}) gefangen sind.
Somit kreuzt jedes {\sc Landau}-Niveau infolge des
starken Confining-Potentials die {\sc Fermi}-Energie am Rand der Probe,
an dem somit ein Kanal nicht verschwindender
Zu\-stands\-dich\-te entsteht.
\par
Die Bewegung der Elektronen am Rand der Probe kann man sich
aus Teilkreisen
(engl.\ {\it skipping orbits\/})
zusammengesetzt denken.
Eine Streuung nach innen oder gar von Probenrand zu Probenrand
besitzt eine verschwindende Wahrscheinlichkeit, da ein vom Rand
in die Probe hineinkreisendes Elektron,
wenn es auf eine St\"orstelle trifft,
nach einem weiteren Umlauf wieder
an den Rand zur\"uckgestreut wird.
Offensichtlich haben {\it skipping orbits\/}
eine h\"ohere Frequenz als die Vollkreise,
das hei\ss t, ihre Energie ist gr\"o\ss er.
Dies erkl\"art auf semiklassischen Niveau,
warum die Randzust\"ande eine h\"ohere
Energie haben m\"ussen als die Zust\"ande
im Innern.
\par
Wir fassen zusammen:
Solange wir uns also nicht im Zentrum eines {\sc Landau}-Niveaus befinden,
verschwindet die Leif\"ahigkeit im Probeninnern: Dort stehen keine Zust\"ande
f\"ur den Stromflu\ss\ zur Verf\"ugung.
F\"ur jedes {\sc Landau}-Niveau bildet sich ein Randkanal aus,
wobei der dem niedrigsten {\sc Landau}-Level zugeordnete Kanal am
weitesten au\ss en liegt.
\par
Beschreiben wir nun den Stromtransport
am Rand der Probe im Rahmen einer 1-dimensionalen Transporttheorie
von
{\sc Landauer} und {\sc B\"uttiker}
\cite{Buettiker90, Landauer87}.
Eine solche Theorie kann in Analogie zur W\"armeleitung gesehen
werden. Dort wird die Frage behandelt, wie Energie von einer
hei\ss en zu einer kalten Stelle transportiert wird. Energie
und Temperatur sind in gleicher Weise duale oder konjugierte
Variablen wie Teilchenzahl (oder besser Ladung) und chemisches Potential.
Stellen wir uns also die auf dem Rand kontaktierte Probe als ein System
vor, in dessen Rand an bestimmten Stellen das chemische Potential
lokal fixiert ist. Zur Berechnung der gemessenen Leitf\"ahigkeiten
m\"ussen wir einen Ausdruck f\"ur einen 1-dimensionalen Strom
in Termen der 1-dimensionalen Zu\-stands\-dich\-te angeben.
\par
Diese ist f\"ur den allgemeinen Fall gegeben durch
\begin{equation}
D_{1D}(E) = g_s g_v \cdot \frac{1}{h}\sqrt{\frac{m}{2E}}.
\end{equation}
Es ist hervorzuheben, da\ss\ es zu jedem Energieeigenwert
$E$ {\it zwei\/} Gruppengeschwindigkeiten $v_g$ gibt,
die sich durch ihr Vorzeichen unterscheiden.
Wir k\"onnen schreiben:
\begin{equation}
v_g(E) = \frac{1}{\hbar}
           \left. \frac{dE}{dk} \right|_E                  \nonumber\\
       = \frac{1}{\hbar}                                 %
           \left. \frac{d(\hbar^2k^2/2m)}{dk} \right|_E    \nonumber\\
       = \left. \frac{\hbar k}{m}           \right|_E    \nonumber\\
       = \pm \sqrt{ \frac{2E}{m} }.
\end{equation}
In der Berechnung des elektrischen Stroms in einem 1D-Elektronensystem,
der sich aus dem Produkt von Gruppengeschwindigkeit $v_g$ und
Elektronenladung $e$ gewichtet mit der Zu\-stands\-dich\-te $D_{1D}(E)$
und der {\sc Fermi}-{\sc Dirac}-Verteilungsfunktion
\begin{equation}
f(E)=\frac{1}{\exp\,(E-\mu)/kT+1}
\end{equation}
ergibt, darf nur ein Zweig ber\"ucksichtigt werden. Somit ist
\begin{eqnarray}
I &=& \int_0^\infty \frac{1}{2}\,ev_g D_{1D}(E) f(E) \,dE \nonumber\\
  & &                                                     \nonumber\\
  &=& \frac{e}{2} \cdot \int_0^\infty
        \sqrt{\frac{2E}{m}} \cdot g_sg_v \cdot
        \frac{1}{h}
        \sqrt{\frac{2m}{E}} \cdot f(E) \,dE               \nonumber\\
  & &                                                     \nonumber\\
  &=& g_sg_v \cdot \frac{e}{h} \cdot
        \int_0^\infty
        f(E) \,dE                                         \nonumber\\
  & &                                                     \nonumber\\
  &=& g_sg_v \cdot
        \frac{e}{h} \, \mu.
\end{eqnarray}
Man erinnere sich, da\ss\ im Grenzfall $T \rightarrow 0$
die {\sc Fermi}-{\sc Dirac}-Verteilungsfunktion die Form
\begin{equation}
f(E)=\left\{
            \begin{array}{ccl}
              1, & \phantom{123} &  \mbox{ {\rm falls $E<\mu$,} }\\
              0, & \phantom{123} &  \mbox{ {\rm falls $E>\mu$,} }
            \end{array}
     \right.
\end{equation}
hat. Durch Einsetzen dieses Ausdrucks in das obige
Integral verifiziert man leicht das Ergebnis der
Rechnung f\"ur den Fall verschwindender Temperatur.
\par
Betrachten wir nun einen direkten Halbleiter mit $g_v=1$,
so folgt aus der zweifachen Spinentartung $g_2=2$ f\"ur den
Transport in 1D-Kan\"alen mit
\begin{equation}
\mu=eU
\end{equation}
die Beziehung
\begin{equation}
I=i \cdot \frac{2e^2}{h} \cdot U=:R_{ball}\cdot U,
\end{equation}
f\"ur $i$ besetzte Subb\"ander mit
\begin{equation}
R_{ball}=\frac{U}{I}
        =\frac{h}{2e^2}
        =12.9064 {\dots} \, k\Omega.
\end{equation}
Eine solche Quantisierung der elektrischen Leitf\"ahigkeit
im Regime des ballistischen Transports
(in dem die Bewegung der Elektronen streufrei erfolgt)
wurde 1988 erstmalig entdeckt
\cite{Wees88, Wharam88a, Wharam88b, Wharam90},
siehe auch
\cite{Khurana8811}.
In starken Magnetfeldern (und auch in starken elektrischen
Feldern) wird die Spinentartung me\ss bar aufgehoben, und
es entwickeln sich f\"ur die Leitf\"ahigkeit zus\"atzlich
Halb-Plateaux, d.\,h.\  ungerade Vielfache von $e^2/h$.
Neuere Forschungen in Bochum und Cambridge haben gezeigt,
da\ss\ unterhalb von $2e^2/h$ ein Zwischenplateau
auch ohne \"au\ss eres Magnetfeld auftritt,
wobei noch nicht ganz klar ist, ob es sich dabei
um eine $0.5\cdot(2e^2/h)$- oder
um eine $0.7\cdot(2e^2/h)$-{\it Struktur\/}
handelt.
Unabh\"angig davon ist {\it schon die Struktur an sich\/}
ein Hinweis auf eine spontane Spin-Polarisation bzw.\
einen spontanen Magnetismus
infolge einer Elektron-Elektron-Wechselwirkung
\cite{Thomas96,TscheuschnerWieck96}.
\par
Im folgenden wollen wir die Quantisierung der {\sc Hall}-Leitf\"ahigkeit
durch die Quantisierung der 1D-Randzust\"ande verstehen. Dabei k\"onnen
wir davon ausgehen, da\ss\ die soeben vorgestellte Spinentartung aufgehoben
ist. Wir setzen also $g_s=1$ und setzen ferner $g_v=1$
an, was z.\,B.\ f\"ur GaAs-Heterostrukturen gilt.
\par
Wir nehmen im folgenden an, da\ss\ unsere Probe die Form
eines typischen {\sc Hall}-Stabes
(engl.\ {\it Hall bar\/})
habe und
mit sechs Kontakten $\mu_1,\dots \mu_6$
(im Uhrzeigersinn gez\"ahlt) ver\-se\-hen ist,
welche die chemischen Potentiale festlegen. {\it Per conventionem\/}
sei $\mu_1$ die Source (hier flie\ss t der Strom $j_x$ hinein)
und $\mu_4$ der Drain (hier flie\ss t der Strom $j_x$ ab).
Die {\sc Hall}-Spannung kann zwischen den jeweils gegen\"uberliegenden
Kontakten $\mu_2$ und $\mu_6$ bzw.\ $\mu_3$ und $\mu_5$ abgenommen
werden. Entsprechend messen wir den longitudinalen Spannungsabfall
zwischen $\mu_2$ und $\mu_3$ bzw.\ $\mu_5$ und $\mu_6$. \"Uber
die Kontakte $\mu_2,\mu_3,\mu_5,\mu_6$ sollten keine Str\"ome
ab- oder zuflie\ss en, so da\ss\ $\mu_1$ und $\mu_4$ den Gesamtstrom
tragen m\"ussen. Dies ist gleichbedeutend mit
der Bedingung, da\ss\ \loq spannungsrichtig gemessen wird\hiq,
das hei\ss t, der Innenwiderstand unseres Voltmeters gegen
unendlich geht.
\par
Der totale Strom eines Kontaktes oder Reservoirs ist die Differenz
der Str\"ome, die von einlaufenden und auslaufenden Kan\"alen getragen
werden. Wir m\"ussen also den ankommenden Strom von dem injizierten
Strom abziehen, um den totalen Strom zu erhalten.
\eject
\bild{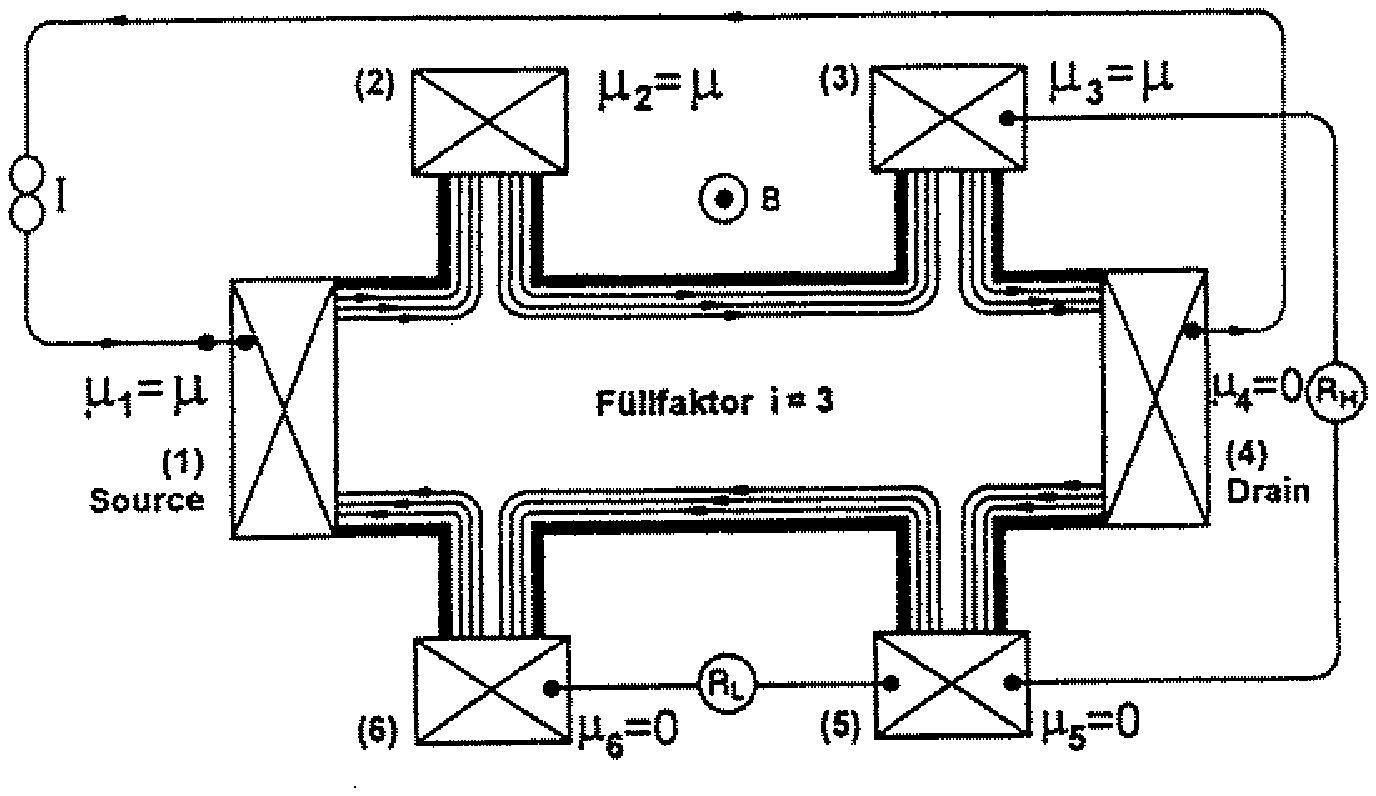}{Randstromkan\"ale}{16}
Unter der Annahme,
da\ss\ der Transportstrom sich im Uhrzeigersinn bewegt, erhalten
wir
\begin{equation}
\begin{array}{lc}
\mbox{Source reservoir}      \phantom{123} &
   \mu_1:\phantom{x}I_{tot}=\phantom{-}I=i\cdot\frac{e}{h}(\mu_1-\mu_6) \\
\mbox{Potential reservoir}   \phantom{123} &
   \mu_2:\phantom{x}I_{tot}=\phantom{-}0=i\cdot\frac{e}{h}(\mu_2-\mu_1) \\
\mbox{Potential reservoir}   \phantom{123} &
   \mu_3:\phantom{x}I_{tot}=\phantom{-}0=i\cdot\frac{e}{h}(\mu_3-\mu_2) \\
\mbox{Drain reservoir}       \phantom{123} &
   \mu_4:\phantom{x}I_{tot}=         - I=i\cdot\frac{e}{h}(\mu_4-\mu_3) \\
\mbox{Potential reservoir}   \phantom{123} &
   \mu_5:\phantom{x}I_{tot}=\phantom{-}0=i\cdot\frac{e}{h}(\mu_5-\mu_4) \\
\mbox{Potential reservoir}   \phantom{123} &
   \mu_6:\phantom{x}I_{tot}=\phantom{-}0=i\cdot\frac{e}{h}(\mu_6-\mu_5)
\end{array}
\end{equation}
Der {\sc Hall}-Widerstand ergibt sich somit als
\begin{equation}
R_H=\frac{U_H}{I}
   =\frac{(\mu_3-\mu_5)/e}{I}
   =\frac{(\mu_3-\mu_5)}{i(\mu_3-\mu_5)e^2/h}
   =\frac{h}{ie^2},
\end{equation}
der longitudinale Magnetowiderstand als
\begin{equation}
R_{xx}=\frac{U_{xx}}{I}
      =\frac{(\mu_2-\mu_3)/e}{I}
      =0.
\end{equation}
Man beachte, da\ss\ die Leitf\"ahigkeit pro Randkanal gerade
\begin{equation}
\delta\sigma=\frac{e^2}{h}
\end{equation}
ist, die Summe somit
\begin{equation}
\sigma=i\cdot\frac{e^2}{h} \, .
\end{equation}
\par
Das Randkanal-Bild ber\"ucksichtigt, im Gegensatz zum Lokalisierungsbild,
die Geometrie der Probe und das Vorhandensein von Kontakten. Es erkl\"art
die Quantisierung des {\sc Hall}-Widerstandes als Folge der 1D-Quantiserung
der Leitf\"ahigkeit. Als versteckte Annahme enth\"alt es aber die
Lokalisierung im Bulk und kann daher nicht f\"ur sich isoliert betrachtet
werden. Wohl aber mu\ss\ eine Theorie der Edge-States vertr\"aglich sein
mit einer solchen, welche den Bulk beschreibt. Das Randkanal-Bild erkl\"art
allerdings {\it nicht\/} die Form der \"Uberg\"ange zwischen den Plateaux und
deren Breite.
\vfill\eject\noindent%
\section{Aufgaben} 
Es ist f\"ur das Verst\"andnis des Versuches {\it sehr n\"utzlich\/},
die theoretischen Aufgaben schon im Rahmen der Vorbereitungen
zu l\"osen. Die experimentellen Aufgaben sollten im Rahmen
der Durchf\"uhrung und Auswertung des Versuchs {\it vollst\"andig\/}
bearbeitet werden. Die relevanten Daten der Ger\"ate-Eichung und
der MBE-Wachstumsprotokolle erfragen Sie bitte beim Assistenten.
\subsection{Theorie}
Die folgenden Aufgaben sollten einen Anhaltspunkt daf\"ur geben,
was von den Studierenden im Rahmen der Vorbereitung erwartet wird.
\par
Eine Bemerkung zum Literaturstudium: Obwohl dieses Skript so
{\it self-contained\/} wie m\"oglich gehalten wurde, sei zumindest
eine Lekt\"ure der Quellen
\cite{Klitzing80}, \cite{Klitzing86} und \cite{Klitzing90}
em\-pfoh\-len. Wer noch ein wenig \"uber unseren
Horizont blicken m\"ochte, sei auf
\cite{Kivelson96}
verwiesen.
\par
W\"ahrend des Versuchs besteht ausreichend Zeit, \"uber den theoretischen
Hintergrund des Experiments zu diskutieren und offene Fragen zu kl\"aren.
Die relevanten Daten der Probe, das hei\ss t, die MBE-Wachstumsprotokolle,
und die Eichung der Ger\"ate erfragen sie bitte beim Assistenten.
\par
Transportmessungen im Quantenregime sind Pr\"azisionsmessungen.
Besondere Sorgfalt mu\ss\ daher auf eine wirksame Unterdr\"uckung
von St\"oreinfl\"ussen (Netzbrummen, St\"orungen durch Radio-
und Fernsehsender, thermisches Hintergrund-Rauschen etc.) gelegt werden.
Eine sehr effektive Methode in diesem Zusammenhang ist die Lock-In-Technik.
Im Addendum zu dieser Anleitung finden Sie einen Auszug aus dem einf\"uhrenden
Kapitel der Anleitung eines kommerziellen Lock-In-Verst\"arkers,
in dem das Prinzip auch f\"ur den elektronischen Laien verst\"andlich
erkl\"art wird \cite{Stanford}.
\begin{enumerate}
\item
In der obigen Einf\"uhrung wurde die Formel f\"ur den
klassischen {\sc Hall}-Effekt in einem Einteilchen-Bild
hergeleitet. Verifizieren Sie noch einmal die Formel
im Rahmen der {\sc Drude}-Theorie des elektrischen Transports
in Termen der inelastischen
{\sc Drude}-Impulsrelaxationszeit $\tau$.
\item
Verifizieren Sie die Inversion des Leitf\"ahigkeitstensors
zum Widerstandstensor (das hei\ss t: suche die inverse
Matrix $\varrho_{ij}$ von $\sigma_{ij}$) und diskutieren
Sie den Fall sehr hoher Magnetfelder.
\item
Sch\"atzen  Sie die 2-dimensionale Ladungstr\"agerdichte $n_{2D}$
und die elektrische Feld\-st\"ar\-ke $F_s$ an der Grenzfl\"ache einer
$\mbox{Ga}\mbox{As}$-$\mbox{Al}_{0.3}\mbox{Ga}_{0.7}\mbox{As}$-%
Heterostruktur ab, die sich durch eine Bandkantendiskontiuit\"at
von $300 \;\mbox{meV}$ auszeichnet.
\item
F\"ur ein unendlich hohes Kastenpotential der Breite $L$
ergeben sich die Energieeigenwerte
(mit $m_z$ als effektive Masse
der Ladungstr\"ager in z-Richtung)
\begin{equation}
E_j = \frac{\hbar^2k_j^2}{2m_z},
\phantom{^123}
k_j = \pi\,\frac{j+1}{L},
\phantom{123}
(j=0,1,\dots),
\end{equation}
das hei\ss t als
\begin{equation}
E_j=\frac{h^2(j+1)^2}{2m_zL^2},
\phantom{123}
(j=0,1,\dots).
\end{equation}
F\"ur ein Dreieckpotential ergeben sich die Energieeigenwerte
n\"aherungsweise als
\begin{equation}
E_j \approx
\left( \frac{\hbar^2}{2m_z} \right) ^{1/3}
\left[ \frac{3\pi e F_s}{2} \left(j+\frac{3}{4}\right) \right] ^{2/3},
\phantom{123}
(j=0,1,\dots).
\end{equation}
Berechnen Sie mit der elektrischen Feld\-st\"ar\-ke $F_s$ aus Aufgabe 3 und
$m_z=0.07\,m_0$ die Subbandenergien $E_0$ und $E_1$.
\item
Sei $m_x=m_y=0.07\,m_0$.%
\footnote{Die effektive Masse kann in verschiedene Richtungen
          verschiedene Werte annehmen.}
Berechnen Sie die {\sc Fermi}-Energie $E_F$ f\"ur
\begin{equation}
n = 3 \times 10^{11} {\rm cm}^{-2}.
\end{equation}
Wieviele Subb\"ander sind bei $T=0\,K$ und der
2-dimensionalen Ladungstr\"agerdichte $n_{2D}$ aus Aufgabe 3
besetzt?
\item
Machen Sie sich mit der expliziten L\"osung
der {\sc Schr\"odinger}-Gleichung im konstanten
\"au\ss eren Magnetfeld vertraut
\cite{LandauLifshitz}.
Was geschieht im Falle einer {\sc Vogt}-Geometrie
und in Falle eines schr\"agen
(engl.\ {\it tilted\/})
Magnetfeldes (qualitativ)
\cite{Clausnitzer85}?
\item
Zeichnen Sie qualitativ den Verlauf
von $\varrho_{xx}$ und $\varrho_{yx}$
als Funktion des Magnetfeldes f\"ur folgende
F\"alle:
\begin{itemize}
   \item[(a)] klassischer Grenzfall;
   \item[(b)] $T=0$ ohne lokalisierte Zust\"ande;
   \item[(c)] $T=0$ mit lokalisierten Zust\"anden.
\end{itemize}
Kennzeichnen Sie auf der Magnetfeldachse
bei b) und c) die Stelle $\mu B=1$.
\item
Erkl\"aren Sie das Prinzip der Lock-In-Me\ss technik,
und entwerfen Sie einen Aufbau f\"ur das {\sc Hall}-Experiment.
\end{enumerate}
\vfill\eject\noindent%
\subsection{Experiment}
Das Experiment sollte in etwa wie folgt ablaufen:
\begin{enumerate}
\item
Zusammen mit dem Assistenten machen sich die
Studenten mit der Me\ss apparatur vertraut
(Helium-Kanne bzw.\ Helium-Kryostat,
Heliumgas-R\"uckf\"uhrleitungssystem,
Magnet und Netzger\"at, Probenhalter,
{\sc Hall}-Bar und seine Geometrie etc.).
\item
Sodann wird der Probenhalter mit der Probe beladen und
in die Helium-Kanne vorsichtig (!) eingebracht.
\item
Die Messungen sind zun\"achst bei der Temperatur des fl\"ussigen
Heliums unter Normaldruck durchzuf\"uhren und mehrmals zu wiederholen.
\item
Mit Hilfe des vom Autor ersonnenen \loq {\sc von\,Klitzing}\,izer\hiq,
einem Steckfeld bestehend aus einer Kollektion von Wendelpotis,
die auf die Werte der quantisierten {\sc Hall}-Widerst\"ande
justiert sind, pr\"ufe man das grunds\"atzliche Funktionieren
der Me\ss\-ein\-rich\-tung. So k\"onnen grobe Fehler ausgeschlossen werden.
\item
Erst jetzt wird begonnen, den Gasdruck \"uber dem
fl\"ussigen He-Spiegel langsam zu abzusenken (\loq Abpumpen\hiq).
\item
W\"ahrend dieses Abpumpens soll sowohl $\varrho_{yx}$
als auch $\varrho_{xx}$ mehrfach in $B$-sweeps gemessen
werden, um die Temperaturabh\"angigkeit zu studieren.
\item
Nachdem die niedrigstm\"ogliche Temperatur
erreicht ist, werden die Messungen wiederholt.
\item
Sodann wird die Probe durch die eingebaute LED
mit IR-Lichtblitzen unterschiedlicher Dauer
beleuchtet. Die Messungen sind zu wiederholen,
um den Einflu\ss\ der Beleuchtung zu studieren.
Es soll analysiert werden, wie sich das Verhalten
der Probe nach Belichtung im Laufe der Zeit
\"andert. Dazu ist es notwendig, die Messungen
noch mehrfach zu wiederholen
(typ.\ 10 sec., 2 min., 20 min., 2 h nach
Beleuchtung).
\item
Die abschlie\ss enden Messungen
von $\varrho_{yx}$ und $\varrho_{xx}$
werden am darauffolgenden Tag durchgef\"uhrt.
\item
({\it Fakultativ:\/})
W\"ahrend des Versuchs gibt es Totzeiten.
Diese sollen nicht nur genutzt werden zur
Diskussion der Theorie des Effektes,
sondern auch - wenn m\"oglich - zur Hospitation
an der MBE-Anlage der Arbeitsgruppe w\"ahrend des
Wachsens einer Probe (\loq wachsen\hiq\
hier verstanden als transitives Verb),
die Sie selbst am zweiten Tag
kontaktieren und messen d\"urfen.
\end{enumerate}
\vspace*{1.0cm}
\bild{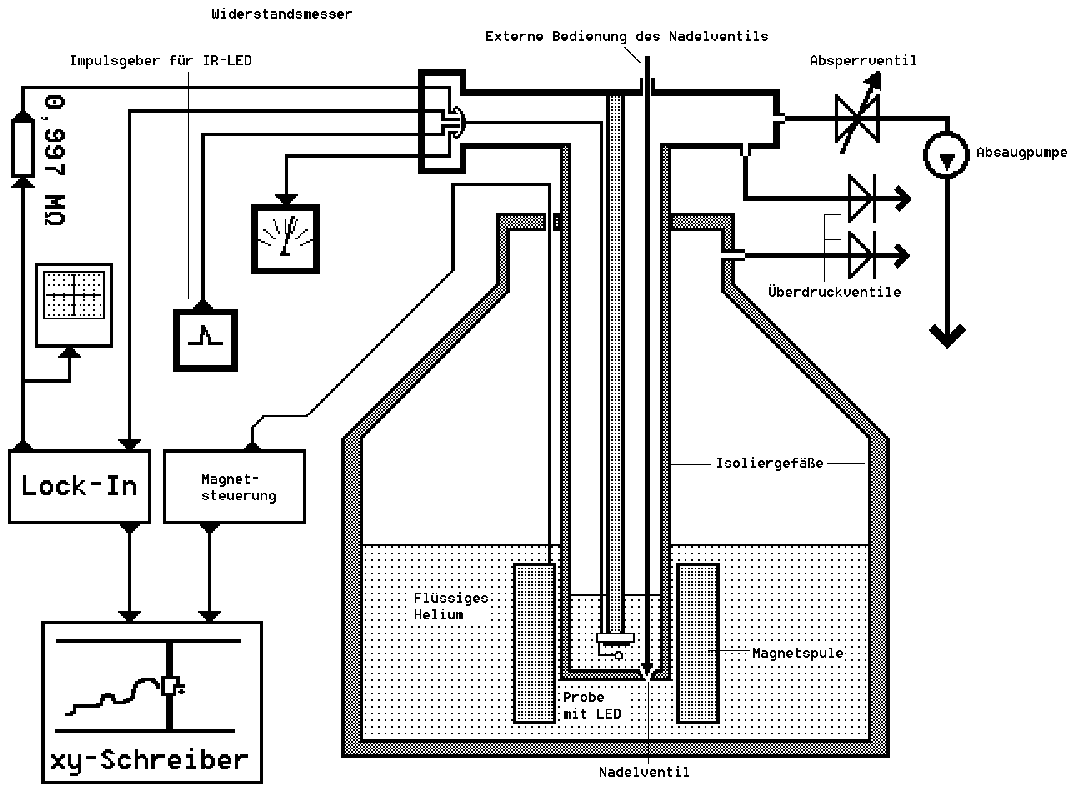}{Helium-Kanne (schematisch) \cite{Hensel96}}{16}
\bild{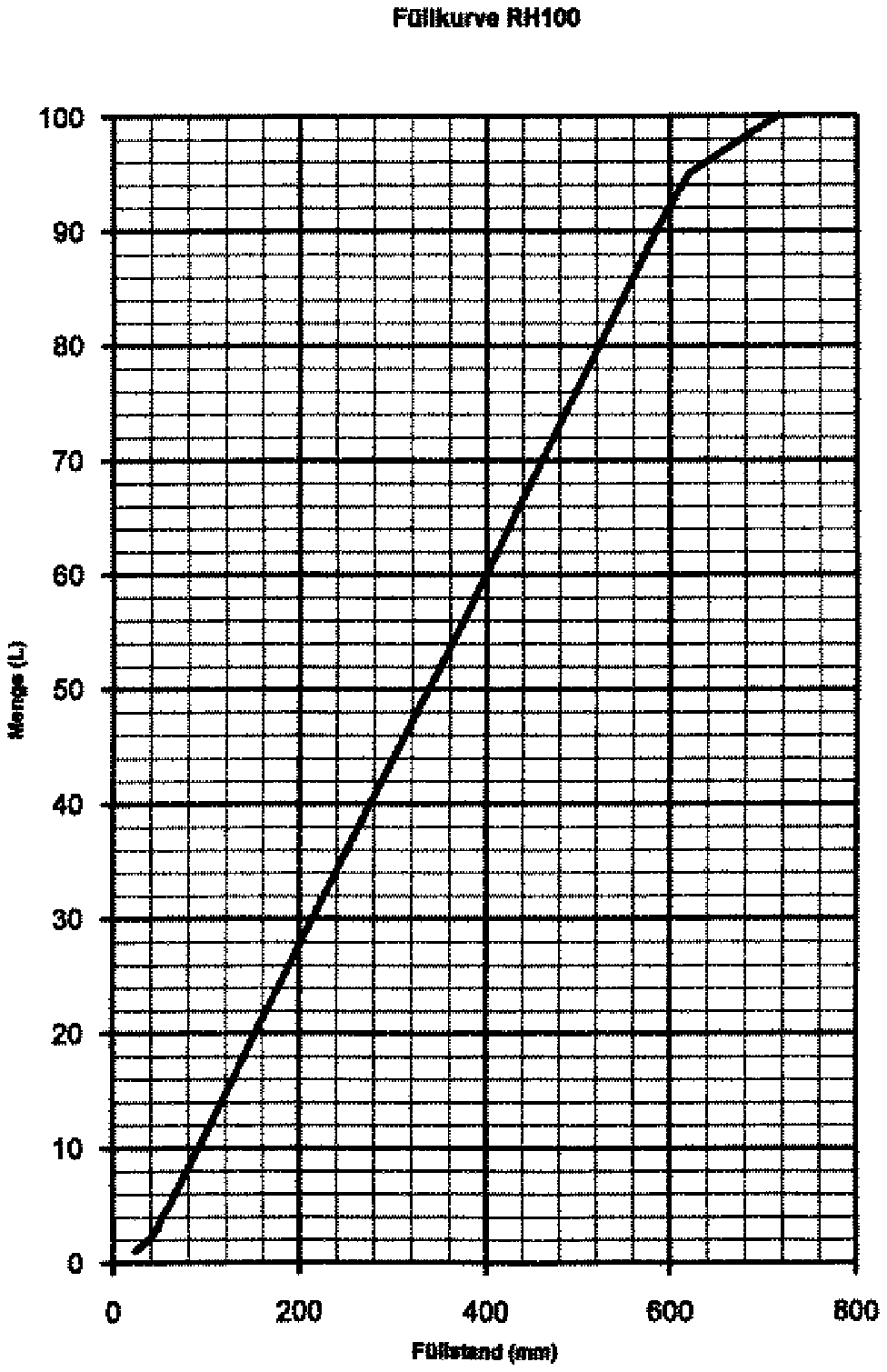}{F\"ullkurve der Helium-Kanne}{12}
\bild{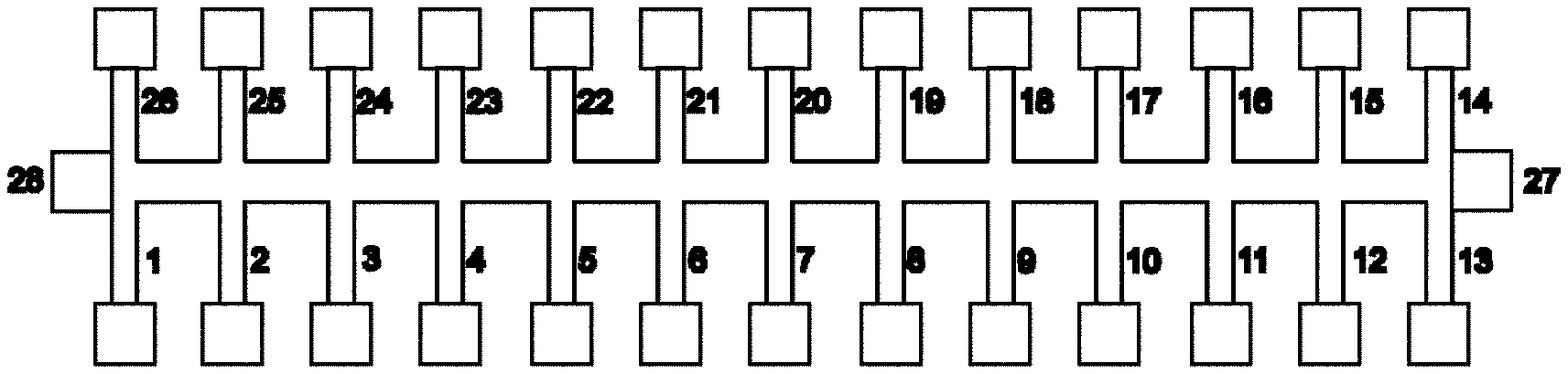}{{\sc Hall}-Streifen (Skizze)}{12}
\bild{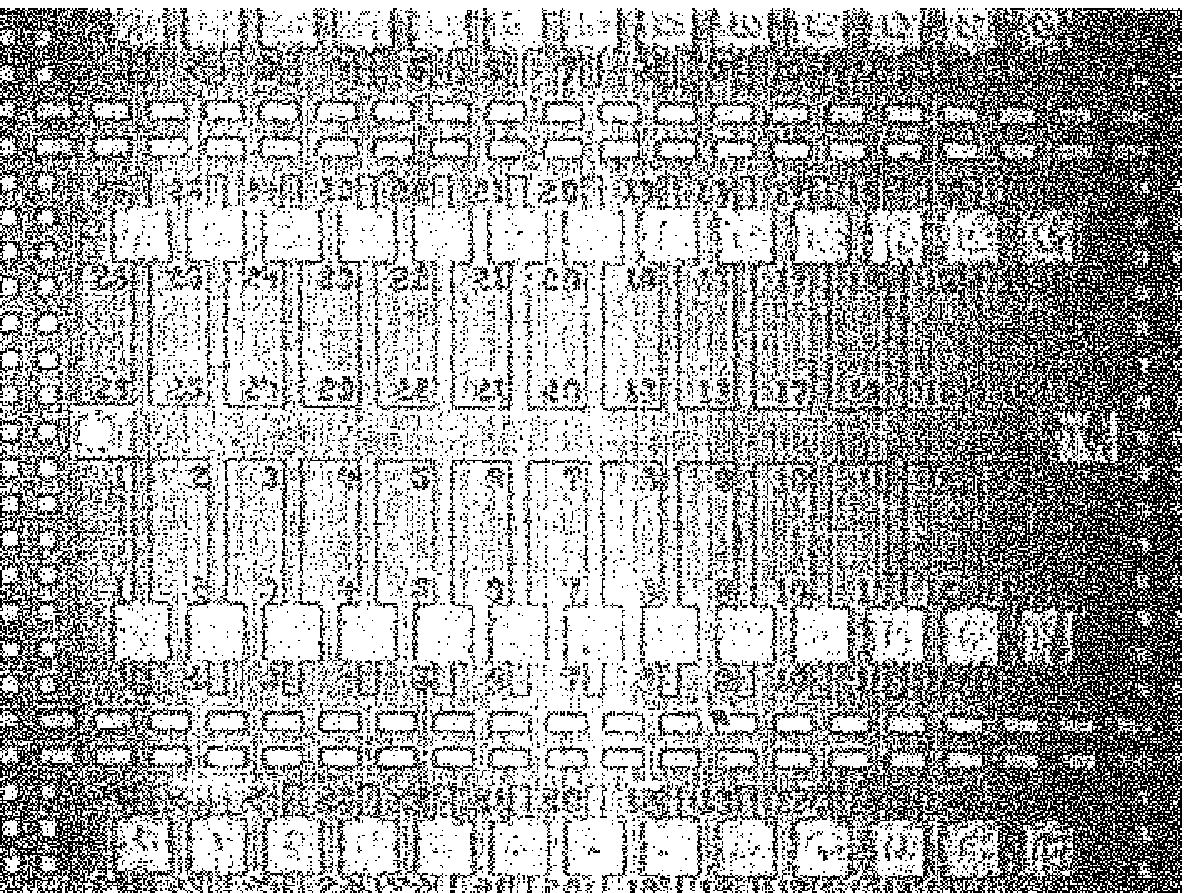}{{\sc Hall}-Streifen (Photo)}{12}
\bild{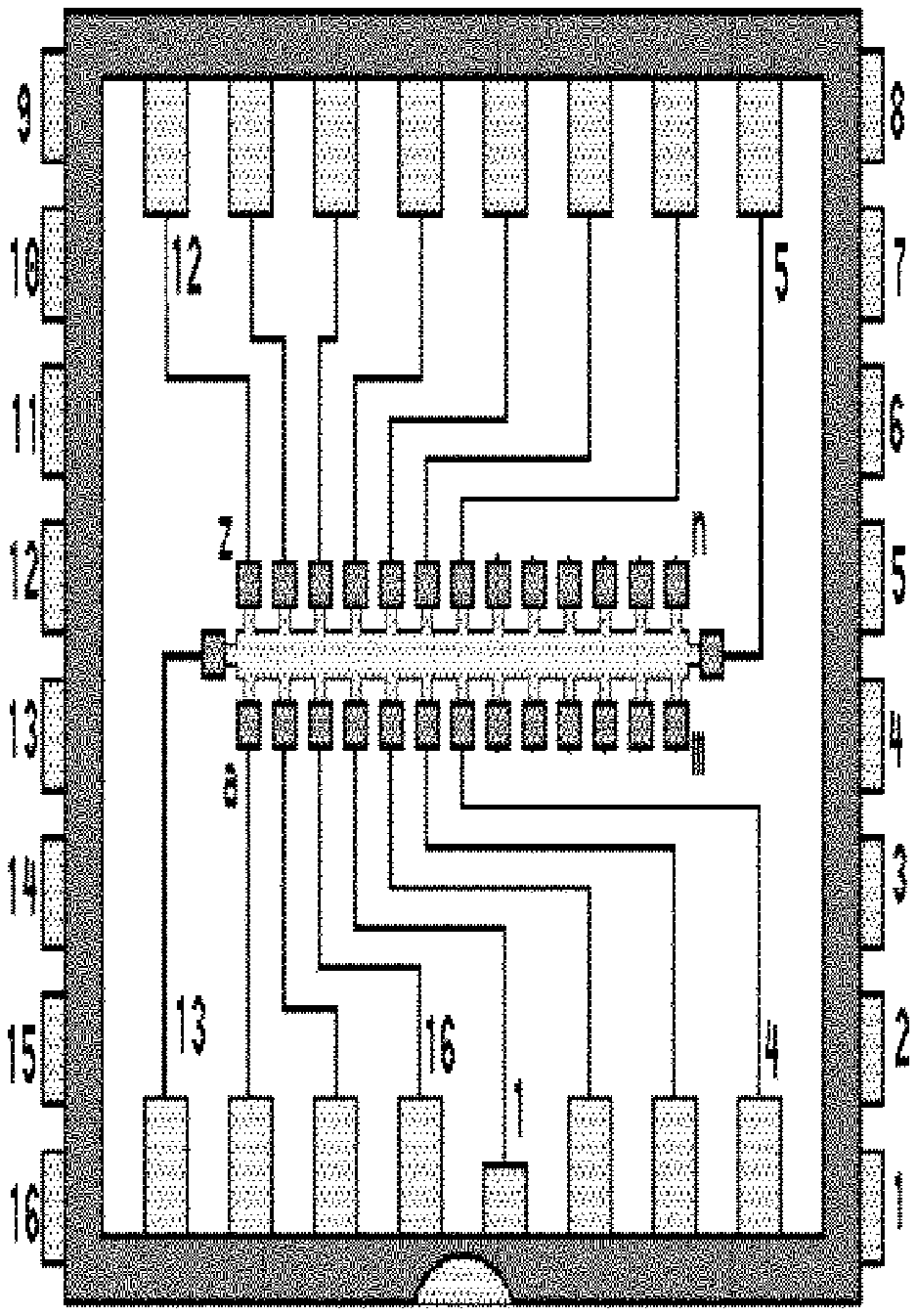}{Bondplan f\"ur die {\sc Hall}-Probe
              (schematisch) \cite{Hensel96}}{12}
\vfill\eject\noindent%
\subsection{Auswertung}
Wenn Sie Schritt f\"ur Schritt die folgenden Punkte durchgehen,
d\"urfte die Auswertung Ihnen keine besonderen Probleme bereiten.
Halten Sie sich daher bitte an das folgende {\it Curriculum\/}:
\begin{enumerate}
\item
Machen Sie sich eine Zeichnung der {\sc Hall}-Probe und
tragen Sie in diese ein, in welcher Richtung der Strom flie\ss t,
in welcher Richtung das Magnetfeld wirkt und wo welche
Spannungen abgegriffen werden.
\item
Machen Sie sich eine Zeichnung Ihrer Verschaltung
(Lock-In-Verst\"arker, Trenntrafo, Vorwiderstand, XY-Schreiber)
und tragen Sie in diese ein, wo welcher Strom flie\ss t
und wo welche Spannung anliegt.
\item
Machen Sie eine Bestandsaufnahme der Bereichseinstellungen
der verwendeten Ge\-r\"a\-te (Lock-In-Verst\"arker, XY-Schreiber).
Bereiten Sie Ihre Graphen so auf, da\ss\ alle Achsen beschriftet sind.
\item
Bestimmen Sie die in Abwesenheit eines Magnetfeldes
vorliegende Leitf\"ahigkeit $\sigma_0$ aus dem \"uber
den Vorwiderstand in die Probe injizierten Strom und
der gemessenen Spannung:
\begin{equation}
\sigma_0 = \frac{j_x}{E_x} = \frac{I_x}{U_y}.
\end{equation}
\item
Bestimmen Sie $\varrho_{xx}$
(klassisch gegeben durch $m/n_{2D}e^2\tau$)
und $\varrho_{yx}$
(klassisch gegeben durch $B/n_{2D}e$)
als Funktion des Magnetfeldes
aus den von Ihnen aufgenommenen Me\ss kurven.
\par
{\it Hinweis:\/} Auf dem XY-Schreiber k\"onnen Sie die gemessenen
Spannungen direkt ablesen, wenn Sie alle Umrechnungsfaktoren
aus den Regler- und Bereichseinstellungen ber\"ucksichtigen.
Es ist
\begin{equation}
R_H = \frac{U_y}{I_x} = \frac{h}{\nu e^2}.
\end{equation}
Um herauszufinden, bei welchem $\nu$ man jeweils liegt,
tr\"agt man entsprechend der Relation
\begin{equation}
\frac{1}{R_H} = \nu \cdot \frac{e^2}{h} = \frac{e\,n_{2D}}{B}
\end{equation}
die Kehrwerte der gemessenen Plateau-Widerst\"ande gegen
$1/B$ auf. Man sieht sofort, da\ss\ die einzelnen
inversen Widerstandsstufen einen konstanten Abstand haben und
kann sogleich die Frage beantworten, welches die
kleinste Stufe f\"ur den inversen {\sc Hall}-Widerstand ist.
\item
Extrapolieren Sie auf die Magnetfeldst\"arke, bei welcher der
F\"ullfaktor $\nu=1$ erreicht ist.
\item
Bestimmen Sie die Elektronendichte nach drei verschiedenen Methoden:
\begin{itemize}
\item[(a)]
klassisch aus der Steigung der gemittelten Kurve
entsprechend der Beziehung
\begin{equation}
n_{2D} = \frac{B}{\varrho_{yx}e} ;
\end{equation}
\item[(b)]
aus den Plateaux des quantisierten {\sc Hall}-Effektes,
das hei\ss t, aus den Plateaux von $\varrho_{yx}$
und den ermittelten F\"ullfaktoren $\nu$,
entsprechend der Beziehung
\begin{equation}
n_{2D} = \frac{B}
              {\left(
               {\displaystyle \frac{h}{\nu e^2} }
               \right)
               \,e}
       = \frac{\nu e B}{h} ;
\end{equation}

\item[(c)]
aus den {\sc Shubnikov}-{\sc de Haas}-Oszillationen.
Tragen Sie dabei $1/B_i$ als Funktion ganzer Zahlen $i$ auf
und identifizieren Sie die Minima von $\varrho_{xx}$
mit dem zu\-ge\-h\"o\-ri\-gen {\sc Landau}-Niveau-Index $i$
bzw.\ F\"ullfaktor $\nu$.
\end{itemize}
Vergleichen Sie die drei unabh\"angigen Resultate inklusive
einer Fehlerabsch\"atzung (!). Decken sich die Werte?
\item
Berechnen Sie die Beweglichkeit
\begin{equation}
\mu=\frac{e\tau}{m}
\end{equation}
und die
{\sc Drude}-Impulsrelaxationszeit $\tau$.
aus $\sigma_0$
und der von Ihnen bestimmten
2-dimensionalen La\-dungs\-tr\"a\-ger\-dich\-te $n_{2D}$.
Vergleichen Sie $\tau$, $\mu$ und $m$ mit typischen Werten
f\"ur einen 3-dimensionalen Halbleiter und f\"ur ein
Metall bei $4.2\,K$.
\item
Berechnen Sie die {\sc Fermi}-Energie $E_F$
aus der 2-dimensionalen Zu\-stands\-dich\-te%
\linebreak
$D_{2D}(E)$
und
2-dimensionalen Ladungstr\"agerdichte $n_{2D}$.
\item
Bestimmen Sie $h/e^2$ aus dem {\sc Hall}-Plateau mit dem kleinsten
Index $i$.
\item
Mit der Beleuchtung der Probe haben Sie die
2-dimensionale Ladungstr\"agerdichte $n_{2D}$
in der Probe erh\"oht.
Was ist dabei der relevante physikalische Mechanismus?
Die Beleuchtung bleibt erhalten,
auch wenn das Licht ausgeschaltet wird,
solange die Probe nicht erw\"armt wird.
(Dieses Ph\"anomen h\"angt mit den sogenannten DX-Zentren%
\footnote{%
DX steht f\"ur {\it donor complex\/}, siehe z.\,B.\ \cite{LangLogan79}.
         }
in $\mbox{Al}_x\mbox{Ga}_{1-x}\mbox{As}$ zusammen, die
sich durch die $\mbox{Si}$-Dotierung bilden, sowie mit
Tunnelprozessen im allgemeinen.)
Bestimmen Sie die 2-dimensionale Ladungs\-tr\"a\-ger\-dich\-te
$n_{2D}$ und die Beweglichkeit $\mu$ {\it nach\/}
Abschalten der Beleuchtung $t$$=$$0$.
Tragen Sie dazu beide Gr\"o\ss en gegen $\log\,t$ auf.
Beantworten Sie durch Extrapolation der Daten die Frage:
Wie lange m\"u\ss ten wir warten, bis sich der Zustand
vor Beleuchtung wieder eingestellt hat?
\item
Die inverse {\sc Sommerfeld}sche Feinstrukturkonstante
ist - unabh\"angig von den Einheiten - gegeben durch
\begin{equation}
\alpha^{-1}=(h/e^2)(2/\mu_0c)=137,036 \, \dots\;,
\end{equation}
wobei wir f\"ur die Permeabilit\"at des Vakuums
\begin{equation}
\mu_0=4\pi*10^{-7}\,\mbox{Hm}
\end{equation}
und f\"ur die Lichtgeschwindigkeit im Vakuum
\begin{equation}
c=2.9979*10^{8}\,\mbox{m/s}
\end{equation}
angesetzt haben.
Berechnen Sie diese aus dem Me\ss wert f\"ur
die {\sc von\,Klitzing}-Kon\-stan\-te
\begin{equation}
R_{vK}=\frac{h}{e^2}.
\end{equation}
\item
Diskutieren Sie die Rolle der Temperatur f\"ur das Auftreten
des quantisierten {\sc Hall}-Effektes. Vergleichen Sie dabei
$kT$ mit den charakteristischen Energien des Systems, n\"amlich
\begin{itemize}
\item
der {\sc Fermi}-Energie $E_F$,
\item
der {\sc Landau}-Aufspaltung $\hbar\omega_c$,
\item
der Breite der ausgedehnten Zust\"ande $\hbar/\tau$.
\end{itemize}
\item
({\it Fakultativ:\/})
Zeichnen Sie die \"Aquipotentiallinien und die
Strompfade bei einer {\sc Hall}-Geometrie in einem
{\sc Hall}-Plateau.
\item
({\it Fakultativ:\/})
Warum tritt der quantisierte {\sc Hall}-Effekt
nicht in einem $3$-di\-men\-si\-o\-na\-len Elektronengas auf?
Wie k\"onnen Sie sich experimentell den \"Ubergang
(engl.\ {\it crossover\/}) von $2D$ nach $3D$ vorstellen
(Hinweis: Multilayer)?
\end{enumerate}
\vfill\eject\noindent%
\section{Anhang: Physikalische Formelsammlung}
Nach \cite{Barnett96}.
\par
Alle Gr\"o\ss en sind N\"aherungsgro\ss en,
wenn es nicht ausdr\"ucklich anders angegeben ist.
\begin{equation}
a = 0.815\,(10)
\end{equation}
bedeutet, da\ss\ mit $70\,\%$ \loq{\it confidence level\/}\hiq\
der wahre Wert zwischen $0.805$ und $0.825$ liegt.
\par
Alle Gr\"o\ss en sind in SI-Einheiten angegeben,
wenn es nicht ausdr\"ucklich anders spezifiziert wird.%
\footnote{Die Idee, die Formeln in dieser Weise zusammenzustellen,
          geht zwar nicht auf {\sc A.D.\ Wieck} zur\"uck;
          aber er war es, der auf die Idee kam, sich eine
          kleine Karte anzufertigen, die er in das Etui seines
          wissenschaftlichen Taschenrechners {\sc Casio} FX-50F,
          welcher zudem die Naturkonstanten festverdrahtet hat,
          hineinschieben konnte. {\sc Dirk de\,Vries} hat die
          Formelsammlung weiter verbessert. Die vorliegende
          Version enth\"alt zus\"atzlich noch einige weitere,
          f\"ur die Praxis n\"utzliche Fakten. \"Ahnliche
          Formelsammlungen - zugeschnitten auf die Anwendung
          innerhalb der experimentellen Festk\"orperphysik -
          findet man an vielen Stellen, zum Beispiel auch
          in der \loq Bibel\hiq\ von {\sc Ando}, {\sc Fowler}
          und {\sc Stern} \cite{AndoFowlerStern82}.}
\subsection{Allgemeine Physik} 
\subsubsection{Naturkonstanten}
{\sc Boltzmann}-Konstante
\begin{equation}
k=1.380\,658\,(12) \times 10^{-23}\;{\rm J}/{\rm K}
\end{equation}
{\sc Planck}sches Wirkungsquantum
\begin{eqnarray}
h      &=&  6.626\,075\,5\,(40) \times 10^{-34}\;{\rm J}{\rm s}
\end{eqnarray}
Mit
\begin{equation}
\hbar = \frac{h}{2\pi}
\end{equation}
ist
\begin{eqnarray}
\hbar  &=&  1.054\,576\,6\,(63) \times 10^{-34}\;{\rm J}{\rm s}  \nonumber\\
       &=&  6.582\,122\,0\,(20) \times 10^{-22}\;{\rm MeV}\,{\rm s}
\end{eqnarray}
Lichtgeschwindigkeit im Vakuum (exakter Wert)
\begin{equation}
c=\frac{1}{\sqrt{\varepsilon_0\mu_0}}
 =2.997\,924\,58 \times 10^8\;{\rm m}/{\rm s}
\end{equation}
{\sc Newton}sche Gravitationskonstante
\begin{eqnarray}
G_N &=& 6.672\,59\,(85) \times 10^{-11}
        \;{\rm m}^3/{\rm k}{\rm g}\,{\rm s}^2                    \nonumber\\
    &=& 6.707\,11\,(86) \times 10^{-39}
        \;\hbar c \, ({\rm GeV}/c^2)^{-2}
\end{eqnarray}
\subsubsection{Eigenschaften von Elementarteilchen}
Elektrische Elementarladung
\begin{equation}
e=1.602\,177\,33\,(49) \, \times 10^{-19}\;{\rm C}
\end{equation}
Elektronenmasse
\begin{eqnarray}
m_e &=& 9.109\,389\,7\,(54)  \, \times 10^{-31}\;{\rm k}{\rm g} \nonumber\\
    &=& 0.510\,999\,06\,(15) \, {\rm MeV}/c^2
\end{eqnarray}
Protonenmasse
\begin{eqnarray}
m_p &=& 1.672\,623\,1\,(10)  \, \times 10^{-27}\;{\rm k}{\rm g}  \nonumber\\
    &=& 938.272\,31\,(28)    \, {\rm MeV}/c^2
\end{eqnarray}
Neutronenmasse
\begin{eqnarray}
m_n &=& 1.674\,928\,6\,(10)  \, \times 10^{-27}\;{\rm k}{\rm g}  \nonumber\\
    &=& 939.565\,63\,(28)    \, {\rm MeV}/c^2
\end{eqnarray}
\subsubsection{Quantisierter Hall Effekt}
{\sc von\,Klitzing}-Konstante
\begin{equation}
R_{vK} = \frac{h}{e^2} = 25.812\,805\,{\dots}{\rm k}\Omega
\end{equation}
\begin{table}\vspace{0.5cm}
\begin{center}
%
\begin{tabular}{|l|r|}                                              \hline
$ h/e^2  $              & $ 25.812\,805\, {\dots} {\rm k}\Omega $ \\ \hline
$ h/2e^2 $              & $ 12.906\,402\, {\dots} {\rm k}\Omega $ \\ \hline
$ h/3e^2 $              & $  8.604\,268\, {\dots} {\rm k}\Omega $ \\ \hline
$ h/4e^2 $              & $  6.453\,201\, {\dots} {\rm k}\Omega $ \\ \hline
$ h/5e^2 $              & $  5.162\,561\, {\dots} {\rm k}\Omega $ \\ \hline
$ h/6e^2 $              & $  4.302\,134\, {\dots} {\rm k}\Omega $ \\ \hline
$ h/7e^2 $              & $  3.687\,543\, {\dots} {\rm k}\Omega $ \\ \hline
$ h/8e^2 $              & $  3.226\,600\, {\dots} {\rm k}\Omega $ \\ \hline
\end{tabular}
\normalsize
\end{center}
\vspace{0.5cm}
\caption{Quantisierte {\sc Hall}-Widerst\"ande}
\vspace{0.75cm}\end{table}
\subsubsection{Energie {\it versus\/}
               Frequenz {\it versus\/}
               Temperatur etc.}
\begin{equation}
E=eU=\frac{hc}{\lambda}=h\nu=\hbar\omega=kT
\end{equation}
Um ein Gef\"uhl f\"ur Gr\"o\ss enordnungen zu bekommen,
sei ein Blick in die beigef\"ugte Tabelle empfohlen.-
Ferner ist es manchmal n\"utzlich, die folgenden
Beziehungen zu kennen:
\begin{eqnarray}
1\,{\rm J}  &=&   6.241  \,{\dots}\,
                  \times 10^{ 18}\,{\rm eV}                      \\
h           &=&   4.136  \,{\dots}\,
                  \times 10^{-15}\,{\rm eV}\cdot{\rm s}          \\
1\,{\rm kg} &=&   0.8988 \,{\dots}\,
                  \times 10^{ 17}\,{\rm J}  \,/\, c^2
            \,=\, 5.609  \,{\dots}\,
                  \times 10^{ 35}\,{\rm eV} \,/\, c^2            \\
1\,{\rm eV} &=&   1.602\,177\,33\,(49)\,\times 10^{-19}\,{\rm J} \\
1\,{\rm eV}/c^2
            &=&   1.782\,662\,70\,(54)\,\times 10^{-36}\,{\rm kg}\
\end{eqnarray}
\begin{table}\vspace{0.5cm}
\begin{center}
%
\begin{tabular}{|c|c|c|}              \hline
$eU$                       & $\nu$            & $T$                \\ \hline
                                                                      \hline
${\bf 1\,eV}$              & $242\,{\rm THz}$ & 11\,600\,{\rm K}   \\ \hline
${\bf 1\,meV}$             & $242\,{\rm GHz}$ & 11.6\,{\rm K}      \\ \hline
${\bf 1}\,\boldmu{\bf eV}$ & $242\,{\rm MHz}$ & 11.6\,{\rm mK}     \\ \hline
                                                                      \hline
$26\,{\rm meV}$            & $6.24\,{\rm THz}$ & ${\bf 300\,K}$    \\ \hline
$86\,\mu{\rm eV}$          & $20.8\,{\rm GHz}$ & ${\bf 1\,K}$      \\ \hline
$8.6\,\mu{\rm eV}$         & $2.08\,{\rm GHz}$ & ${\bf 100\,mK}$   \\ \hline
$0.86\,\mu{\rm eV}$        & $208 \,{\rm MHz}$ & ${\bf 10\,mK}$    \\ \hline
$86\,{\rm neV}$            & $20.8\,{\rm MHz}$ & ${\bf 1\,mK}$     \\ \hline
\end{tabular}
\normalsize
\end{center}
\vspace{0.5cm}
\caption{Beispiele f\"ur Energie versus Frequenz versus Temperatur etc.}
\vspace{0.75cm}\end{table}
\subsubsection{Elektrodynamik}
Magnetische Vakuum-Permeabilit\"at (\loq Nachgiebigkeit\hiq)
\begin{equation}
\mu_0=4\pi \times 10^{-7}\;{\rm H}/{\rm m}
     =12.566\,370\,614\,\times  10^{-7}\;{\rm H}/{\rm m}
\end{equation}
Elektrische Vakuum-Permittivit\"at (\loq Durchl\"assigkeit\hiq)
\begin{equation}
\varepsilon_0=1/\mu_0c^2
             =8.854\,187\,817\,\times 10^{-12}\;{\rm F}/{\rm m}
\end{equation}
\subsubsection{Geophysikalische Konstanten}
Nomineller atmosph\"arischer Druck auf der Erdoberfl\"ache
\begin{equation}
1\;\mbox{{\rm atm}}=1.01325 \times 10^5\;{\rm N}/{\rm m}^2
\end{equation}
Historische Einheit f\"ur den Druck
\begin{equation}
1\,\mbox{{\rm Torr}}
     =(1/760)\,\mbox{{\rm atm}}
     =133.322 \,{\dots}\, {\rm N}/{\rm m}^2
\end{equation}
Nominelle Fallbeschleunigung auf der Erdoberfl\"ache
\begin{equation}
g=9.80665\;{\rm m}/{\rm s}^2
\end{equation}
\subsubsection{Thermodynamik der Hohlraumstrahlung}
{\sc Stefan}-{\sc Boltzmann}-Gesetz:
\begin{equation}
\int_0^\infty F(\lambda)\,d\lambda = \sigma T^4
\end{equation}
mit
\begin{equation}
\sigma=5.6697 \times 10^{-8}\;{\rm W}/{\rm m}^2{\rm K}^4
\end{equation}
Leistung der Strahlung eines schwarzen K\"orpers
\begin{equation}
P(\mbox{{\rm pW}})
    = 5.67\,A({{\rm cm}}^{2}) \cdot b \cdot (T_1^4-T_2^4)
\end{equation}
wobei
\begin{eqnarray}
b=1.00 & & \mbox{f\"ur ideal schwarzen K\"orper} \nonumber\\
b=0.05 & & \mbox{f\"ur poliertes Metall}
\end{eqnarray}
{\sc Wien}sches Verschiebungsgesetz
\begin{equation}
\lambda_{Pmax} \cdot T = \mbox{{\it const}}
\end{equation}
insbesondere haben wir
\begin{equation}
\lambda_{Pmax}=\frac{2898\,\mu{\rm m}}{T}
\end{equation}
\subsubsection{Atomphysik}
Elektronenmasse
\begin{equation}
m_e = 9.109\,389\,7\,(54) \times 10^{-31}\;{\rm k}{\rm g}
\end{equation}
{\sc Bohr}scher Atomradius
\begin{equation}
a_0=4\pi\varepsilon_0
    \cdot
    \frac{\hbar^2}{m_e e^2}
   =\frac{\varepsilon h^2}
         {2\pi m_e e^2}
   =5.292 \times 10^{-11}\,\mbox{{\rm m}}
\end{equation}
{\sc Rydberg}-Konstante
\begin{equation}
1\,\,\mbox{\rm Ry}=\frac{1}{4\pi\varepsilon_0}\cdot
                   \frac{e^2}{2a_0}
                  =\frac{e^4m_e}{(4\pi\varepsilon_0)^2\hbar^2}
                  =13.61 \, {\rm eV}
\end{equation}
{\sc Bohr}sche Energienieveaus im Atom
\begin{equation}
E_n=\frac{h^2n^2}{8 m_e a_0^2}
\end{equation}
\subsubsection{Molek\"ulphysik}
{\sc Avogadro}-Konstante
\begin{equation}
N_{Avo}=6.022045 \times 10^{23}\;\mbox{{\rm mol}}^{-1}
\end{equation}
\subsection{Physik der Halbleiter, insbesondere GaAs} 
\subsubsection{Galliumarsenid}
\lq\lq built-in potential\rq\rq\ in GaAs
\begin{equation}
V_{bi}=1.24\,{\rm V}
\end{equation}
Dielektrische Konstante im Vakuum versus dielektrische Konstante in GaAs
\begin{equation}
\varepsilon_*:=\varepsilon_{\rm GaAs}=13.1\,\varepsilon_0
\end{equation}
Elektronenmasse versus effektive Elektronenmasse
\begin{eqnarray}
m_e   &=& 0.910{\dots} \times 10^{-30}\;{\rm kg} \\
m_e^* &=& 0.637{\dots} \times 10^{-31}\;{\rm kg} \,=\, 0.07\,m_e
\end{eqnarray}
{\sc Bohr}scher Atomradius versus k\"unstlichem {\sc Bohr}schen Atomradius
\begin{eqnarray}
a_0    &=&  4\pi\varepsilon_0\cdot\frac{\hbar^2}{m_e e^2}
      \,=\, \frac{\varepsilon_0 h^2}{2\pi m_e e^2}
      \,=\, 5.292 \times 10^{-11}\,\mbox{{\rm m}}     \\
a_0^*  &=&  4\pi\varepsilon_*\cdot\frac{\hbar^2}{m_e^* e^2}
      \,=\, \frac{\varepsilon_* h^2}{2\pi m_e^* e^2}
      \,=\, 9.90 \,\mbox{{\rm nm}}
      \,=\, \frac{13.1}{0.07}\,a_0
\end{eqnarray}
{\sc Rydberg}-Konstante versus effektive {\sc Rydberg}-Konstante
\begin{eqnarray}
1\,\,\mbox{\rm Ry}    &=&  \frac{1}{4\pi\varepsilon_0}\cdot
                           \frac{e^2}{2a_0}
                     \,=\, \frac{e^4m_e}{(4\pi\varepsilon_0)^2\hbar^2}
                     \,=\, 13.61 \, {\rm eV} \\
1\,\,\mbox{\rm Ry}^*  &=&  \frac{1}{4\pi\varepsilon_0^*}\cdot
                           \frac{e^2}{2a_0^*}
                     \,=\, \frac{e^4m_e^*}{(4\pi\varepsilon_0^*)^2\hbar^2}
                     \,=\, 5.552 \, {\rm meV}
                     \,=\, \frac{0.07}{(13.1)^2}\,\mbox{\rm Ry}
\end{eqnarray}
\subsubsection{Energieniveaus im Kastenpotential}
Diese Beziehung ist wichtig f\"ur {\it quantum wells\/} etc.\
\begin{equation}
E_n=\frac{h^2n^2}{8 m_e^* {a_0^*}^2}
   =5377\,{\rm meV} \cdot n^2\,/a\,({\rm nm})^2
\end{equation}
\subsubsection{Zu\-stands\-dich\-ten und Fermi-Energien}
Zu\-stands\-dich\-te
\begin{eqnarray}
    & &                                                 \nonumber\\
D_{nD} &=& \frac{\mbox{Zu\-stands\-dich\-te in $n$
                    Raumdimensionen incl.\ Spin}}{m^n}
\end{eqnarray}
wobei
\begin{eqnarray}
    & &                                \nonumber\\
D_{3D}(E) &=& \frac{8\pi m\sqrt{2mE}}{h^3}            \\
    & &                                \nonumber\\
D_{2D}(E) &=& \frac{4\pi m}          {h^2}            \\
    & &                                \nonumber\\
D_{1D}(E) &=& \frac{2}{h}\sqrt{\frac{2m}{E}}
\end{eqnarray}
{\sc Fermi}-Energie
\begin{eqnarray}
  & &                                               \nonumber\\
E_{F_3}
  &=&   \frac{h^2     }{8    m} \left( \frac{3 n_{3D} }{\pi} \right)^{2/3}
 \,=\, 52.1  \; meV \cdot
        \left( n_{3D}(10^{18}\;cm^{-3}) \right)^{2/3}           \\
  & &                                               \nonumber\\
E_{F_2}
  &=&   \frac{h^2 n_{2D}  }{4\pi m} \phantom{xxxxxii}
 \,=\, 34.2  \; meV \cdot
        \left( n_{2D} (10^{12}\;cm^{-2}) \right)                 \\
  & &                                               \nonumber\\
E_{F_1}
  &=&   \frac{h^2n_{1D}^2}{32m   } \phantom{xxxxxii}
 \,=\, 13.43 \; meV \cdot
        \left( n_{1D} (10^{ 6}\;cm^{-1}) \right)^2               \\
  & &
\end{eqnarray}
{\sc Fermi}-Wellenl\"ange
\begin{eqnarray}
\lambda_{F_3} &=& 2\sqrt[3]{\frac{ \pi}{3 n_{3D} }}                \\
\lambda_{F_2} &=&  \sqrt   {\frac{2\pi}{  n_{2D} }}                \\
\lambda_{F_1} &=&          {\frac{4   }{  n_{1D} }}
\end{eqnarray}
\subsubsection{Ladungstransport im klassischen Regime}
$n$-dimensionale Ladungstr\"agerdichte
\begin{eqnarray}
    & &                                                 \nonumber\\
n_{nD} &=& \frac{\mbox{Anzahl der Ladungstr\"ager}}
             {\mbox{$n$-dimensionales Volumen}}
\end{eqnarray}
Spezifischer Widerstand (Resistivit\"at) in $n$ Raumdimensionen
\begin{eqnarray}
           & &                                                    \nonumber\\
\varrho_n  &=& R \,\cdot\,
               \frac{\mbox{$n$$-$$1$-dimensionale Querschnittsfl\"ache}}
                    {\mbox{$1$-dimensionale L\"ange}}
          \,=\, \frac{1}{e \mu n_{nD}}
\end{eqnarray}
Beweglichkeit in $n$ Raumdimensionen
\begin{eqnarray}
          & &                                                    \nonumber\\
\mu       &=&  \frac{\mbox{mittlere Driftgeschwindigkeit}}
                    {\mbox{angelegte elektrische Feldst\"arke}}  \nonumber\\
          &\phantom{=}&                                          \nonumber\\
          &=&  \frac{1}{e} \,\cdot\,
               \frac{\mbox{$n$-dimensionale Leitf\"ahigkeit}}
                    {\mbox{$n$-dimensionale Ladungstr\"agerdichte}}
         \,=\, \frac{1}{e \varrho_n n_{nD}}
\end{eqnarray}
Elastischer mittlerer freier Weg
\begin{eqnarray}
  & &                                               \nonumber\\
l_{3D}
  &=&   \frac{\mu h}{2e}
        \sqrt[3]{\frac{3 n_{3D} }{ \pi}}
       =203.6\,\mbox{{\rm nm}}
        \cdot\mu\cdot
        \sqrt[3]{ n_{3D} \,(10^{18}\,\mbox{{\rm cm}}^{-3})}       \\
  & &                                               \nonumber\\
l_{2D}
  &=&   \frac{\mu h}{ e}
        \sqrt   {\frac{ n_{2D} }{2\pi}}\phantom{i}
       =165.0\,\mbox{{\rm nm}}
        \cdot\mu\cdot
        \sqrt   { n_{2D} \,(10^{12}\,\mbox{{\rm cm}}^{-2})}       \\
  & &                                               \nonumber\\
l_{1D}
  &=&   \frac{\mu h}{4e}
                        n_{1D}        \phantom{xii}
       =103.4\,\mbox{{\rm nm}}
        \cdot\mu\cdot
                { n_{1D} \,(10^{ 6}\,\mbox{{\rm cm}}^{-1})}
\end{eqnarray}
{\sc Drude}-Relaxationszeit
\begin{equation}
\tau=\frac{m}{e}\cdot\mu
    =398\,\mbox{{\rm fs}}\cdot\mu
\end{equation}
\subsubsection{Hall-Messungen nach van der Pauw}
Wir setzen eine quadratische Geometrie der Probe voraus,
die wir an ihren vier Ecken kontaktieren.
Dann erhalten wir im klassischen Regime
nach {\sc van\,der\,Pauw} \cite{Pauw1958}:
\vspace*{0.5cm}
\par\noindent%
2-dimensionale Ladungstr\"agerdichte
\begin{equation}
n_{2D} = \frac{BI}{eV_H}
\end{equation}
Beweglichkeit in $2$ Raumdimensionen
\begin{equation}
\mu=\frac{1}{\varrho n_{2D} e}
   =\frac{{\rm ln}\,2}{\pi} \frac{I}{V} \frac{1}{ n_{2D} e}
   =\frac{{\rm ln}\,2}{\pi} \frac{V_H}{V} \frac{I}{ n_{2D} eV_H}
   =\frac{{\rm ln}\,2}{\pi} \frac{V_H}{\,VB}
   =0.2206\,\frac{V_H}{V\,B}
\end{equation}
Umrechnung von Laboreinheit auf SI
\begin{equation}
  \mu(\mbox{cm}^2\mbox{(Vs)}^{-1})
= 10\,000\cdot\mu(\mbox{ m}^2\mbox{(Vs)}^{-1})
= 10\,000\cdot\mu
\end{equation}
\begin{table}\vspace{0.5cm}
\begin{center}
%
\begin{tabular}{|c|c|}                                              \hline
$\mu(\mbox{cm}^2\mbox{(Vs)}^{-1})$   & Qualit\"at der Probe      \\ \hline
                                                                    \hline
$200\,000$             & immerhin                                \\ \hline
$600\,000$             & gut                                     \\ \hline
$1\,000\,000$          & sehr gut                                \\ \hline
$2\,000\,000$          & exzellent                               \\ \hline
$10\,000\,000$         & Weltspitze (1/2 Jahr ausheizen usw.)    \\ \hline
\end{tabular}
\normalsize
\end{center}
\vspace{0.5cm}
\caption{Probenqualit\"aten in Termen ihrer Beweglichkeit}
\vspace{0.75cm}\end{table}
\subsubsection{Gated 2DEG}
Mit $V_g$ als Gate-Spannung und $d$ als isolierender Dicke
ist die 2-dimensionale La\-dungs\-tr\"a\-ger\-dich\-te in GaAs
\begin{equation}
n_{2D} = \frac{\varepsilon_*V_g}{ed}
       = 72.4 \times 10^{12}\,\mbox{cm}^{-2}\,V_g(V)/d(nm)
\end{equation}
\subsubsection{Elektronen im Magnetfeld}
Zentralkraft versus {\sc Lorentz}-Kraft
\begin{equation}
\frac{mv^2}{r} = evB
\end{equation}
Zyklotronfrequenz
\begin{equation}
m \omega^2 r = e \cdot \omega r \cdot B,
\phantom{xxxx}\mbox{daraus:}\phantom{x}
\omega_c := \omega = \frac{eB}{m}
\end{equation}
klassischer magnetischer Radius
\begin{equation}
r_{class}=\frac{mv^2}{evB}
         =\frac{\sqrt{2m \cdot mv^2/2}}{eB}
         =\frac{\sqrt{2mE}}{eB}
         =28.21\,\mbox{{\rm nm}}\cdot
          \sqrt{E\,(\mbox{{\rm meV}})}\,/\,B
\end{equation}
magnetische L\"ange%
\footnote{%
Radius der Zyklotron-\loq Nullpunktsbahn\hiq\
entsprechend der Nullpunktsenergie $E=\hbar\omega_c/2$.
{\it Vorsicht:\/} Semiklassisch zu denken, kann in die
Irre f\"uhren!}
\begin{equation}
l_{B} = \frac{\sqrt{2m\hbar\omega_c/2}}{eB}
      = \sqrt{\frac{\hbar}{eB}},
\phantom{12345}
2\pi \cdot l_{B}^2 = \frac{h}{eB}
\end{equation}
Wegen $n_{2D}=\nu\cdot eB/h$ ist
\begin{equation}
\nu=\frac{ n_{2D} \,h}{eB}
   =41.36 \cdot \frac{ n_{2D} \,(10^{12}\,\mbox{{\rm cm}}^{-2})}{B}
\end{equation}
Der quantenmechanische magnetische Radius
\begin{equation}
r_{quant} =       \sqrt{\frac{h(\nu+1)}{2\pi eB}}
          =       25.656\,\mbox{{\rm nm}}\cdot
                  \sqrt{\frac{\nu+1}{B}}
\end{equation}
erf\"ullt n\"aherungsweise
\begin{equation}
r_{quant} \approx \frac{h}{eB} \,
                  \sqrt{\frac{ n_{2D} }{2\pi}}
          =       165\,\mbox{{\rm nm}}\cdot
                  \sqrt{ n_{2D}\,(10^{12}\,\mbox{{\rm cm}}^{-2})}\,/\,B
\end{equation}
Zyklotronenergie
\begin{equation}
\hbar\omega_c=\frac{\hbar eB}{m}
             =1.654\,\mbox{{\rm meV}}\cdot B
\end{equation}
\subsubsection{Verarmung (Depletion)}
Verarmungsl\"ange ({\it depletion length\/})
\begin{equation}
l_{depl}=\sqrt{\frac{2\varepsilon}{e n_{3D} }
               (V_{bi}-\frac{2kT}{e})   }
        \approx
        \frac{1.31\,\mu m}
             {\sqrt{ n_{3D} \,(10^{15}\,\mbox{{\rm cm}}^{-3})}}
\end{equation}
Das hei\ss t:
F\"ur eine geringe Dotierung von
$10^{15}\,\mbox{cm}^{-3}$
betr\"agt sie typischerweise $1\,\mu\mbox{m}$,
f\"ur eine gro\ss e Dotierung einige $nm$.
\subsection{Mesa-\"Atze f\"ur GaAs}
Zur Pr\"aparierung bzw.\ Strukturierung eines
{\sc Hall}-Streifens ({\it engl.\/} {\sc Hall} bar)
mu\ss\ man die Parameter kennen, die f\"ur das Heraus\"atzen
eines {\it Mesas\/} ({\it in Alaska:\/} Tafelberg)
entscheidend sind.
\begin{equation}
1\,\mbox{Teil}\,{\rm H}_2{\rm SO}_4\,:\,
8\,\mbox{Teile}\,{\rm H}_2{\rm O}_2\,:\,
1000\,\mbox{Teile}\,{\rm H}_2{\rm O}
\end{equation}
\"atzen
\begin{equation}
150\,{\rm nm}\,\,\,\,{\it in}\,\,\,\,3.5\,{\rm min}
\end{equation}
\subsection{Weitere n\"utzliche Formeln}
\vspace*{0.3cm}
\begin{eqnarray}
\frac{k}{k'}
&\approx& 0.737\,\log\left( \frac{0.1}{1-\frac{a}{b}} \right) + 1.384,
                                                            \nonumber\\
&=&       0.647 - 0.737\,\log\left( 1-\frac{a}{b} \right),
\phantom{1234}
0.7 < \frac{a}{b} < 1
\end{eqnarray}
\vspace*{0.3cm}
\begin{equation}
\frac{k}{k'} \approx
0.844\,\frac{a}{b} + 0.36,
\phantom{1234}
0.1 < \frac{a}{b} < 0.7
\end{equation}
\vspace*{0.3cm}
\begin{equation}
    \frac{k}{k'} \approx
- \,\frac{0.57}{\log\frac{a}{b}} =
    \frac{0.57}{\log\frac{b}{a}},
    \phantom{1234}
0 < \frac{a}{b} < 0.01
\end{equation}
%
%
\begin{table}\vspace{0.5cm}
\begin{center}
\renewcommand{\baselinestretch}{1.33}
\scriptsize
\begin{tabular}{|l|}              \hline
All quantities in SI, unless specified otherwise\\
$D_n$=$n$-dimens.density of states$/\mbox{m}^n$ incl.spin \\
$m$=effective charge carrier mass, here $0.07\, \sbm{m}{e}$\\
$\epsilon$=dielectric constant, here $13.1\, \epsilon _\circ$\\
$\sbm{V}{bi}$=built-in potential, here $\du{1.24}{V}$
    \hfill $a_\circ$=Bohr's radius\\
$l_n$=$n$-dimens.elast.mean free path
    \hfill$\tau$=momen.scatt.time\\
$N_n$=$n$-dimens. carrier density
    \hfill $E$=energy\\
$\mu$=charge carrier mobility
    \hfill $\sbm{E}{F}$=Fermi-energy\\
$\varrho_n$=$n$-dimens. resistivity
    \hfill $\lambda$=wavelength\\
$T$=temperature
    \hfill $\epsilon_\circ=\du{8.854188\times 10^{-12}}{F/m}$\\
$\sbm{l}{\scriptstyle depl}$=depletion length
    \hfill $\sbm{V}{g}$=gate voltage\\
$\sbm{r}{\scriptstyle class}$=classical magnetic length
    \hfill $B$=magnetic field\\
$\sbm{r}{\scriptstyle quant}$=quant.mech.magn.length
    \hfill $A$=area\\
$\nu$=2-dimens.spin-split filling factor
    \hfill $d$=insulat.thickness\\
$\sbm{\omega}{c}$=cyclotron frequency
    \hfill $e$=electron charge\\
$P$=black body radiation power
    \hfill $h$=Planck's quantum\\
$b$=black body coeff: 1 (black) $\ldots$ 0.05 (pol.metal)\\
$N_2=\frac{\epsilon \sbm{V}{g}}{ed}=72.4\times 10^{12}\,
    \mbox{cm}^{-2}\, \sbm{V}{g}(\mbox{V})/d\,(\mbox{nm})$\\
$          D_1=\frac{2}{h}\sqrt{\frac{2m}{E}}
    \hfill D_2=\frac{4\pi m}{h^2}
    \hfill D_3=\frac{8\pi m\sqrt{2mE}}{h^3}$\\
$E_{\mbox{F}_1}=\frac{h^2N_1^2}{32m}=\du{13.43}{meV}\,
    (N_1(\du{10^6}{cm}^{-1}))^2$\\
$E_{\mbox{F}_2}=\frac{h^2N_2}{4\pi m}=\du{34.2}{meV}\,
     N_2(\du{10^{12}}{cm}^{-2})$\\
$E_{\mbox{F}_3}=\frac{h^2}{8m}(\frac{3N_3}{\pi})^{2/3}=\du{52.1}{meV}\,
    (N_3(\du{10^{18}}{cm}^{-3}))^{2/3}$\\
$       \lambda_{\mbox{F}_1}=\frac{4}{N_1}
    \hfill \lambda_{\mbox{F}_2}=\sqrt{2\pi/N_2}
    \hfill \lambda_{\mbox{F}_3}=2\sqrt[3]{\frac{\pi}{3N_3}}$\\
$l_1=\frac{\mu h}{4e}N_1=\du{103.4}{nm}\,\mu N_1(\du{10^6}{cm}^{-1})
    \hfill \varrho_n=\frac{1}{e\mu N_n}$\\
$l_2=\frac{\mu h}{e}\sqrt{\frac{N_2}{2\pi}}=
    \du{165}{nm}\,\mu\sqrt{N_2(\du{10^{12}}{cm}^{-2})}$\\
$l_3=\frac{\mu h}{2e}\sqrt[3]{\frac{3N_3}{\pi}}=
    \du{203.6}{nm}\,
    \mu \sqrt[3]{N_3(\du{10^{18}}{cm}^{-3})}$\\
$\sbm{l}{depl}=\sqrt{\frac{2\epsilon}{eN_3}
               (\sbm{V}{bi}-\frac{2kT}{e})}\approx
               \frac{\du{1.31}{$\mu$m}}{\sqrt{N_3(\du{10^{15}}{cm}^{-3})}}$\\
$\sbm{r}{class}=\frac{\sqrt{2mE}}{eB}
               =\du{28.21}{nm}\, \sqrt{E(\mbox{meV})}/B$\\
$\sbm{r}{quant}\approx\frac{h}{eB}\sqrt{\frac{N_2}{2\pi}}
               =\du{165}{nm}\, \sqrt{N_2(\du{10^{12}}{cm}^{-2})}/B$\\
$\sbm{r}{quant}=\sqrt{\frac{h(\nu +1)}{2\pi eB}}
               =\du{25.656}{nm}\sqrt{\frac{\nu +1}{B}}$\\
$\nu =\frac{N_2h}{eB}=41.36\,\frac{N_2(\du{10^{12}}{cm}^{-2})}{B}
               \hfill \tau=\frac{m}{e}\mu=\du{398}{fs}\,\mu$\\
$\hbar\sbm{\omega}{c}=\frac{\hbar eB}{m}=\du{1.654}{meV}\,B
               \hfill a_\circ=\frac{\epsilon h^2}{2\pi e^2 m}=\du{9.90}{nm}$\\
$P(\mbox{pW})=5.67\, A(\mbox{cm}^2)b(T_1^4-T_2^4)
               \hfill \lambda_{\sbm{P}{max}}
               =\frac{\du{2898}{$\mu$m}}{T}$\\ \hline
\end{tabular}
\renewcommand{\baselinestretch}{1.33}
\normalsize
\end{center}
\vspace{0.5cm}
\caption{{\it Zum Kopieren und Ausschneiden:\/} Vorderseite der Formelkarte}
\vspace{0.75cm}\end{table}
%
%
\begin{table}\vspace{0.5cm}
\begin{center}
\renewcommand{\baselinestretch}{1.33}
\scriptsize
\begin{tabular}{|l|}              \hline
$k=\du{1.380\,658(12)\times 10^{-23}}{J/K}$\\
$h=\du{6.626\,0755(40)\times 10^{-34}}{Js}$\\
$c=\du{2.997\,924\,58\times 10^8}{m/s}$\\
$e=\du{1.602\,177\,33(49)\times 10^{-19}}{C}$\\
$\sbm{m}{e} =\du{9.109\,534(54)\times 10^{-31}}{kg}$\\
$\sbm{m}{p}=\du{1.672\,6231(10)\times 10^{-27}}{kg}$\\
$\sbm{m}{n}=\du{1.674\,9286(10)\times 10^{-27}}{kg}$\\
$\mu_\circ =\du{4\pi\times 10^{-7}}{H/m}=\du{1.256\,637\ldots\times 10^{-6}}{H/m}$\\
$\varepsilon_\circ =1/\mu_\circ c^2=\du{8.854\,1878\ldots\times 10^{-12}}{F/m}$\\
$\du{1}{atm}=\du{1.01325\times 10^5}{N/m}^2$\\
$\du{1}{Torr}=\du{(1/760)}{atm}=\du{133.322\ldots}{N/m}^2$\\
$g=\du{9.80665}{m/s}^2$\\
$G =\du{6.672\,59(85)\times 10^{-11}}{m}^3\mbox{kg}^{-1}\mbox{s}^{-2}$\\
$\sbm{N}{\scriptstyle Avo}=\du{6.022\,1367(36)\times 10^{23}}{mol}^{-1}$\\
$\sigma =\du{5.670\,51(19)\times 10^{-8}}{W}\mbox{m}^{-2}\mbox{K}^{-4}$\\
$E_n=\frac{h^2n^2}{8ma^2}$\\
$N_2=\frac{BI}{eV_H}$\\
$\mu=\frac{\ln2\,V_H}{\pi\,VB}=0.2206\frac{V_H}{VB}$\\
\phantom{%
$l_n$=$n$-dimens.elast.mean free path
    \hfill$\tau$=momen.scatt.time}\\[182pt]
\hspace*{\fill}A.D.\,Wieck\,/\,D.K.\,de\,Vries\,/\,R.D.T.\,1992-1997\\
\hline
\end{tabular}
\renewcommand{\baselinestretch}{1.00}
\normalsize
\end{center}
\vspace{0.5cm}
\caption{{\it Zum Kopieren und Ausschneiden:\/} R\"uckseite der Formelkarte}
\vspace{0.75cm}\end{table}
%
%
\vfill\eject\noindent%
\end{document}